\documentclass[article,aps,twocolumn,showpacs,amsmath,amssymb,superscriptaddress,nofootinbib,floatfix]{revtex4-2}

\usepackage{graphicx}
\usepackage{dcolumn}
\usepackage{bm}
\usepackage[colorlinks=true,linkcolor=blue,citecolor=blue,urlcolor=blue]{hyperref}

\usepackage{array,booktabs,tabularx}
\usepackage[caption=false]{subfig}
\usepackage[USenglish]{babel}
\usepackage{braket}
\usepackage{xcolor}
\usepackage{amsfonts}
\usepackage{amssymb}
\usepackage[normalem]{ulem}
\usepackage{amsmath}
\usepackage{cleveref}

\crefname{equation}{Eq.}{Eqs.}
\crefname{figure}{Fig.}{Figs.}

\begin{document}

\title{Frustrated charge density wave and quasi-long-range bond-orientational order in the magnetic kagome FeGe}

\author{D. Subires}
\affiliation{Donostia International Physics Center (DIPC), Paseo Manuel de Lardizábal. 20018, San Sebastián, Spain}
\affiliation{University of the Basque Country (UPV/EHU), Basque Country, Bilbao, 48080 Spain}

\author{A. Kar}
\affiliation{Donostia International Physics Center (DIPC), Paseo Manuel de Lardizábal. 20018, San Sebastián, Spain}

\author{A. Korshunov}
\affiliation{Donostia International Physics Center (DIPC), Paseo Manuel de Lardizábal. 20018, San Sebastián, Spain}

\author{C. A. Fuller}
\affiliation{Swiss-Norwegian BeamLines at European Synchrotron Radiation Facility}

\author{Y. Jiang}
\affiliation{Donostia International Physics Center (DIPC), Paseo Manuel de Lardizábal. 20018, San Sebastián, Spain}

\author{H. Hu}
\affiliation{Donostia International Physics Center (DIPC), Paseo Manuel de Lardizábal. 20018, San Sebastián, Spain}

\author{D. C\u{a}lug\u{a}ru}
\affiliation{Department of Physics, Princeton University, Princeton, NJ 08544, USA}

\author{C. McMonagle}
\affiliation{Swiss-Norwegian BeamLines at European Synchrotron Radiation Facility}

\author{C. Yi}
\affiliation{Max Planck Institute for Chemical Physics of Solids, 01187 Dresden, Germany}

\author{ S. Roychowdhury}
\affiliation{Max Planck Institute for Chemical Physics of Solids, 01187 Dresden, Germany}
\affiliation{Department of Chemistry, Indian Institute of Science Education and Research Bhopal, Bhopal-462 066, India}

\author{C. Shekhar}
\affiliation{Max Planck Institute for Chemical Physics of Solids, 01187 Dresden, Germany}

\author{J. Strempfer}
\affiliation{Advanced Photon Source, Argonne National Laboratory, Lemont, IL 60439} 

\author{A. Jana}
\affiliation{CNR-Istituto Officina dei Materiali (CNR-IOM), Strada Statale 14, km 163.5, 34149 Trieste, Italy}
\affiliation{International Center for Theoretical Physics (ICTP), Strada Costiera 11, 34151 Trieste, Italy}

\author{I. Vobornik}
\affiliation{CNR-Istituto Officina dei Materiali (CNR-IOM), Strada Statale 14, km 163.5, 34149 Trieste, Italy}

\author{J. Dai}
\affiliation{ALBA Synchrotron Light Source, 08290 Barcelona, Spain}

\author{M. Tallarida}
\affiliation{ALBA Synchrotron Light Source, 08290 Barcelona, Spain}

\author{D. Chernyshov}
\affiliation{Swiss-Norwegian BeamLines at European Synchrotron Radiation Facility}

\author{A. Bosak}
\affiliation{European Synchrotron Radiation Facility (ESRF), BP 220, F-38043 Grenoble Cedex, France}

\author{C. Felser}
\affiliation{Max Planck Institute for Chemical Physics of Solids, 01187 Dresden, Germany}

\author{B. Andrei Bernevig}
\affiliation{Donostia International Physics Center (DIPC), Paseo Manuel de Lardizábal. 20018, San Sebastián, Spain}
\affiliation{Department of Physics, Princeton University, Princeton, NJ 08544, USA}
\affiliation{IKERBASQUE, Basque Foundation for Science, 48013 Bilbao, Spain}

\author{S. Blanco-Canosa}
\email{sblanco@dipc.org}
\affiliation{Donostia International Physics Center (DIPC), Paseo Manuel de Lardizábal. 20018, San Sebastián, Spain}
\affiliation{IKERBASQUE, Basque Foundation for Science, 48013 Bilbao, Spain}

\date{June 2024}

\begin{abstract}
The intrinsic frustrated nature of a kagome lattice is amenable to the realization of exotic phases of matter, such as quantum spin liquids or spin ices, and more recently the multiple-$\mathrm{\textbf{q}}$ charge density waves (CDW) in the kagome metals. Despite intense efforts to understand the mechanism driving the electronic modulations, its origin is still unknown and hindered by competing interactions and intertwined orders. Here, we identify a dimerization-driven 2D hexagonal charge-diffuse precursor in the antiferromagnetic kagome metal FeGe and demonstrate that the fraction of dimerized/undimerized states is the relevant order parameter of the multiple-$\mathrm{\textbf{q}}$ CDW of a continuous phase transition. The pretransitional charge fluctuations with propagation vector $\mathrm{\textbf{q}=\textbf{q}_M}$ at T$_{\mathrm{CDW}}$$<$T$<$T$^*$(125 K) are anisotropic, hence holding a quasi-long-range bond-orientational order. The broken translational symmetry emerges from the anisotropic diffuse precursor, akin to the Ising scenario of antiferromagnetic triangular lattices. The temperature and momentum dependence of the critical scattering show parallels to the stacked hexatic $\mathrm{B}$-phases reported in liquid crystals and transient states of CDWs and highlight the key role of the topological defect-mediated melting of the CDW in FeGe.

\end{abstract}
\maketitle
The ground state of the strongly degenerated frustrated lattices is a fertile ground for emergent phenomena driven by the competing interactions \cite{Moessner_2006,Savary_2017}. For instance, the magnetic ground state and the long-range order of a frustrated network of spins are often a consequence of a subtle balance among the second (or higher order) nearest-neighbor and spatially anisotropic interactions \cite{Metcalf_1974,Jiang_2006}, spin-orbit coupling \cite{Iaconis_2018}, defects and disorder \cite{Houtappel_1950,Shokef_2011}. In strongly correlated electron systems with high degree of frustration, Coulomb repulsion introduces interactions between spin, charge and orbital degrees of freedom, providing grounds for the study of competing intertwined orders \cite{Weber_2006,Tsunetsugu_2009}.    

Of particular interest is the phase transition in 2D triangular antiferromagnetic lattices \cite{Bernu_1992,Capriotti_1999}, where spins are aligned $120^\circ$ from each other in the basal plane, that unveils unconventional correlated diffuse patterns characteristic of frustrated magnetism \cite{Janas_2021,Tosic_2023}, Kosterlitz-Thouless phases \cite{Sato_2003} or spin ices \cite{Castelnovo_2008}. In the charge sector, the Kosterlitz-Thouless-Halperin-Nelson-Young (KTHNY) theory predicts that a 2D  phase transition is topological \cite{Kosterlitz_1973,Halperin_1978,Nelson_1979}, described by the continuous unbinding of topological defects, and the transition from an ordered solid to an isotropic liquid is commonly preceded by an intermediate state characterized by short-range positional but quasi-long-range bond orientational (BO) order \cite{Birgeneau_1986,Dai_1992,Aharony_1986}.

The kagome lattice, a geometrically frustrated fabric of corner-sharing triangles \cite{Wilson2024}, has recently emerged as a platform to study the phase transition from an electronic crystal (charge density wave, CDW) to an isotropic liquid. Due to the particular geometry of the kagome net that features van Hove (VHS) singularities, Dirac cones and dispersionless flat bands \cite{Meier_2020}, theory proposed the appearance of many body phases, allowing for the observation of anomalous and fractional Hall effect \cite{Liu2018,Yu2021,Yang2020,Fujishiro2021}, chiral CDWs \cite{Guo2022,Mielke2022,Jiang2021}, superconductivity \cite{Neupert2022,Yin2022}, loop currents \cite{Denner_2021,Li2024,Dong2023,Christensen2022,Tazai2023} and heavy fermion physics \cite{Ronny_PRB_2012,Ronny_PRL_2013,Huang2020,Kourris2023}. At particular filling fractions, the Fermi surface is perfectly nested by a wavevector $\mathrm{q_M}$=($\frac{1}{2}$ 0), resulting in a 2$\times$2 CDW. Examples of multiple-$\mathrm{\textbf{q}}$ CDW orders have been observed in the AV$_3$Sb$_5$ \cite{Ortiz_2019,Ortiz_2020} (A=K, Rb, Cs; hereafter AVS) and ScV$_6$Sn$_6$ (hereafter SVS) \cite{Arachchige_2022} series of the kagome family. In the weakly correlated AVS \cite{Tan_2021,Subedi_2022}, the first order phase transition is achieved without phonon softening \cite{Miao_2021}, pointing to a prominent role of order-disorder scenarios \cite{Ratcliff_2021,Subires_2023}. On the other hand, the ground state of SVS displays a different lattice landscape, with the collapse of a high temperature soft mode at $\mathrm{\textbf{q}^*}$=($\frac{1}{3}\ \frac{1}{3}\ \frac{1}{2}$) \cite{Korshunov_2023,Hu_2023} that competes with the low temperature ordered phase at $\mathrm{\textbf{q}_{CDW}}$=($\frac{1}{3}\ \frac{1}{3}\ \frac{1}{3}$) \cite{Cao_2023,Tan_2023,Lee_2023,Kang_2023,Cheng_2023,Tuniz_2023}. 

\begin{figure*}
    \centering
    \includegraphics[width=1\linewidth]{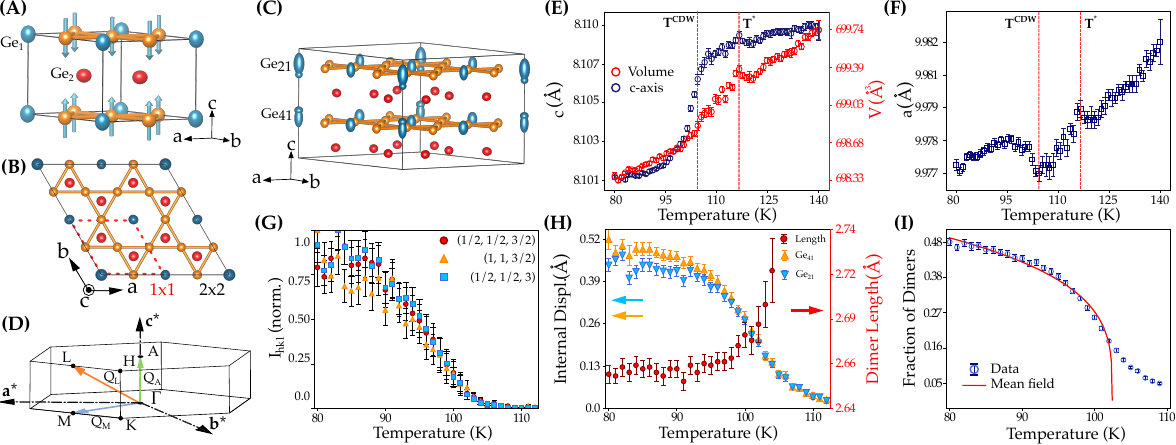}
    \caption{\textbf{x-ray diffraction and analysis of the order parameter.} (\textbf{A}) Normal state structure (non-CDW) of FeGe and the spin polarization of the kagome planes. Orange symbols are the Fe atoms, the blue symbols represent the Ge atoms in the kagome plane (trigonal Ge$_1$) and red symbols are the Ge atoms in the honeycomb layer (Ge$_2$). Arrows stand for the spin-up and spin-down in the Fe sites. (\textbf{B}) Top view of the dimerized CDW structure. (\textbf{C}) Dimerized CDW structure obtained from the hard x-ray refinement, highlighting the dimerization of the trigonal Ge$_1$ (here Ge$_{21}$ and Ge$_{41}$ move in opposite ways). The oval shape of Ge$_{21}$ and Ge$_{41}$ stand for their average site occupancy. (\textbf{D}) High symmetry points in the non-magnetic Brillouin zone. (\textbf{E}) Temperature dependence of the \textit{c}-axis lattice parameter and volume, V. (\textbf{F}) Temperature dependence of the \textit{a} lattice parameter, highlighting the anomalies at T$^*$ and T$_{\mathrm{CDW}}$. (\textbf{G}) Normalized temperature dependence of a set of CDW reflections. (\textbf{H}) Temperature dependence of the internal displacements of the trigonal Ge (Ge$_{21}$ and Ge$_{41}$) in the dimerized phase, showing a continuous 2$^{\mathrm{nd}}$ order-like transition. (\textbf{I}) Temperature dependence of the dimer length (Ge$_{21}$-Ge$_{41}$ distance) in the dimerized phase. (\textbf{J}) Definition of fraction of dimers (\textit{fd}) as the relevant order parameter and its fitting to a mean field power law, yielding a T$_{\mathrm{CDW}}\sim$102 K. The departure from the mean field at T$>$T$_\mathrm{CDW}$ is a consequence of the charge density fluctuations.}
    \label{Fig1}
\end{figure*}

A different scenario is devised in the antiferromagnetic FeGe, holding the same lattice symmetry as AVS and SVS. Whereas AVS and SVS are non magnetic \cite{Kenney_2021}, FeGe orders antiferromagnetically (AFM) below $\sim$400 K, with the magnetic moments polarized along the \textit{c}-axis within each kagome layer and antiferromagnetically between planes (A-type antiferromagnetic order) \cite{Teng_2022,Yin_2022}. A multiple-$\mathrm{\textbf{q}}$ CDW strongly intertwined with the magnetic order develops below $\sim$100 K with propagation vectors connecting the VHS at $\mathrm{M}$, $\mathrm{L}$ and the AFM $\mathrm{A}$ points of the BZ, emphasizing the complex entanglement between charge, spin and lattice degrees of freedom \cite{Chen_Yi_2023,Teng_2024,jiang2023kagome}. The phase transition does not involve a phonon collapse at either $\mathrm{M}$ or $\mathrm{L}$ at T$>$T$_\mathrm{CDW}$, but a sizable spin-phonon coupling of the low energy mode at $\mathrm{A}$ \cite{Miao_2023,teng2024spinchargelattice} and a moderate hardening of an optical mode below T$_\mathrm{CDW}$ at $\mathrm{M}$ \cite{Teng_2022}. On the other hand, several angle resolved photoemission spectroscopy (ARPES) experiments \cite{Teng_2023,Oh2024}, not yet reproduced by density functional theory (DFT), highlighted the important role of both the orbital dependent saddle points close to the Fermi level \cite{Teng_2023} and trigonal Ge (Ge$_1$ in figure \ref{Fig1}(C)) dimerization \cite{Wang_2023,Zhao_2023,Miao_2023,Teng_2023}. This debate has been further fueled by DFT calculations that pointed out a divergence of the electronic susceptibility, which correlates with the nesting function, at the $\mathrm{K}$ point of the BZ \cite{Wu2023}. X-ray diffraction reported a dimerization of Ge$_1$ across the CDW \cite{Shi_2023} that is supported by the partial softening of the flat \textit{k$_z$}=$\pi$ plane in the DFT+U calculations \cite{Miao_2023,Chen_2023}. Furthermore, neutron and Raman scattering \cite{Wu_2023} found an enhancement of the crystalline symmetry upon cooling through the CDW transition and describe the phase transition as an interplay between the $\mathrm{L}$ and $\mathrm{A}$ order parameters. Recent scanning tunneling microscopy (STM) experiments locates the CDW in the strong coupling regime \cite{Chen_2023}, where phase fluctuations destroy the long range charge order and order-disorder scenarios might play a prominent role. However, following symmetry arguments, the dimerized scenario is also consistent with the presence of a primary order parameter at $\mathrm{M}$ and $\mathrm{L}$ and the nature of the multiple-$\mathrm{\textbf{q}}$ CDW and its dynamics is far from being settled. 

Here, we use x-ray diffraction, diffuse scattering (DS), ARPES, DFT and Monte Carlo simulations to solve the symmetry and the electronic band structure of the low temperature CDW phase of FeGe and to reveal a high temperature quasi-2D hexagonal diffuse precursor localized along the $\mathrm{M-L}$ direction. The dimerization-induced lattice frustration demonstrates that the fraction of dimers can be considered as the relevant order parameter of the continuous phase transition. At intermediate temperature, $\mathrm{T_{CDW}}<\mathrm{T}<\mathrm{T^{*}(125\,K)}$, we identify a state where the critical scattering is anisotropic around the $\mathrm{M}$ point, showing short-range positional but quasi-long-range bond orientational order, akin to the stacked \textit{hexatic} phases observed in lyotropic liquid crystals \cite{Huang_2006}. Our results suggest that the phase transition fits within an order-disorder scenario captured by the Ising model of triangular lattices \cite{Wannier_1950} and the melting of the CDW is driven by the unbinding of topological defects, such as dislocation pairs and shear of domain walls \cite{Halperin_1978,Mermin_1979,Brinkman_1982}.

\section{Results}
\subsection{x-ray diffraction}

Hexagonal FeGe (\textit{P6/mmm}, space group No=191) consists of individual FeGe kagome layers within the unit cell, with trigonal Ge$_1$, separated by honeycomb Ge$_2$ atoms (Figure \ref{Fig1}(A-B)) \cite{Kang_2020}. 
Figures \ref{Fig1}(E-F) show the temperature dependence of the unit cell parameters obtained from the refinement of the x-ray diffraction patterns (supplementary information S2). The unit cell volume shrinks $\sim$0.2\% between 140 and 80 K and identifies two critical temperatures; T$^*$ $\sim$125 K, more clearly visible in the thermal evolution of in-plane lattice parameter \textit{a}, and the CDW transition at T$_\mathrm{CDW}\sim$105 K (figure \ref{Fig1}(F)). The change in volume shows a gradual crossover between two phases and is mostly driven by shortening of the \textit{c}-axis lattice parameter ($\sim$0.12\%) (figure \ref{Fig1}(E)) that smoothly varies through the transition, while the in-plane lattice parameters only undergo small structural variations at T$^*$ and T$_\mathrm{CDW}$, figure \ref{Fig1}(F). The shortening of the \textit{c}-axis is consistent with a dimerization of trigonal Ge in the kagome plane, as previously reported on a basis of magnetic energy saving and x-ray diffraction \cite{Miao_2023,Shi_2023,Wang_2023}.  

With further cooling below T$_\mathrm{CDW}$, a multiple-\textbf{q} CDW develops with propagation vectors \textbf{q}$_\mathrm{M}$=($\frac{1}{2}\ 0\ 0$), \textbf{q}$_\mathrm{L}$=($\frac{1}{2}\ 0\ \frac{1}{2}$) and \textbf{q}$_\mathrm{A}$=($0\ 0\ \frac{1}{2}$). The temperature dependence of several normalized CDW peaks is summarized in figure \ref{Fig1}(G). The rather linear T-dependence down to 80 K evidences that the phase transition cannot be properly identified as first-order. This growth in intensity following a continuous phase transition is also consistent with the small release of entropy measured in the specific heat \cite{Shi_2023,Teng_2022} (supplementary figure S1).
We have indexed the low-temperature CDW phase within the non-centrosymmetric \textit{P6mm} space group with a partial dimerization of the trigonal Ge$_1$, resulting in a disordered composite final structure - an overlap of dimerized and undimerized regions \cite{Shi_2023}. We also point out that the low temperature CDW can also be indexed within the high temperature centrosymmetric space group \textit{P6/mmm} with similar figure of merit (\textit{R$_1$}) (supplementary information S2), but, however, non-centrosymmetric space groups were inferred to explain the double cone magnetic transition at low temperature \cite{Zhou2023}. 
At T=80 K, we find that the dimerized trigonal Ge atoms (labelled as Ge$_{21}$ and Ge$_{41}$ in figure \ref{Fig1}(C)) are lifted $\sim\pm$0.5 \r{A} from the kagome plane, figure \ref{Fig1}(H), in agreement with the DFT calculations \cite{Wang_2023} and previous diffraction studies \cite{Miao_2023,Shi_2023}, and a dimerization fraction (the occupancy ratio of Ge$_1$ in a dimerized and undimerized position) of 50\%, larger than the previously reported. Figure \ref{Fig1}(H) displays the dimer length (Ge$_{21}$ and Ge$_{41}$ distances between adjacent kagome planes) that smoothly decreases below T$_\mathrm{CDW}$ and reaches a constant value of 2.66 \r{A} below 95 K. 
Following these experimental evidences, we define the fraction of dimers (\textit{fd}), given by the occupancy of corresponding Ge$_1$ sites, as a relevant order parameter of the continuous phase transition, figure \ref{Fig1}(I). The gradual growth of \textit{fd} can be fitted to a mean field behavior, returning a critical temperature T$_\mathrm{CDW}\sim$102 K and a long tail of critical fluctuations, characteristic of a reduced dimensionality. The presence of short-range
charge correlations at T$>$T$_\mathrm{CDW}$ is reminiscent to the magnetic critical scattering in Ho thin films undergoing a
dimensionality crossover \cite{Ott_2004}. 

\subsection{ARPES and DFT calculations}

\begin{figure*}
    \centering
    \includegraphics[width=0.9\linewidth]{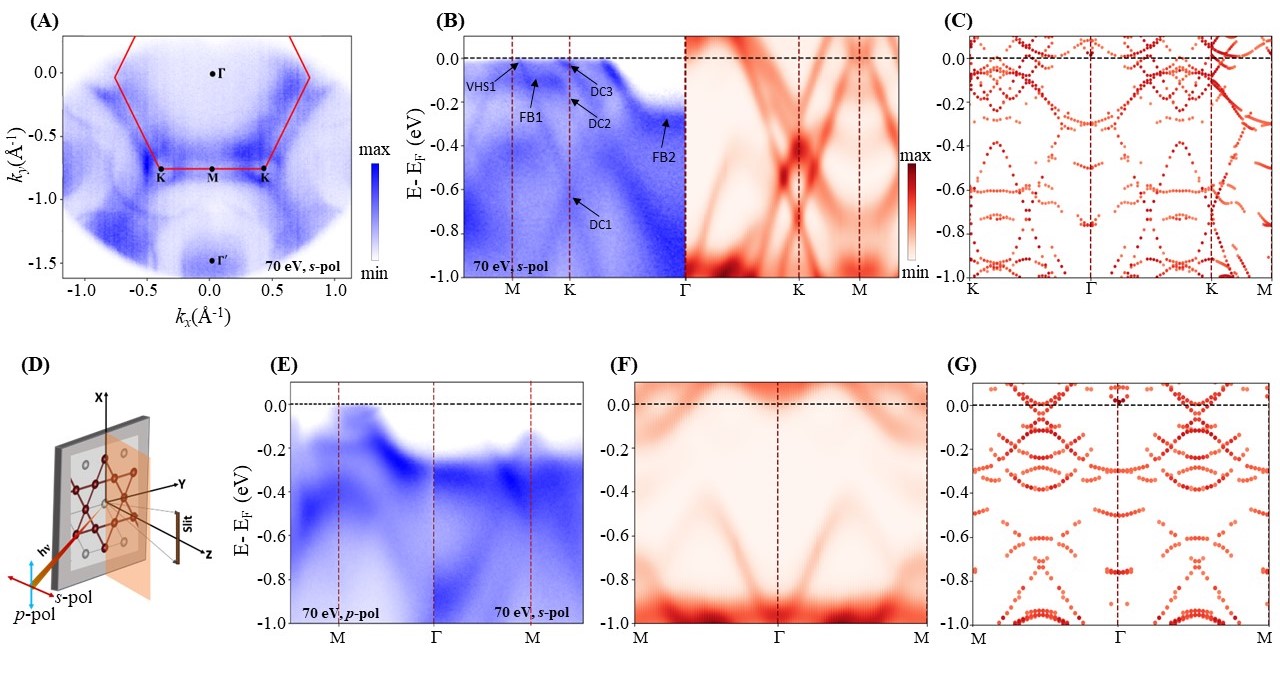}
    \caption{\textbf{ARPES and DFT calculations.}(\textbf{A}) Fermi surface map taken with E$_\mathrm{in}$=70 eV (k$_\mathrm{z}$=0) and $s$-polarized ($s$-pol) light at T=10 K. (\textbf{B}) A comparative ARPES (blue) and DFT calculated bulk band structure (red) in CDW phase along $\Gamma-\mathrm{K-M}$ symmetry direction (calculated bulk band structures are Gaussian broadened with a broadening parameter of 0.06 eV). The ARPES spectra was taken with 70 eV photon energy with $s$-pol light. (`DC', `VHS', and `FB' represent the Dirac cone, Van Hove singularity, and flat band, respectively). (\textbf{C}) DFT calculated folded surface band structure of FeGe along $\mathrm{K}-\Gamma-\mathrm{K-M}$ with honeycomb surface termination in the CDW phase. (\textbf{D}) ARPES experimental geometry with the $s$-pol and $p$-pol light vectors. (\textbf{E}) $\mathrm{M}-\Gamma-\mathrm{M}$ valence band spectra taken with 70 eV and $p$-pol and $s$-pol light (left and right panel, respectively). (\textbf{F}) Bulk calculated band structure along $\mathrm{M}-\Gamma-\mathrm{M}$ with CDW phase. (\textbf{G}) $\mathrm{M}-\Gamma-\mathrm{M}$ honeycomb surface terminated folded band structure of FeGe in CDW phase.
    }
    \label{Fig2}
\end{figure*}

Having structurally characterized the FeGe crystals, we now move on to its electronic structure. In figure \ref{Fig2}(A), we show the Fermi map of FeGe obtained for \textit{k}$_z$=0 ($\mathrm{E_i}$=70 eV, T=10 K) that partially covers both the first and second Brillouin zones. The band structure is in agreement with the previous experimental reports \cite{Oh2024,Teng_2022,Teng_2023,Zhao_2023}, showing a hexagonal Fermi surface typical of the kagome metals \cite{Kang_2020,Kang_2022}.
Our first observation is the sizable photoemission matrix element effects in different momentum spaces, highlighted by the surface state pocket emerging at $\Gamma$´ of the neighboring BZs (figure \ref{Fig2}(A)). The constant energy contours have hexagonal symmetry with rounded triangular electron pockets surrounded by a larger circular hole pocket at each $\mathrm{K}$ point of the BZ. The triangular pockets result in Dirac crossings (DC3) around the Fermi level, while the circular hole pocket corresponds to van Hove singularity (VHS1), as labeled in figure \ref{Fig2}(B) and supplementary figures S3 and S4. The DFT orbital projections of the bands (supplementary information S8 and S9) show that the Fermi surface presents mainly Fe-3$\mathrm{d}$ character and reproduces the quasi-flat bands, VHS, and Dirac points close to the Fermi level \cite{Yi_2023,Wu2023,Teng_2023,Teng_2024}. The dimerized trigonal  \textit{p}$_z$ orbital of Ge$_1$, which mainly contributes to the electron pocket at the Brillouin zone center $\overline{\Gamma}$ in the normal state, is pushed down to 0.5-1.0 eV below the Fermi level in the CDW state (supplementary information S10 and S11) and is not visible in Fermi surface contour. 

Focusing on the \textit{k}$_z$=0 plane, in figure \ref{Fig2}(B, E), we plot the energy-momentum band dispersion of FeGe along $\Gamma$-K-M and $\Gamma$-M high symmetry directions, respectively. In Fig. \ref{Fig2} (B) (left panel), three different Dirac crossings (DC1, DC2, and DC3) are identified and located at -0.05, -0.15, and -0.65 eV below the Fermi level at the corner of the BZ. The measured Dirac velocity for DC1 is $\nu_{DC1}$= 2.7 $\times$10$^{5}$ m/s, which is comparable to AVS \cite{Kang_2022}. A flat band at 0.12 eV below the Fermi level appears along the $\mathrm{K-M}$ direction (FB1), consistent with the AFM spin majority band reported by DFT calculations \cite{Teng_2023}. Another \textit{flattish} band behavior was observed around the $\Gamma$ point along the $\Gamma$-K symmetry direction, 0.28 eV below E$_\mathrm{F}$ (FB2), assumed to be responsible for the correlated electronic properties in FeGe \cite{Yi_2023}. In Fig. \ref{Fig2} (E), a comparative ARPES spectrum taken with $p$-polarized ($p$-pol, left panel) and $s$-polarized ($s$-pol, right panel) light along $\Gamma$-M high symmetry direction is shown for \textit{k}$_z$=0 plane. The U-shaped band around $\Gamma$ is visible for $p$-pol light, indicating an in-plane orbital contribution to the band structure. Additionally, the `$\wedge$' shaped band observed at M the point, 0.1 eV below E$_F$ (Fig. \ref{Fig2}(E), right panel) is visible for $s$-pol light and is attributed to the $\mathrm{d_{xz}}$ and $\mathrm{d_{yz}}$ orbitals (supplementary information S10). The valence band along the $\Gamma$-M direction, the two VHS were identified at $\mathrm{M}$ close to the Fermi level, more clearly visible along the $\mathrm{K-M-K}$ direction using circularly polarized light (supplementary information S3). Moreover, our high-resolution photoemission data do not find any signatures of either a gap opening at the $\mathrm{E_F}$, around the $\mathrm{M}$ and $\mathrm{L}$ points \cite{Teng_2023} or an electron pocket at $\mathrm{A}$ \cite{Zhao_2023}. Overall, the kagome features are consistent with the previous ARPES experiments but its comparison with the \textit{ab initio} DFT simulations had relied on individual renormalization factors of the electronic band dispersion \cite{Teng_2022,Teng_2023}, which were not explained or derived by DFT calculations. 

Aided with the \textit{P6mm} symmetry of the CDW state, figure \ref{Fig2} (and supplementary information S3 and S4) compares the energy dispersion of FeGe bands for k$_z$=0 plane with the DFT calculations considering the bulk (figures \ref{Fig2} (B right panel) and (F)) and folded CDW surface bands (figures \ref{Fig2}(C, G)) with honeycomb termination along the $\Gamma$-K-M and $\Gamma$-M high symmetry directions, respectively. From figure \ref{Fig2}, we observe that the ARPES band structure of FeGe is an admixture of the bulk and surface states. Without any adjustable parameter nor the inclusion of the Hubbard term (U), the band dispersion along the $\Gamma$-K-M direction agrees with the DFT simulations, both in electron velocity and bandwidth, resulting in Dirac crossings (DC1) mainly deriving from the $\mathrm{d_{xy}}-\mathrm{d_{x^2-y^2}}$ orbitals of Fe (Fig. \ref{Fig2} (B) and supplementary figures S10 and S13). The V-shaped band observed at -1 eV binding energy at $\Gamma$, matched by a CDW bulk band, originates mainly from the $\mathrm{p_z}$ orbitals of trigonal Ge$_1$ that hybridize with the $\mathrm{d_{xz}}-\mathrm{d_{yz}}$ orbitals of Fe and is dragged down from -0.5 to -1 eV in the CDW phase. Furthermore, the U-shaped band at -0.28 eV at $\Gamma$ is attributed to CDW honeycomb surface folded bands ($\mathrm{d_{xz}}-\mathrm{d_{yz}}$ orbitals of Fe) with contribution from the CDW bulk bands ($\mathrm{d_{xz}}-\mathrm{d_{yz}}$ orbital character of Fe) on k$_z$=$\pi$ plane after folding, i.e., folded from L ($\frac{1}{2}$ 0 $\frac{1}{2}$) (supplementary information S5 and S10). Therefore, surface U-shaped bands can also be seen as k$_z$-projected bulk bands with some surface reconstructions. Moreover, the `$\wedge$' shaped band 0.1 eV below E$_\mathrm{F}$ observed at M, figure \ref{Fig2}(E), is a result of a combination of bulk (close to $\Gamma$) and surface bands (close to M).  
The larger agreement between the experimental and \textit{ab initio} band structure and the complete orbital description of the kagome bands, without the need of any renormalization factor, downplay the correlation effects to describe the electronic band structure of FeGe.

\subsection{Diffuse scattering and Monte Carlo simulations}

With the crystal and electronic structure of CDW-FeGe solved in detail, we now pay all attention to the role of the $\mathrm{M}$, $\mathrm{L}$ and $\mathrm{A}$ high symmetry points of the BZ in the formation of the charge density wave. In particular, we aim at searching for diffuse signals at T $>$T$_\mathrm{CDW}$ that are characteristic fingerprints of local (short-range) pretransitional fluctuations of a CDW phase transition, crucial to identify the leading instabilities.        

\begin{figure*}
    \centering
    \includegraphics[width=0.85\linewidth]{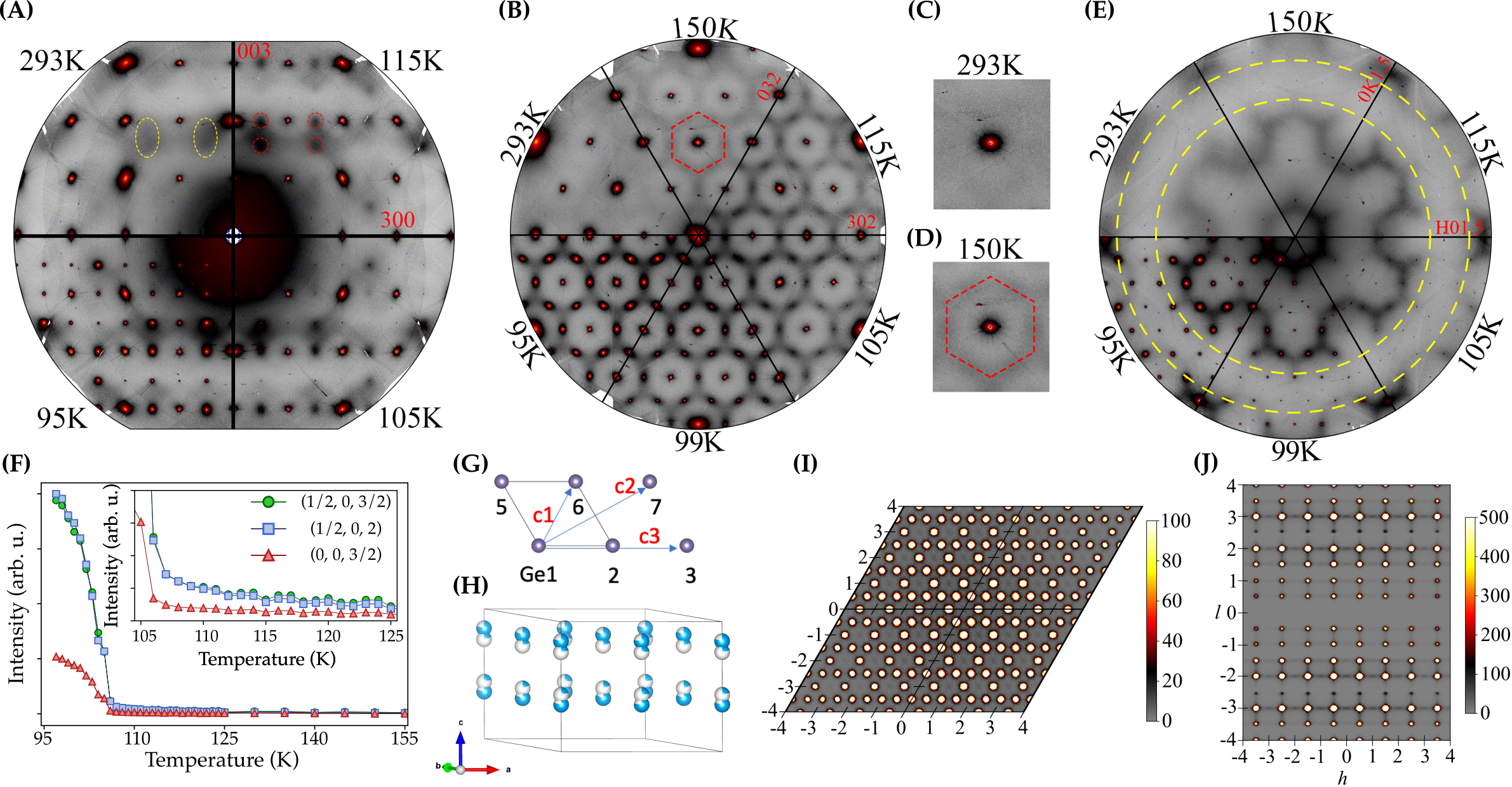}
    \caption{\textbf{Diffuse scattering and Monte Carlo modeling}. (\textbf{A}) (h 0 l) diffuse maps at different temperatures. The yellow ellipse at 300 K highlights the DS along the M-L direction and demonstrates the uncorrelated charge scattering between kagome layers. No charge precursor is detected at the A point. (\textbf{B}) (h k 2) diffuse maps as a function of temperature. The red hexagon underlines the 2D hexagonal shape of the DS. At 115 K, the anisotropic charge scattering concentrates at the M points (blue ellipse). (\textbf{C})-(\textbf{D}) Zoom-in of the hexagonal DS at 300 K and 150 K. (\textbf{E}) (h k\ $\frac{3}{2}$) diffuse maps. The absence of scattering enclosed in between the yellow circumferences is a result of the small in-plane Fe displacements. (\textbf{F}) Temperature dependence of the DS as a function of temperature obtained from the integration of a ROI defined as red circles in (\textbf{A}). (\textbf{G}) Schematics of the model used for the MC simulations. \textit{c}$_{i's}$ describe the interaction energies taken into account in the Ising Hamiltonian. \textit{c}$_4$ stands for the out-of-plane nearest-neighbor interaction. (\textbf{H}) Real space configuration of the average structure of the Ge$_1$ included in the MC simulation. (\textbf{I}-\textbf{J}) (h k 2) and (h 0 l) DS maps obtained from MC, respectively, indexed according to the room temperature disordered unit cell.
    }
    \label{Fig3}
\end{figure*}

First, let us focus on the ($\mathrm{h \ 0 \ l}$) plane of the kagome lattice, whose temperature dependence diffuse map is displayed in figure \ref{Fig3}(A). Clear diffuse clouds, already visible at room temperature (RT), are present at $\mathrm{h}\pm\frac{1}{2}$ and $\mathrm{l}\pm\frac{1}{2}$ for h and l integer. The diffuse scattering (DS) develops sizable intensity around the l=2 region of the reciprocal space, with a rod-like diffuse intensity along the M-L direction (yellow ellipses in figure \ref{Fig3}A), resulting in a stack of nearly uncorrelated kagome layers and a very short-range 2$\times$2 order. 
The diffuse intensities are strongly modulated along the l-direction, due to the negative correlation of the atomic displacements along the \textit{c}-axis, presumably associated to the Ge$_1$ dimerization. The diffuse precursor is nearly temperature independent upon cooling down to T$^*$=125 K. Below T$^*$, the spectral weight starts to accumulate at the $\mathrm{L}$ point, namely at ($\frac{1}{2}\ 0 \ \frac{3}{2}$) and ($\frac{3}{2} \ 0 \ \frac{3}{2}$), and smoothly increases, diverging at 105 K, in agreement with the transition temperature observed by x-ray diffraction in Fig. \ref{Fig1}(F). The diffuse precursor is completely absent in the in-plane polarized AFM FeSn \cite{Kang_2019} (supplementary figure S25), demonstrating that the origin of the correlated rod-like diffuse signals is a result of the dimerization-driven short-range charge fluctuations arising from the out-of-plane AFM-coupled spins in a triangular lattice.

The charge precursor also localizes at the the $\mathrm{M}$ point below $\sim$125 K, following the same temperature dependence as at the $\mathrm{L}$ point. The DS down to 105 K, fitted to a Lorentzian functions, returns a value of the out-of-plane correlation length of less than one unit cell for both M and L. 
Remarkably, no DS is detected at (0 0 $\frac{1}{2}$) ($\mathrm{A}$ point in the BZ) from RT down to T$_\mathrm{CDW}$. Following the Ge$_1$ dimerization and the spin-phonon coupling mechanisms reported experimentally and theoretically \cite{Miao_2023,Wang_2023} as the main driving force for the CDW formation, the absence of any pretransitional scattering at the $\mathrm{A}$ point of the BZ seems to contradict those scenarios. On general grounds, one would expect an enhancement of the charge density correlation function driven by a progressive magnetostriction-driven dimerization of the kagome planes that eventually collapses in a superlattice reflection at T$_\mathrm{CDW}$, as observed experimentally in spin-Peierls compounds \cite{Schoeffel1996,Gama1993}.    
With further cooling, CDWs with propagation vectors (0 0 $\frac{1}{2}$) (A), ($\frac{1}{2}$ 0 0) (M) and ($\frac{1}{2}$ 0 $\frac{1}{2}$) (L) appear at T$_\mathrm{CDW}$=105 K with an in- (out-of) plane correlation lengths of $\mathrm{A}= 40\pm3\,\mathrm{\AA}$ ($40\pm\,\mathrm{\AA}$), $\mathrm{M}= 43\pm2\,\mathrm{\AA}$ ($41\pm2\,\mathrm{\AA}$) and $\mathrm{L}= 40\pm3\,\mathrm{\AA}$ ($38\pm2\,\mathrm{\AA}$), respectively, in good agreement with the values reported in the literature \cite{Chen_2023}.

Next, we move our attention to the temperature dependence diffuse maps in the ($\mathrm{h\ k}$\ 2) plane in figure \ref{Fig3}(B)) that shows even richer diffuse features (supplementary figure S18 for the ($\mathrm{h\ k\ 3}$) plane). At RT and down to $\sim$125 K, distinctive DS holding a 6-fold symmetry emerges in form of hexagonal diffuse pattern with a diameter of $\mathbf{a^*}\sim0.72\,\mathrm{\AA^{-1}}$, surrounding the Bragg reflections. The diffuse intensity does not vary either with the azimuthal angle or from ring to ring, but with l. Besides, the independence of the diffuseness with ($\mathrm{h\ k}$) for any $\mathrm{l}$ indicates a highly ordered crystalline structure. This structured DS in momentum space is a hallmark of the strong geometric frustration of a triangular lattice \cite{Tosic_2023,Janas_2021}, arising from the out-of-plane AFM-coupled kagome planes, hence imaging a fabric of emergent dimerized/non-dimerized clusters in the normal state. 

Between 125 K and $\mathrm{T_{CDW}}$, an anisotropic diffuse signal starts to condense at the M point of the BZ ($\frac{1}{2}$\ 0\ 0) together with a strong dependence of the hexagonal diffuse intensity in the ($\mathrm{h\ k\ 2}$) plane, characteristic of occupational (or substitutional) disorder (dimerized-non dimerized states), rather than driven by atomic vibrations, whose scattering goes to zero at low-$\mathrm{q}$ (supplementary information S7).
With further cooling, the anisotropic diffuse signal smoothly grows in intensity down to 105 K, thus preserving the C$_6$-symmetry of the lattice, in agreement with the x-ray diffraction data. 
Focusing on half integer values of $\mathrm{l}$, the ($\mathrm{h\ k}\ \frac{3}{2}$) plane in figure \ref{Fig3}(E), also reveals a honeycomb diffuse pattern, whose intensity strongly varies with the azimuthal angle and $\mathrm{q}$ (supplementary figure S18 for the ($\mathrm{h\ k}\ \frac{3}{2}$) and ($\mathrm{h\ k}\ \frac{5}{2}$) cuts), presumably driven by the small in-plane displacements of the Fe atoms \cite{shi2024}. Some quasi-circular hexagonal shape of the DS is still present below $\mathrm{T_{CDW}}$, hence differentiating from Brazovskii scenario reported in isotropic systems \cite{Brazovskii}, skyrmion lattices \cite{Jano2013} and hole doped cuprates \cite{Boschini_2021}. 

Visualizing the hexagonal DS as a composite of binary disorder that try to pack in a triangular lattice of \textit{P6mm} symmetry \cite{Aebischer2006}, the substitutional disorder can be modelled by a triangular Ising antiferromagnet lattice.
To fully understand the 3D diffuse pattern, we have carried out Monte Carlo (MC) simulations to achieve a microscopic realization of a large 2$\times$2$\times$2 unit cell (supplementary information S8), starting from a negative nearest neighbour correlation and assuming that the Ge$_1$ at (0.5 0 \textit{z}), (0.5 0.5 \textit{z}) and (0 0.5 \textit{z}) are disordered. In the MC simulation, dimerized and non-dimerized Ge$_1$ were modelled according to the Ising Hamiltonian:

\begin{equation}
\begin{split}
    H = \sum_{<i,j>:NN}c_1\sigma_i\sigma_j + \sum_{<i,j>:NNN}c_2\sigma_i\sigma_j +  \\ \sum_{<i,j>:4NN}c_3\sigma_i\sigma_j + \sum_{<i,j>:z-NN}c_4\sigma_i\sigma_j + \sum_{i}h\sigma_i + E_0,
    \end{split}
\label{MC}
\end{equation}

- where \textit{c}$_{i=1,2,3}$ are in-plane nearest-neighbour (NN), next nearest-neighbour (NNN) and 4th-NN coupling, \textit{c}$_4$ is \textit{z}-directional NN coupling, \textit{h} is the magnetic field and \textit{E}$_0$ is a constant. $\sigma$ is a ‘spin’ value (-/+)1 representing dimerized and non-dimerized Ge$_1$, respectively. The DS simulations for the ($\mathrm{h\ k\ 2}$) and ($\mathrm{h\ 0\ l}$) planes are displayed in figure \ref{Fig3}(I-J), nicely matching the experimental diffuse maps. Furthermore, the 3D model also reproduces the Bragg nodes at the A point of the BZ as a result of the doubling of the unit cell (supplementary figure S28). The high degree of frustration that emerges upon cooling and the absence of DS at the A point indicates that the dynamics of the CDW in FeGe is of order-disorder type and further confirms that the fraction of dimerized states can be considered as the relevant order parameter. We note that our model (eq. \ref{MC}) is \emph{derived} from ab-initio studies of the dimerization energetics, and \emph{not} just fitted to the data. This is a distinct new method, presented in the supplementary information S8, which allows a further check on the consistency of our results.

\subsection{Anisotropic DS and quasi-long-range BO order}

\begin{figure*}
    \centering
\includegraphics[width=0.95\linewidth]{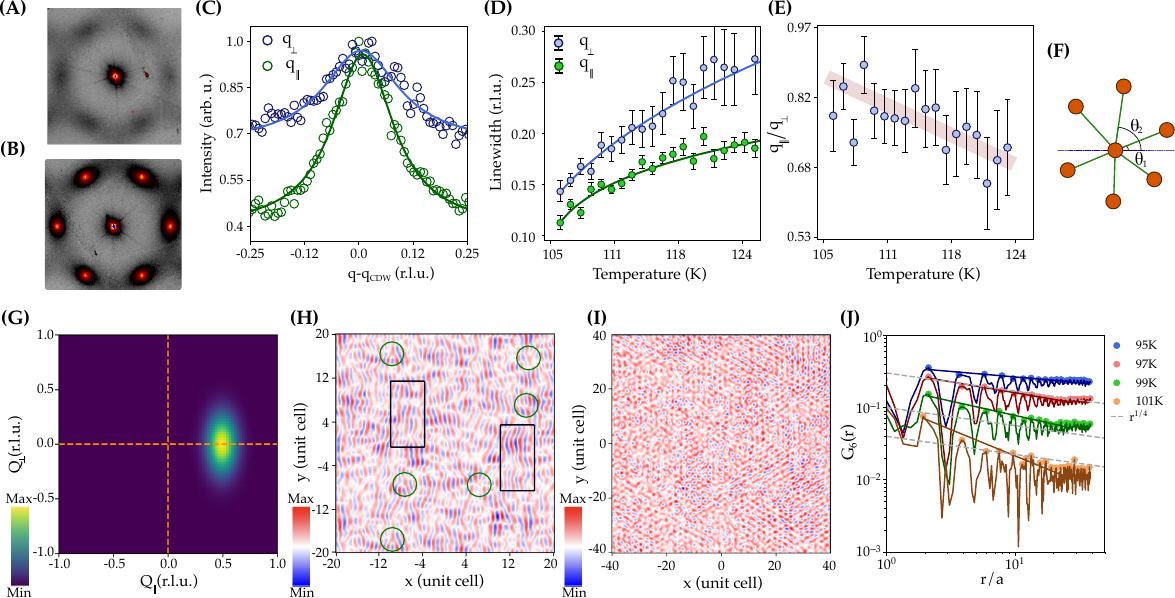}
    \caption{\textbf{Anisotropic DS and quasi-long-range BO order}. (\textbf{A}) DS map around the [1 1 2] Bragg peak at T=115 K. (\textbf{B}) T= 95 K. (\textbf{C}) Profile of the diffuse signal along the $\Gamma$-M-$\Gamma$ (parallel to $\mathrm{q_M}$, q$_{\parallel}$, radial direction) and the K-M-K (perpendicular to $\mathrm{q_M}$, q$_{\bot}$, azimuthal directions). (\textbf{D}) Temperature dependence of the DS peak width, q$_{\bot}$ and q$_{\parallel}$. The solid line is the fitting to a critical power law. (\textit{E}) Temperature dependence of the peak width ratio, q$_{\parallel}$/q$_{\bot}$. The faded line is a guide to eye. (\textit{F}) Sketch illustrating the bond orientational order parameter $\Psi_6$(\textbf{r}$_i$) defined by the eq.(\ref{order_parameter}). (\textbf{G}) Simulated DS of a single-q CDW in a triangular lattice with the experimental peak width at 95 K. (\textbf{H}) Fourier Transform (FT) of the simulated single-q DS of (\textbf{G}). The melting of the CDW is driven by dislocation unbinding (green circles) and shear (black boxes). (\textbf{I}) FT of the experimental triple-q CDW of FeGe at 95 K, showing a real space reconstruction of the charge density. The charge density map was discretized by the Voronoi tessellation to find the neighbor statistics of the kagome lattice (supplementary information S9). (\textbf{J}) The bond-order correlation function, \textit{G}$_6(r)$ shows a constant value $\sim1$ at $T=95\,\mathrm{K}$, as expected in the solid phase and an algebraic decay at 95K$<$T$<$T$_\mathrm{CDW}$, that locates the solid to hexatic transition. Solid lines are fits of the upper envelopes of the data to an algebraic decay $\sim r^b$. A universal critical value of $\eta_6\rightarrow\frac{1}{4}$ (grey line) is predicted by the KTHNY theory on approaching the hexatic to liquid phase.}
    \label{Fig4}
\end{figure*}

Focusing on the intermediate phase between T$^*$(125 K)$<$T$<$T$_\mathrm{CDW}$, the DS at the $\mathrm{M}$ point develops a sizable anisotropic peak broadening, see figure \ref{Fig4}(A, B), not observed at the $\mathrm{L}$ point. The diffuse profile at 125 K was fitted to a Lorentzian lineshape \cite{Aeppli1984,Zaluz2017,Peterson1994} and its correlation length extends to 5 \r{A} along q$_{\parallel}$ ($\Gamma$-M-$\Gamma$ direction, radial peak width) and less than one unit cell along q$_{\bot}$ (K-M-K direction, azimuthal peak width) directions, respectively, indicating a short-range positional but quasi-long-range bond orientational order of the sixfold director. This anisotropic scattering is characteristic of melting of the CDW by topological defects that appear due to thermal fluctuations in the 2D kagome plane, akin to the stacked hexatics observed in multilayer smectic $\mathrm{B}$ phases of liquid crystals \cite{Birgeneau_1978,Pindak_1981}. 
The temperature dependence of the peak widths (figure \ref{Fig4}(D)) follows a mean field behavior, while the peak width ratio (q$_{\parallel}$/q$_{\bot}$) (fig. \ref{Fig4}(E)), which is proportional to the mean dislocation distance, approaches to $\sim$1 at T$_\mathrm{CDW}$, as expected in proximity to a broken translational symmetry state. 
It is also worth noting that some anisotropy is still present below $\mathrm{T_{CDW}}$, suggesting a coexistence of regions with short and long-range translational order \cite{shi2024}. This hints at the fragility of the low temperature CDW state in FeGe \cite{Chen_2023}, further supported by the drastic reduction of the anisotropy of DS in annealed crystals (supplementary figures S19 and S20).
In figure \ref{Fig4} (G), we simulate the DS in the reciprocal space for the anisotropic broadening of the single-q diffuse signal observed experimentally. The anisotropic spatial profile of the q$_\mathrm{M}$ charge scattering is consistent with the melting of the CDW by defects (dislocations in the green-dashed circles) and shear (black boxes) in fig. \ref{Fig4} (H) with a reduced coherence of the 1D chain. However, when the CDW peak broadens parallel to $\mathrm{q_M}$, the real space maps describe a relative phase change along the 1D domains walls (supplementary figure S29), akin to the formation topological defects in transition metal dichalcogenides and superconductors \cite{Cheng_2023,Domrose_2023,Aishwarya_2024,Guillamon_2009}.

\section{Discussion}
The results presented here bring important information for the understanding of the multiple-q CDW transition in FeGe and in kagome metals, in general. First of all, within both the equally plausible low temperature centrosymmetric \textit{P6/mmm} and non-centrosymmetric \textit{P6mm} space groups, we are able to achieved a better description of the band structure at the DFT level without either the assumption of renormalization factors or the Hubbard term, U, in DFT. This has important consequences for the lattice dynamics calculations and puts constraints on the U values used to soften a rather flat phonon mode in the \textit{k}$_z$=$\pi$ plane \cite{Miao_2023}. It appears that the electronic structure of FeGe is less correlated than previously assumed, despite the strong intertwining of charge and spin orders. Indeed, the comprehensive description of the band structure must rely on the precise treatment of dimerized/undimerized crystal structure and AFM within the DFT framework \cite{shi2024,Zhou2023,Wu2023,Zhao_2023}.

Moreover, we have identified a quasi-2D hexagonal diffuse scattering along the M-L line of the BZ at high temperature that evolves towards a localized charge precursor at M and L points at low temperature. Although, in principle, this would discard charge fluctuations at A as the leading instability and the driving force of the CDW, the strong geometric frustration of the triangular lattice introduced by the dimer formation actually causes the DS at M and L points, demonstrating an order-disorder transformation that follows a 2$^\mathrm{nd}$ order-like phase transition. 
Nevertheless, we point out that the order-disorder scenario in FeGe differs from the AVS, where the phase transition is achieved by the freezing, without softening, of a transverse phonon mode \cite{Miao_2021,Subires_2023}. More importantly, the short-range charge fluctuations are also reminiscent of a fragile metastable CDW phase or some \textit{glassiness} that would explain the extreme sensitivity to external perturbations \cite{Chen_2023} and the effect of annealing \cite{Oh2024}. 

On the other hand, although the profile and correlation length of the diffuse spots demonstrate the presence of phase fluctuations, bearing a strong resemblance to critical scattering due melting of dislocation pairs and shear, we caution about setting the phase transition within the KTHNY theory. The continuous topological melting via unbinding of defect-pairs is predicted to occur in 2-dimensional systems \cite{Nelson_1979}, thus the precise character of the phase transition requires a topological analysis of the real space charge density. This is further justified since FeGe is a 3D system, although the FeGe$_1$ kagome plane is purely 2D, and the melting should occur as a single, first order transition. 

Within the KTHNY theory \cite{Halperin_1978}, the bond orientational (BO) order parameter is characterized by a local ordering field describing orientation between neighboring sites:

\begin{equation}
    {\Psi}_{6}(\textbf{r}_k)=\frac{1}{N_{k}} \sum_{j=1}^{N_{k}}\mathrm{e}^{i6\theta_{kj}},
    \label{order_parameter}
\end{equation}

- where \textit{N}$_{k}$ represents the maxima of the charge density distribution (equivalent to the number of particles in colloidal systems) around a reference point located at position \textbf{r}$_k$ and $\theta_{kj}$ defines the angle the \textit{k}-\textit{j} bond, fig. \ref{Fig4}(F). 
The bond-order correlation function \textit{G}$_6$(\textit{r}) is defined as:

\begin{equation} \label{orientational}
    G_6(r)=\frac{1}{N_{r}} \sum_{<i,j>}^{N_{r}}{\Psi}_{6}(\textbf{r}_i){\Psi}^{*}_{6}(\textbf{r}_j),
\end{equation}

- where \textit{N}$_r$ is the charge density at a distance \textit{r}. The KTHNY theory states that \textit{G}$_6$(\textit{r}) decays algebraically in the hexatic phase, \textit{G}$_6$(\textit{r})$\propto$\textit{r}$^{-\eta_6}$, indicating quasi-long-range bond order, and exponentially in the liquid phase, \textit{G}$_6$(\textit{r})$\propto\mathrm{e}^{-r/\xi_6}$, where $\xi_6$ is the orientational correlation length. 

Figure \ref{Fig4}(I) shows the Fourier transform (FT) of the corresponding structure factor \textit{S}($\mathrm{\textbf{q}}$) at T=95 K \cite{Rcomin_Science,Rcomin_Comment,Kang_2019,Rcomin_Answer}. The FT, representing the real space distribution of the charge density, is further discretized by performing a Voronoi tessellation (supplementary information S9). 
As displayed in figure \ref{Fig4}(J), \textit{G}$_6$(\textit{r}) is close to 1 at 95 K and its upper envelope can be fitted to a constant, as expected in the solid phase. Increasing the temperature results in a faster algebraic decay demonstrating the quasi-long-range orientational order up to 101 K that would define the solid-to-hexatic transition, but, nevertheless, does not follow with an exponential decay, as the KTHNY theory predicts on approaching the liquid phase. This is a consequence of the Voronoi construction that does not capture in detail the charge density discretization in real space at T$>$105 K and prevents the extraction of the Frank´s constant, \textit{K}$_A$, that describes the effective stiffness of the BO field, and a comprehensive analysis of the BO order parameter.

In conclusion, we have carried out a comprehensive experimental survey to show that the CDW in the antiferromagnetic FeGe is of order-disorder type that fits within the Ising model of a triangular lattice. The order-disorder transformation is a direct consequence of the strong frustration introduced by the dimerization of the trigonal Ge$_\mathrm{1}$ on the kagome plane that double the unit cell along \textbf{a}, \textbf{b} and \textbf{c} directions of the crystal. 
Moreover, we observe an anisotropic diffuse signal at T$>$T$_\mathrm{CDW}$, characteristic of the melting of the CDW by topological defects that resembles the hexatic $\mathrm{B}$ phases in liquid crystals and TMDs \cite{Domrose_2023,Cheng_2023,Aishwarya_2024}. Microscopically, the anisotropic DS could be a consequence of the fragility of the CDW or driven by the small in-plane Fe displacements, as inferred by x-ray diffraction \cite{shi2024} and infrared spectroscopy \cite{Wenzel_2024}. Although our correlation analysis is consistent with a quasi-long-range bond orientational order at T$\leq$T$_\mathrm{CDW}$, it cannot guarantee that FeGe fits within the KTHNY theory. However, it opens new perspectives to look at the melting of the charge modulations in kagome lattices and the possibility to study them in detail in the 2D limit \cite{Song2023} with more advanced diffraction techniques.

\section{Methods}
Single crystals of FeGe were grown by the chemical vapor transport method using iodine as transport agent. High-purity Fe and Ge powders were mixed together with a molar ratio of Fe:Ge = 1:1. The mixture and iodine ($\sim$10 mg/cm$^3$) were loaded into a quartz tube and vacuum sealed. The tube was placed in a two-zone furnace, $\Delta$T= 620-560ºC \cite{Wenzel_2024}. 

Single crystal diffraction was carried out at the Swiss-Norwegian beamline (SNBL) BM01, European Synchrotron Radiation Facility (ESRF), with incident energy E$_\mathrm{i}$= 20 keV and a Pilatus 2 M detector \cite{Dyadkin_2016}. The raw data were processed with SNBL toolbox (SNBL Toolbox) and CrysAlis Pro software. The refinements were carried out with SHELXL2018/1 using SHELXLE as the GUI. Single crystal diffuse scattering were performed at the ID28 beamline at ESRF with \textit{E$_i$}=17.8 keV and a Dectris PILATUS3 1M X area detector. We use the CrysAlis software package for the orientation matrix refinement and the ID28 software ProjectN for reconstruction of the  reciprocal space maps and plotted in Albula. The components ($h$ $k$ $l$) of the scattering vector are expressed in reciprocal lattice units (r.l.u.), ($h$ $k$ $l$)= $h \mathbf{a}^*+k \mathbf{b}^*+l \mathbf{c}^*$, where $\mathbf{a}^*$, $\mathbf{b}^*$, and $\mathbf{c}^*$ are the reciprocal lattice vectors. 

Hard x-ray resonant scattering experiments at the Fe-K edge (E$_\mathrm{i}$= 7.115 keV) were performed at at the beamline 4ID-D of the Advanced Photon Source at Argonne National Laboratory.

Angle Resolved Photoemission spectroscopy experiments (ARPES) were performed at the LOREA beamline (MBS electron analyzer, base pressure of 10$^{-10}$, angular resolution of 0.2$^{\circ}$, energy resolution 10 meV) of ALBA and APE-LE beamline (DA30 electron analyzer, base pressure of 10$^{-11}$ mbar, angular resolution of 0.2$^{\circ}$, energy resolution 10 meV) of ELETTRA research facilities. 

The first-principle calculations in this work use the Vienna ab initio Simulation Package (VASP)\cite{kresse1996efficiency, kresse1993ab1, kresse1993ab2, kresse1994ab, kresse1996efficient} with generalized gradient approximation of Perdew-Burke-Ernzerhof (PBE) exchange-correlation potential\cite{perdew1996generalized}. A $8\times8\times8$ ($5\times5\times5$) $k$-mesh for non-CDW (CDW) phase and an energy cutoff of 500 eV are used. 
The maximally localized Wannier functions are obtained using WANNIER90\cite{marzari1997maximally, souza2001maximally, marzari2012maximally, pizzi2020wannier90}. A local coordinate system at the kagome site is adopted in order to decompose $d$ orbitals when construct MLWFs, the same as the one used in Ref.\cite{jiang2023kagome}.
The Wannier tight-binding models are symmetrized using \textit{Wannhr\_symm} in \textit{WannierTools} \cite{wu2018wanniertools}. The unfolding of CDW bands is performed using \textit{VaspBandUnfolding} package \cite{zheng2018vaspbandunfolding, popescu2012extracting}. The Fermi surface is computed using \textit{WannierTools} \cite{wu2018wanniertools} and visualized using \textit{Fermisurfer} \cite{kawamura2019fermisurfer}. 

Monte Carlo (MC) simulations were performed in order to generate real-space realizations of possible atomic configurations which could be used to model the observed diffuse scattering. 
The average FeGe unit cell at 80 K, solved from the single crystal diffraction experiments on BM01, was taken and expanded to a 32$\times$32$\times$32 supercell with all trigonal Ge atoms set to be non-dimerized (represented in the simulation as the Ising variable, $\sigma= +1$). A small number ($\sim$5\%) of the Ge was randomly converted to dimers (represented with $\sigma= -1$).
In each MC move, an Ising variable was randomly inverted, and the moves were accepted or rejected following the Metropolis condition with the Hamiltonian expressed in eq.\ref{MC} and the MC temperature set to 80K.
Four key interactions between neighboring Ising variables were considered: the nearest neighbor (NN), next nearest neighbor (NNN), and the third nearest neighbors (3NN) in the \textit{ab}-plane ($c_1, c_2$ and $c_3$, respectively), and the NN along the \textit{c}-axis, $c_4$. The parameter \textit{h} is a magnetic consideration that accounts for the overall proportion of dimerized Ge in the model. The numerical values for $c_1,\ c_2,\ c_3,\ c_4,\ h$ and $E_0$ were derived from DFT calculations.
Five MC simulations were run in this way, all starting from different randomised configurations to try to minimise the possibility of getting stuck in a local energy minimum. Each simulation was run until the energy converged, and then the resulting atomic configuration was used as input to the program Scatty \cite{Paddison:vk5031}, which calculates the average diffuse scattering from all given simulations. The small in-plane displacement of the Fe atoms are not included in the MC simulations, hence the DS does not reproduce the absence of diffuse intensity within the dashed-yellow area in figure \ref{Fig3}(E). 

\section{Data availability}
Source data are provided with this paper. The scattering, ARPES, DS and IXS data generated in this study will be deposited in the Figshare database.
\bibliography{FeGe}

\begin{thebibliography}{122}%
\makeatletter
\providecommand \@ifxundefined [1]{%
 \@ifx{#1\undefined}
}%
\providecommand \@ifnum [1]{%
 \ifnum #1\expandafter \@firstoftwo
 \else \expandafter \@secondoftwo
 \fi
}%
\providecommand \@ifx [1]{%
 \ifx #1\expandafter \@firstoftwo
 \else \expandafter \@secondoftwo
 \fi
}%
\providecommand \natexlab [1]{#1}%
\providecommand \enquote  [1]{``#1''}%
\providecommand \bibnamefont  [1]{#1}%
\providecommand \bibfnamefont [1]{#1}%
\providecommand \citenamefont [1]{#1}%
\providecommand \href@noop [0]{\@secondoftwo}%
\providecommand \href [0]{\begingroup \@sanitize@url \@href}%
\providecommand \@href[1]{\@@startlink{#1}\@@href}%
\providecommand \@@href[1]{\endgroup#1\@@endlink}%
\providecommand \@sanitize@url [0]{\catcode `\\12\catcode `\$12\catcode
  `\&12\catcode `\#12\catcode `\^12\catcode `\_12\catcode `\%12\relax}%
\providecommand \@@startlink[1]{}%
\providecommand \@@endlink[0]{}%
\providecommand \url  [0]{\begingroup\@sanitize@url \@url }%
\providecommand \@url [1]{\endgroup\@href {#1}{\urlprefix }}%
\providecommand \urlprefix  [0]{URL }%
\providecommand \Eprint [0]{\href }%
\providecommand \doibase [0]{https://doi.org/}%
\providecommand \selectlanguage [0]{\@gobble}%
\providecommand \bibinfo  [0]{\@secondoftwo}%
\providecommand \bibfield  [0]{\@secondoftwo}%
\providecommand \translation [1]{[#1]}%
\providecommand \BibitemOpen [0]{}%
\providecommand \bibitemStop [0]{}%
\providecommand \bibitemNoStop [0]{.\EOS\space}%
\providecommand \EOS [0]{\spacefactor3000\relax}%
\providecommand \BibitemShut  [1]{\csname bibitem#1\endcsname}%
\let\auto@bib@innerbib\@empty
\bibitem [{\citenamefont {Moessner}\ and\ \citenamefont
  {Ramirez}(2006)}]{Moessner_2006}%
  \BibitemOpen
  \bibfield  {author} {\bibinfo {author} {\bibfnamefont {R.}~\bibnamefont
  {Moessner}}\ and\ \bibinfo {author} {\bibfnamefont {A.~P.}\ \bibnamefont
  {Ramirez}},\ }\bibfield  {title} {\bibinfo {title} {Geometrical
  frustration},\ }\href {https://doi.org/10.1063/1.2186278} {\bibfield
  {journal} {\bibinfo  {journal} {Physics Today}\ }\textbf {\bibinfo {volume}
  {59}},\ \bibinfo {pages} {24} (\bibinfo {year} {2006})}\BibitemShut {NoStop}%
\bibitem [{\citenamefont {Savary}\ and\ \citenamefont
  {Balents}(2016)}]{Savary_2017}%
  \BibitemOpen
  \bibfield  {author} {\bibinfo {author} {\bibfnamefont {L.}~\bibnamefont
  {Savary}}\ and\ \bibinfo {author} {\bibfnamefont {L.}~\bibnamefont
  {Balents}},\ }\bibfield  {title} {\bibinfo {title} {Quantum spin liquids: a
  review},\ }\href {https://doi.org/10.1088/0034-4885/80/1/016502} {\bibfield
  {journal} {\bibinfo  {journal} {Reports on Progress in Physics}\ }\textbf
  {\bibinfo {volume} {80}},\ \bibinfo {pages} {016502} (\bibinfo {year}
  {2016})}\BibitemShut {NoStop}%
\bibitem [{\citenamefont {Metcalf}(1974)}]{Metcalf_1974}%
  \BibitemOpen
  \bibfield  {author} {\bibinfo {author} {\bibfnamefont {B.}~\bibnamefont
  {Metcalf}},\ }\bibfield  {title} {\bibinfo {title} {Ground state spin
  orderings of the triangular ising model with the nearest and next nearest
  neighbor interaction},\ }\href
  {https://doi.org/https://doi.org/10.1016/0375-9601(74)90247-3} {\bibfield
  {journal} {\bibinfo  {journal} {Physics Letters A}\ }\textbf {\bibinfo
  {volume} {46}},\ \bibinfo {pages} {325} (\bibinfo {year} {1974})}\BibitemShut
  {NoStop}%
\bibitem [{\citenamefont {Jiang}\ and\ \citenamefont
  {Emig}(2006)}]{Jiang_2006}%
  \BibitemOpen
  \bibfield  {author} {\bibinfo {author} {\bibfnamefont {Y.}~\bibnamefont
  {Jiang}}\ and\ \bibinfo {author} {\bibfnamefont {T.}~\bibnamefont {Emig}},\
  }\bibfield  {title} {\bibinfo {title} {Ordering of geometrically frustrated
  classical and quantum triangular ising magnets},\ }\href
  {https://doi.org/10.1103/PhysRevB.73.104452} {\bibfield  {journal} {\bibinfo
  {journal} {Phys. Rev. B}\ }\textbf {\bibinfo {volume} {73}},\ \bibinfo
  {pages} {104452} (\bibinfo {year} {2006})}\BibitemShut {NoStop}%
\bibitem [{\citenamefont {Iaconis}\ \emph {et~al.}(2018)\citenamefont
  {Iaconis}, \citenamefont {Liu}, \citenamefont {Halász},\ and\ \citenamefont
  {Balents}}]{Iaconis_2018}%
  \BibitemOpen
  \bibfield  {author} {\bibinfo {author} {\bibfnamefont {J.}~\bibnamefont
  {Iaconis}}, \bibinfo {author} {\bibfnamefont {C.}~\bibnamefont {Liu}},
  \bibinfo {author} {\bibfnamefont {G.~B.}\ \bibnamefont {Halász}},\ and\
  \bibinfo {author} {\bibfnamefont {L.}~\bibnamefont {Balents}},\ }\bibfield
  {title} {\bibinfo {title} {{Spin Liquid versus Spin Orbit Coupling on the
  Triangular Lattice}},\ }\href {https://doi.org/10.21468/SciPostPhys.4.1.003}
  {\bibfield  {journal} {\bibinfo  {journal} {SciPost Phys.}\ }\textbf
  {\bibinfo {volume} {4}},\ \bibinfo {pages} {003} (\bibinfo {year}
  {2018})}\BibitemShut {NoStop}%
\bibitem [{\citenamefont {Houtappel}(1950)}]{Houtappel_1950}%
  \BibitemOpen
  \bibfield  {author} {\bibinfo {author} {\bibfnamefont {R.}~\bibnamefont
  {Houtappel}},\ }\bibfield  {title} {\bibinfo {title} {Order-disorder in
  hexagonal lattices},\ }\href
  {https://doi.org/https://doi.org/10.1016/0031-8914(50)90130-3} {\bibfield
  {journal} {\bibinfo  {journal} {Physica}\ }\textbf {\bibinfo {volume} {16}},\
  \bibinfo {pages} {425} (\bibinfo {year} {1950})}\BibitemShut {NoStop}%
\bibitem [{\citenamefont {Shokef}\ \emph {et~al.}(2011)\citenamefont {Shokef},
  \citenamefont {Souslov},\ and\ \citenamefont {Lubensky}}]{Shokef_2011}%
  \BibitemOpen
  \bibfield  {author} {\bibinfo {author} {\bibfnamefont {Y.}~\bibnamefont
  {Shokef}}, \bibinfo {author} {\bibfnamefont {A.}~\bibnamefont {Souslov}},\
  and\ \bibinfo {author} {\bibfnamefont {T.~C.}\ \bibnamefont {Lubensky}},\
  }\bibfield  {title} {\bibinfo {title} {Order by disorder in the
  antiferromagnetic ising model on an elastic triangular lattice},\ }\href
  {https://doi.org/10.1073/pnas.1014915108} {\bibfield  {journal} {\bibinfo
  {journal} {Proceedings of the National Academy of Sciences}\ }\textbf
  {\bibinfo {volume} {108}},\ \bibinfo {pages} {11804} (\bibinfo {year}
  {2011})}\BibitemShut {NoStop}%
\bibitem [{\citenamefont {Weber}\ \emph {et~al.}(2006)\citenamefont {Weber},
  \citenamefont {L\'auchli}, \citenamefont {Mila},\ and\ \citenamefont
  {Giamarchi}}]{Weber_2006}%
  \BibitemOpen
  \bibfield  {author} {\bibinfo {author} {\bibfnamefont {C.}~\bibnamefont
  {Weber}}, \bibinfo {author} {\bibfnamefont {A.}~\bibnamefont {L\'auchli}},
  \bibinfo {author} {\bibfnamefont {F.}~\bibnamefont {Mila}},\ and\ \bibinfo
  {author} {\bibfnamefont {T.}~\bibnamefont {Giamarchi}},\ }\bibfield  {title}
  {\bibinfo {title} {Magnetism and superconductivity of strongly correlated
  electrons on the triangular lattice},\ }\href
  {https://doi.org/10.1103/PhysRevB.73.014519} {\bibfield  {journal} {\bibinfo
  {journal} {Phys. Rev. B}\ }\textbf {\bibinfo {volume} {73}},\ \bibinfo
  {pages} {014519} (\bibinfo {year} {2006})}\BibitemShut {NoStop}%
\bibitem [{\citenamefont {Tsunetsugu}\ \emph {et~al.}(2009)\citenamefont
  {Tsunetsugu}, \citenamefont {Hattori}, \citenamefont {Ohashi}, \citenamefont
  {Kawakami},\ and\ \citenamefont {Momoi}}]{Tsunetsugu_2009}%
  \BibitemOpen
  \bibfield  {author} {\bibinfo {author} {\bibfnamefont {H.}~\bibnamefont
  {Tsunetsugu}}, \bibinfo {author} {\bibfnamefont {K.}~\bibnamefont {Hattori}},
  \bibinfo {author} {\bibfnamefont {T.}~\bibnamefont {Ohashi}}, \bibinfo
  {author} {\bibfnamefont {N.}~\bibnamefont {Kawakami}},\ and\ \bibinfo
  {author} {\bibfnamefont {T.}~\bibnamefont {Momoi}},\ }\bibfield  {title}
  {\bibinfo {title} {Strongly correlated electrons on frustrated lattices},\
  }\href {https://doi.org/10.1088/1742-6596/145/1/012015} {\bibfield  {journal}
  {\bibinfo  {journal} {Journal of Physics: Conference Series}\ }\textbf
  {\bibinfo {volume} {145}},\ \bibinfo {pages} {012015} (\bibinfo {year}
  {2009})}\BibitemShut {NoStop}%
\bibitem [{\citenamefont {Bernu}\ \emph {et~al.}(1992)\citenamefont {Bernu},
  \citenamefont {Lhuillier},\ and\ \citenamefont {Pierre}}]{Bernu_1992}%
  \BibitemOpen
  \bibfield  {author} {\bibinfo {author} {\bibfnamefont {B.}~\bibnamefont
  {Bernu}}, \bibinfo {author} {\bibfnamefont {C.}~\bibnamefont {Lhuillier}},\
  and\ \bibinfo {author} {\bibfnamefont {L.}~\bibnamefont {Pierre}},\
  }\bibfield  {title} {\bibinfo {title} {Signature of n\'eel order in exact
  spectra of quantum antiferromagnets on finite lattices},\ }\href
  {https://doi.org/10.1103/PhysRevLett.69.2590} {\bibfield  {journal} {\bibinfo
   {journal} {Phys. Rev. Lett.}\ }\textbf {\bibinfo {volume} {69}},\ \bibinfo
  {pages} {2590} (\bibinfo {year} {1992})}\BibitemShut {NoStop}%
\bibitem [{\citenamefont {Capriotti}\ \emph {et~al.}(1999)\citenamefont
  {Capriotti}, \citenamefont {Trumper},\ and\ \citenamefont
  {Sorella}}]{Capriotti_1999}%
  \BibitemOpen
  \bibfield  {author} {\bibinfo {author} {\bibfnamefont {L.}~\bibnamefont
  {Capriotti}}, \bibinfo {author} {\bibfnamefont {A.~E.}\ \bibnamefont
  {Trumper}},\ and\ \bibinfo {author} {\bibfnamefont {S.}~\bibnamefont
  {Sorella}},\ }\bibfield  {title} {\bibinfo {title} {Long-range n\'eel order
  in the triangular heisenberg model},\ }\href
  {https://doi.org/10.1103/PhysRevLett.82.3899} {\bibfield  {journal} {\bibinfo
   {journal} {Phys. Rev. Lett.}\ }\textbf {\bibinfo {volume} {82}},\ \bibinfo
  {pages} {3899} (\bibinfo {year} {1999})}\BibitemShut {NoStop}%
\bibitem [{\citenamefont {Janas}\ \emph {et~al.}(2021)\citenamefont {Janas},
  \citenamefont {Lass}, \citenamefont {\ifmmode \mbox{\c{T}}\else
  \c{T}\fi{}u\ifmmode~\mbox{\c{t}}\else \c{t}\fi{}ueanu}, \citenamefont
  {Haubro}, \citenamefont {Niedermayer}, \citenamefont {Stuhr}, \citenamefont
  {Xu}, \citenamefont {Prabhakaran}, \citenamefont {Deen}, \citenamefont
  {Holm-Dahlin},\ and\ \citenamefont {Lefmann}}]{Janas_2021}%
  \BibitemOpen
  \bibfield  {author} {\bibinfo {author} {\bibfnamefont {S.}~\bibnamefont
  {Janas}}, \bibinfo {author} {\bibfnamefont {J.}~\bibnamefont {Lass}},
  \bibinfo {author} {\bibfnamefont {A.-E.}\ \bibnamefont {\ifmmode
  \mbox{\c{T}}\else \c{T}\fi{}u\ifmmode~\mbox{\c{t}}\else \c{t}\fi{}ueanu}},
  \bibinfo {author} {\bibfnamefont {M.~L.}\ \bibnamefont {Haubro}}, \bibinfo
  {author} {\bibfnamefont {C.}~\bibnamefont {Niedermayer}}, \bibinfo {author}
  {\bibfnamefont {U.}~\bibnamefont {Stuhr}}, \bibinfo {author} {\bibfnamefont
  {G.}~\bibnamefont {Xu}}, \bibinfo {author} {\bibfnamefont {D.}~\bibnamefont
  {Prabhakaran}}, \bibinfo {author} {\bibfnamefont {P.~P.}\ \bibnamefont
  {Deen}}, \bibinfo {author} {\bibfnamefont {S.}~\bibnamefont {Holm-Dahlin}},\
  and\ \bibinfo {author} {\bibfnamefont {K.}~\bibnamefont {Lefmann}},\
  }\bibfield  {title} {\bibinfo {title} {Classical spin liquid or extended
  critical range in \textit{h}-$\mathrm{YMnO_3}$?},\ }\href
  {https://doi.org/10.1103/PhysRevLett.126.107203} {\bibfield  {journal}
  {\bibinfo  {journal} {Phys. Rev. Lett.}\ }\textbf {\bibinfo {volume} {126}},\
  \bibinfo {pages} {107203} (\bibinfo {year} {2021})}\BibitemShut {NoStop}%
\bibitem [{\citenamefont {Tošić}\ \emph {et~al.}(2023)\citenamefont
  {Tošić}, \citenamefont {Simonov},\ and\ \citenamefont
  {Spaldin}}]{Tosic_2023}%
  \BibitemOpen
  \bibfield  {author} {\bibinfo {author} {\bibfnamefont {T.~N.}\ \bibnamefont
  {Tošić}}, \bibinfo {author} {\bibfnamefont {A.}~\bibnamefont {Simonov}},\
  and\ \bibinfo {author} {\bibfnamefont {N.~A.}\ \bibnamefont {Spaldin}},\
  }\href@noop {} {\bibinfo {title} {Origin of correlated diffuse scattering in
  the hexagonal manganites}} (\bibinfo {year} {2023}),\ \Eprint
  {https://arxiv.org/abs/2312.07449} {arXiv:2312.07449 [cond-mat.mtrl-sci]}
  \BibitemShut {NoStop}%
\bibitem [{\citenamefont {Sato}\ \emph {et~al.}(2003)\citenamefont {Sato},
  \citenamefont {Lee}, \citenamefont {Katsufuji}, \citenamefont {Masaki},
  \citenamefont {Park}, \citenamefont {Copley},\ and\ \citenamefont
  {Takagi}}]{Sato_2003}%
  \BibitemOpen
  \bibfield  {author} {\bibinfo {author} {\bibfnamefont {T.~J.}\ \bibnamefont
  {Sato}}, \bibinfo {author} {\bibfnamefont {S.~H.}\ \bibnamefont {Lee}},
  \bibinfo {author} {\bibfnamefont {T.}~\bibnamefont {Katsufuji}}, \bibinfo
  {author} {\bibfnamefont {M.}~\bibnamefont {Masaki}}, \bibinfo {author}
  {\bibfnamefont {S.}~\bibnamefont {Park}}, \bibinfo {author} {\bibfnamefont
  {J.~R.~D.}\ \bibnamefont {Copley}},\ and\ \bibinfo {author} {\bibfnamefont
  {H.}~\bibnamefont {Takagi}},\ }\bibfield  {title} {\bibinfo {title}
  {Unconventional spin fluctuations in the hexagonal antiferromagnet
  $\mathrm{YMnO_3}$},\ }\href {https://doi.org/10.1103/PhysRevB.68.014432}
  {\bibfield  {journal} {\bibinfo  {journal} {Phys. Rev. B}\ }\textbf {\bibinfo
  {volume} {68}},\ \bibinfo {pages} {014432} (\bibinfo {year}
  {2003})}\BibitemShut {NoStop}%
\bibitem [{\citenamefont {Castelnovo}\ \emph {et~al.}(2008)\citenamefont
  {Castelnovo}, \citenamefont {Moessner},\ and\ \citenamefont
  {Sondhi}}]{Castelnovo_2008}%
  \BibitemOpen
  \bibfield  {author} {\bibinfo {author} {\bibfnamefont {C.}~\bibnamefont
  {Castelnovo}}, \bibinfo {author} {\bibfnamefont {R.}~\bibnamefont
  {Moessner}},\ and\ \bibinfo {author} {\bibfnamefont {S.~L.}\ \bibnamefont
  {Sondhi}},\ }\bibfield  {title} {\bibinfo {title} {Magnetic monopoles in spin
  ice},\ }\href {https://doi.org/10.1038/nature06433} {\bibfield  {journal}
  {\bibinfo  {journal} {Nature}\ }\textbf {\bibinfo {volume} {451}},\ \bibinfo
  {pages} {42} (\bibinfo {year} {2008})}\BibitemShut {NoStop}%
\bibitem [{\citenamefont {Kosterlitz}\ and\ \citenamefont
  {Thouless}(1973)}]{Kosterlitz_1973}%
  \BibitemOpen
  \bibfield  {author} {\bibinfo {author} {\bibfnamefont {J.~M.}\ \bibnamefont
  {Kosterlitz}}\ and\ \bibinfo {author} {\bibfnamefont {D.~J.}\ \bibnamefont
  {Thouless}},\ }\bibfield  {title} {\bibinfo {title} {Ordering, metastability
  and phase transitions in two-dimensional systems},\ }\href
  {https://doi.org/10.1088/0022-3719/6/7/010} {\bibfield  {journal} {\bibinfo
  {journal} {Journal of Physics C: Solid State Physics}\ }\textbf {\bibinfo
  {volume} {6}},\ \bibinfo {pages} {1181} (\bibinfo {year} {1973})}\BibitemShut
  {NoStop}%
\bibitem [{\citenamefont {Halperin}\ and\ \citenamefont
  {Nelson}(1978)}]{Halperin_1978}%
  \BibitemOpen
  \bibfield  {author} {\bibinfo {author} {\bibfnamefont {B.~I.}\ \bibnamefont
  {Halperin}}\ and\ \bibinfo {author} {\bibfnamefont {D.~R.}\ \bibnamefont
  {Nelson}},\ }\bibfield  {title} {\bibinfo {title} {Theory of two-dimensional
  melting},\ }\href {https://doi.org/10.1103/PhysRevLett.41.121} {\bibfield
  {journal} {\bibinfo  {journal} {Phys. Rev. Lett.}\ }\textbf {\bibinfo
  {volume} {41}},\ \bibinfo {pages} {121} (\bibinfo {year} {1978})}\BibitemShut
  {NoStop}%
\bibitem [{\citenamefont {Nelson}\ and\ \citenamefont
  {Halperin}(1979)}]{Nelson_1979}%
  \BibitemOpen
  \bibfield  {author} {\bibinfo {author} {\bibfnamefont {D.~R.}\ \bibnamefont
  {Nelson}}\ and\ \bibinfo {author} {\bibfnamefont {B.~I.}\ \bibnamefont
  {Halperin}},\ }\bibfield  {title} {\bibinfo {title} {Dislocation-mediated
  melting in two dimensions},\ }\href
  {https://doi.org/10.1103/PhysRevB.19.2457} {\bibfield  {journal} {\bibinfo
  {journal} {Phys. Rev. B}\ }\textbf {\bibinfo {volume} {19}},\ \bibinfo
  {pages} {2457} (\bibinfo {year} {1979})}\BibitemShut {NoStop}%
\bibitem [{\citenamefont {Birgeneau}\ and\ \citenamefont
  {Horn}(1986)}]{Birgeneau_1986}%
  \BibitemOpen
  \bibfield  {author} {\bibinfo {author} {\bibfnamefont {R.~J.}\ \bibnamefont
  {Birgeneau}}\ and\ \bibinfo {author} {\bibfnamefont {P.~M.}\ \bibnamefont
  {Horn}},\ }\bibfield  {title} {\bibinfo {title} {Two-dimensional rare gas
  solids},\ }\href {https://doi.org/10.1126/science.232.4748.329} {\bibfield
  {journal} {\bibinfo  {journal} {Science}\ }\textbf {\bibinfo {volume}
  {232}},\ \bibinfo {pages} {329} (\bibinfo {year} {1986})}\BibitemShut
  {NoStop}%
\bibitem [{\citenamefont {Dai}\ and\ \citenamefont {Lieber}(1992)}]{Dai_1992}%
  \BibitemOpen
  \bibfield  {author} {\bibinfo {author} {\bibfnamefont {H.}~\bibnamefont
  {Dai}}\ and\ \bibinfo {author} {\bibfnamefont {C.~M.}\ \bibnamefont
  {Lieber}},\ }\bibfield  {title} {\bibinfo {title} {Solid-hexatic-liquid
  phases in two-dimensional charge-density waves},\ }\href
  {https://doi.org/10.1103/PhysRevLett.69.1576} {\bibfield  {journal} {\bibinfo
   {journal} {Phys. Rev. Lett.}\ }\textbf {\bibinfo {volume} {69}},\ \bibinfo
  {pages} {1576} (\bibinfo {year} {1992})}\BibitemShut {NoStop}%
\bibitem [{\citenamefont {Aharony}\ \emph {et~al.}(1986)\citenamefont
  {Aharony}, \citenamefont {Birgeneau}, \citenamefont {Brock},\ and\
  \citenamefont {Litster}}]{Aharony_1986}%
  \BibitemOpen
  \bibfield  {author} {\bibinfo {author} {\bibfnamefont {A.}~\bibnamefont
  {Aharony}}, \bibinfo {author} {\bibfnamefont {R.~J.}\ \bibnamefont
  {Birgeneau}}, \bibinfo {author} {\bibfnamefont {J.~D.}\ \bibnamefont
  {Brock}},\ and\ \bibinfo {author} {\bibfnamefont {J.~D.}\ \bibnamefont
  {Litster}},\ }\bibfield  {title} {\bibinfo {title} {Multicriticality in
  hexatic liquid crystals},\ }\href
  {https://doi.org/10.1103/PhysRevLett.57.1012} {\bibfield  {journal} {\bibinfo
   {journal} {Phys. Rev. Lett.}\ }\textbf {\bibinfo {volume} {57}},\ \bibinfo
  {pages} {1012} (\bibinfo {year} {1986})}\BibitemShut {NoStop}%
\bibitem [{\citenamefont {Wilson}\ and\ \citenamefont
  {Ortiz}(2024)}]{Wilson2024}%
  \BibitemOpen
  \bibfield  {author} {\bibinfo {author} {\bibfnamefont {S.~D.}\ \bibnamefont
  {Wilson}}\ and\ \bibinfo {author} {\bibfnamefont {B.~R.}\ \bibnamefont
  {Ortiz}},\ }\bibfield  {title} {\bibinfo {title} {$\mathrm{AV_3Sb_5}$ kagome
  superconductors},\ }\bibfield  {journal} {\bibinfo  {journal} {Nature Reviews
  Materials}\ }\href {https://doi.org/10.1038/s41578-024-00677-y}
  {10.1038/s41578-024-00677-y} (\bibinfo {year} {2024})\BibitemShut {NoStop}%
\bibitem [{\citenamefont {Meier}\ \emph {et~al.}(2020)\citenamefont {Meier},
  \citenamefont {Du}, \citenamefont {Okamoto}, \citenamefont {Mohanta},
  \citenamefont {May}, \citenamefont {McGuire}, \citenamefont {Bridges},
  \citenamefont {Samolyuk},\ and\ \citenamefont {Sales}}]{Meier_2020}%
  \BibitemOpen
  \bibfield  {author} {\bibinfo {author} {\bibfnamefont {W.~R.}\ \bibnamefont
  {Meier}}, \bibinfo {author} {\bibfnamefont {M.-H.}\ \bibnamefont {Du}},
  \bibinfo {author} {\bibfnamefont {S.}~\bibnamefont {Okamoto}}, \bibinfo
  {author} {\bibfnamefont {N.}~\bibnamefont {Mohanta}}, \bibinfo {author}
  {\bibfnamefont {A.~F.}\ \bibnamefont {May}}, \bibinfo {author} {\bibfnamefont
  {M.~A.}\ \bibnamefont {McGuire}}, \bibinfo {author} {\bibfnamefont {C.~A.}\
  \bibnamefont {Bridges}}, \bibinfo {author} {\bibfnamefont {G.~D.}\
  \bibnamefont {Samolyuk}},\ and\ \bibinfo {author} {\bibfnamefont {B.~C.}\
  \bibnamefont {Sales}},\ }\bibfield  {title} {\bibinfo {title} {Flat bands in
  the cosn-type compounds},\ }\href
  {https://doi.org/10.1103/PhysRevB.102.075148} {\bibfield  {journal} {\bibinfo
   {journal} {Phys. Rev. B}\ }\textbf {\bibinfo {volume} {102}},\ \bibinfo
  {pages} {075148} (\bibinfo {year} {2020})}\BibitemShut {NoStop}%
\bibitem [{\citenamefont {Liu}\ \emph {et~al.}(2018)\citenamefont {Liu},
  \citenamefont {Sun}, \citenamefont {Kumar}, \citenamefont {Muechler},
  \citenamefont {Sun}, \citenamefont {Jiao}, \citenamefont {Yang},
  \citenamefont {Liu}, \citenamefont {Liang}, \citenamefont {Xu}, \citenamefont
  {Kroder}, \citenamefont {S{\'u}{\ss}}, \citenamefont {Borrmann},
  \citenamefont {Shekhar}, \citenamefont {Wang}, \citenamefont {Xi},
  \citenamefont {Wang}, \citenamefont {Schnelle}, \citenamefont {Wirth},
  \citenamefont {Chen}, \citenamefont {Goennenwein},\ and\ \citenamefont
  {Felser}}]{Liu2018}%
  \BibitemOpen
  \bibfield  {author} {\bibinfo {author} {\bibfnamefont {E.}~\bibnamefont
  {Liu}}, \bibinfo {author} {\bibfnamefont {Y.}~\bibnamefont {Sun}}, \bibinfo
  {author} {\bibfnamefont {N.}~\bibnamefont {Kumar}}, \bibinfo {author}
  {\bibfnamefont {L.}~\bibnamefont {Muechler}}, \bibinfo {author}
  {\bibfnamefont {A.}~\bibnamefont {Sun}}, \bibinfo {author} {\bibfnamefont
  {L.}~\bibnamefont {Jiao}}, \bibinfo {author} {\bibfnamefont {S.-Y.}\
  \bibnamefont {Yang}}, \bibinfo {author} {\bibfnamefont {D.}~\bibnamefont
  {Liu}}, \bibinfo {author} {\bibfnamefont {A.}~\bibnamefont {Liang}}, \bibinfo
  {author} {\bibfnamefont {Q.}~\bibnamefont {Xu}}, \bibinfo {author}
  {\bibfnamefont {J.}~\bibnamefont {Kroder}}, \bibinfo {author} {\bibfnamefont
  {V.}~\bibnamefont {S{\'u}{\ss}}}, \bibinfo {author} {\bibfnamefont
  {H.}~\bibnamefont {Borrmann}}, \bibinfo {author} {\bibfnamefont
  {C.}~\bibnamefont {Shekhar}}, \bibinfo {author} {\bibfnamefont
  {Z.}~\bibnamefont {Wang}}, \bibinfo {author} {\bibfnamefont {C.}~\bibnamefont
  {Xi}}, \bibinfo {author} {\bibfnamefont {W.}~\bibnamefont {Wang}}, \bibinfo
  {author} {\bibfnamefont {W.}~\bibnamefont {Schnelle}}, \bibinfo {author}
  {\bibfnamefont {S.}~\bibnamefont {Wirth}}, \bibinfo {author} {\bibfnamefont
  {Y.}~\bibnamefont {Chen}}, \bibinfo {author} {\bibfnamefont {S.~T.~B.}\
  \bibnamefont {Goennenwein}},\ and\ \bibinfo {author} {\bibfnamefont
  {C.}~\bibnamefont {Felser}},\ }\bibfield  {title} {\bibinfo {title} {Giant
  anomalous hall effect in a ferromagnetic kagome-lattice semimetal},\ }\href
  {https://doi.org/10.1038/s41567-018-0234-5} {\bibfield  {journal} {\bibinfo
  {journal} {Nature Physics}\ }\textbf {\bibinfo {volume} {14}},\ \bibinfo
  {pages} {1125} (\bibinfo {year} {2018})}\BibitemShut {NoStop}%
\bibitem [{\citenamefont {Yu}\ \emph {et~al.}(2021)\citenamefont {Yu},
  \citenamefont {Wu}, \citenamefont {Wang}, \citenamefont {Lei}, \citenamefont
  {Zhuo}, \citenamefont {Ying},\ and\ \citenamefont {Chen}}]{Yu2021}%
  \BibitemOpen
  \bibfield  {author} {\bibinfo {author} {\bibfnamefont {F.~H.}\ \bibnamefont
  {Yu}}, \bibinfo {author} {\bibfnamefont {T.}~\bibnamefont {Wu}}, \bibinfo
  {author} {\bibfnamefont {Z.~Y.}\ \bibnamefont {Wang}}, \bibinfo {author}
  {\bibfnamefont {B.}~\bibnamefont {Lei}}, \bibinfo {author} {\bibfnamefont
  {W.~Z.}\ \bibnamefont {Zhuo}}, \bibinfo {author} {\bibfnamefont {J.~J.}\
  \bibnamefont {Ying}},\ and\ \bibinfo {author} {\bibfnamefont {X.~H.}\
  \bibnamefont {Chen}},\ }\bibfield  {title} {\bibinfo {title} {Concurrence of
  anomalous hall effect and charge density wave in a superconducting
  topological kagome metal},\ }\href
  {https://doi.org/10.1103/PhysRevB.104.L041103} {\bibfield  {journal}
  {\bibinfo  {journal} {Phys. Rev. B}\ }\textbf {\bibinfo {volume} {104}},\
  \bibinfo {pages} {L041103} (\bibinfo {year} {2021})}\BibitemShut {NoStop}%
\bibitem [{\citenamefont {Yang}\ \emph {et~al.}(2020)\citenamefont {Yang},
  \citenamefont {Wang}, \citenamefont {Ortiz}, \citenamefont {Liu},
  \citenamefont {Gayles}, \citenamefont {Derunova}, \citenamefont
  {Gonzalez-Hernandez}, \citenamefont {Šmejkal}, \citenamefont {Chen},
  \citenamefont {Parkin}, \citenamefont {Wilson}, \citenamefont {Toberer},
  \citenamefont {McQueen},\ and\ \citenamefont {Ali}}]{Yang2020}%
  \BibitemOpen
  \bibfield  {author} {\bibinfo {author} {\bibfnamefont {S.-Y.}\ \bibnamefont
  {Yang}}, \bibinfo {author} {\bibfnamefont {Y.}~\bibnamefont {Wang}}, \bibinfo
  {author} {\bibfnamefont {B.~R.}\ \bibnamefont {Ortiz}}, \bibinfo {author}
  {\bibfnamefont {D.}~\bibnamefont {Liu}}, \bibinfo {author} {\bibfnamefont
  {J.}~\bibnamefont {Gayles}}, \bibinfo {author} {\bibfnamefont
  {E.}~\bibnamefont {Derunova}}, \bibinfo {author} {\bibfnamefont
  {R.}~\bibnamefont {Gonzalez-Hernandez}}, \bibinfo {author} {\bibfnamefont
  {L.}~\bibnamefont {Šmejkal}}, \bibinfo {author} {\bibfnamefont
  {Y.}~\bibnamefont {Chen}}, \bibinfo {author} {\bibfnamefont {S.~S.~P.}\
  \bibnamefont {Parkin}}, \bibinfo {author} {\bibfnamefont {S.~D.}\
  \bibnamefont {Wilson}}, \bibinfo {author} {\bibfnamefont {E.~S.}\
  \bibnamefont {Toberer}}, \bibinfo {author} {\bibfnamefont {T.}~\bibnamefont
  {McQueen}},\ and\ \bibinfo {author} {\bibfnamefont {M.~N.}\ \bibnamefont
  {Ali}},\ }\bibfield  {title} {\bibinfo {title} {Giant, unconventional
  anomalous hall effect in the metallic frustrated magnet candidate,
  $\mathrm{KV_3Sb_5}$},\ }\href {https://doi.org/10.1126/sciadv.abb6003}
  {\bibfield  {journal} {\bibinfo  {journal} {Science Advances}\ }\textbf
  {\bibinfo {volume} {6}},\ \bibinfo {pages} {eabb6003} (\bibinfo {year}
  {2020})}\BibitemShut {NoStop}%
\bibitem [{\citenamefont {Fujishiro}\ \emph {et~al.}(2021)\citenamefont
  {Fujishiro}, \citenamefont {Kanazawa}, \citenamefont {Kurihara},
  \citenamefont {Ishizuka}, \citenamefont {Hori}, \citenamefont {Yasin},
  \citenamefont {Yu}, \citenamefont {Tsukazaki}, \citenamefont {Ichikawa},
  \citenamefont {Kawasaki}, \citenamefont {Nagaosa}, \citenamefont {Tokunaga},\
  and\ \citenamefont {Tokura}}]{Fujishiro2021}%
  \BibitemOpen
  \bibfield  {author} {\bibinfo {author} {\bibfnamefont {Y.}~\bibnamefont
  {Fujishiro}}, \bibinfo {author} {\bibfnamefont {N.}~\bibnamefont {Kanazawa}},
  \bibinfo {author} {\bibfnamefont {R.}~\bibnamefont {Kurihara}}, \bibinfo
  {author} {\bibfnamefont {H.}~\bibnamefont {Ishizuka}}, \bibinfo {author}
  {\bibfnamefont {T.}~\bibnamefont {Hori}}, \bibinfo {author} {\bibfnamefont
  {F.~S.}\ \bibnamefont {Yasin}}, \bibinfo {author} {\bibfnamefont
  {X.}~\bibnamefont {Yu}}, \bibinfo {author} {\bibfnamefont {A.}~\bibnamefont
  {Tsukazaki}}, \bibinfo {author} {\bibfnamefont {M.}~\bibnamefont {Ichikawa}},
  \bibinfo {author} {\bibfnamefont {M.}~\bibnamefont {Kawasaki}}, \bibinfo
  {author} {\bibfnamefont {N.}~\bibnamefont {Nagaosa}}, \bibinfo {author}
  {\bibfnamefont {M.}~\bibnamefont {Tokunaga}},\ and\ \bibinfo {author}
  {\bibfnamefont {Y.}~\bibnamefont {Tokura}},\ }\bibfield  {title} {\bibinfo
  {title} {Giant anomalous hall effect from spin-chirality scattering in a
  chiral magnet},\ }\href {https://doi.org/10.1038/s41467-020-20384-w}
  {\bibfield  {journal} {\bibinfo  {journal} {Nature Communications}\ }\textbf
  {\bibinfo {volume} {12}},\ \bibinfo {pages} {317} (\bibinfo {year}
  {2021})}\BibitemShut {NoStop}%
\bibitem [{\citenamefont {Guo}\ \emph {et~al.}(2022)\citenamefont {Guo},
  \citenamefont {Putzke}, \citenamefont {Konyzheva}, \citenamefont {Huang},
  \citenamefont {Gutierrez-Amigo}, \citenamefont {Errea}, \citenamefont {Chen},
  \citenamefont {Vergniory}, \citenamefont {Felser}, \citenamefont {Fischer},
  \citenamefont {Neupert},\ and\ \citenamefont {Moll}}]{Guo2022}%
  \BibitemOpen
  \bibfield  {author} {\bibinfo {author} {\bibfnamefont {C.}~\bibnamefont
  {Guo}}, \bibinfo {author} {\bibfnamefont {C.}~\bibnamefont {Putzke}},
  \bibinfo {author} {\bibfnamefont {S.}~\bibnamefont {Konyzheva}}, \bibinfo
  {author} {\bibfnamefont {X.}~\bibnamefont {Huang}}, \bibinfo {author}
  {\bibfnamefont {M.}~\bibnamefont {Gutierrez-Amigo}}, \bibinfo {author}
  {\bibfnamefont {I.}~\bibnamefont {Errea}}, \bibinfo {author} {\bibfnamefont
  {D.}~\bibnamefont {Chen}}, \bibinfo {author} {\bibfnamefont {M.~G.}\
  \bibnamefont {Vergniory}}, \bibinfo {author} {\bibfnamefont {C.}~\bibnamefont
  {Felser}}, \bibinfo {author} {\bibfnamefont {M.~H.}\ \bibnamefont {Fischer}},
  \bibinfo {author} {\bibfnamefont {T.}~\bibnamefont {Neupert}},\ and\ \bibinfo
  {author} {\bibfnamefont {P.~J.~W.}\ \bibnamefont {Moll}},\ }\bibfield
  {title} {\bibinfo {title} {Switchable chiral transport in charge-ordered
  kagome metal csv3sb5},\ }\href {https://doi.org/10.1038/s41586-022-05127-9}
  {\bibfield  {journal} {\bibinfo  {journal} {Nature}\ }\textbf {\bibinfo
  {volume} {611}},\ \bibinfo {pages} {461} (\bibinfo {year}
  {2022})}\BibitemShut {NoStop}%
\bibitem [{\citenamefont {Mielke}\ \emph {et~al.}(2022)\citenamefont {Mielke},
  \citenamefont {Das}, \citenamefont {Yin}, \citenamefont {Liu}, \citenamefont
  {Gupta}, \citenamefont {Jiang}, \citenamefont {Medarde}, \citenamefont {Wu},
  \citenamefont {Lei}, \citenamefont {Chang}, \citenamefont {Dai},
  \citenamefont {Si}, \citenamefont {Miao}, \citenamefont {Thomale},
  \citenamefont {Neupert}, \citenamefont {Shi}, \citenamefont {Khasanov},
  \citenamefont {Hasan}, \citenamefont {Luetkens},\ and\ \citenamefont
  {Guguchia}}]{Mielke2022}%
  \BibitemOpen
  \bibfield  {author} {\bibinfo {author} {\bibfnamefont {C.}~\bibnamefont
  {Mielke}}, \bibinfo {author} {\bibfnamefont {D.}~\bibnamefont {Das}},
  \bibinfo {author} {\bibfnamefont {J.-X.}\ \bibnamefont {Yin}}, \bibinfo
  {author} {\bibfnamefont {H.}~\bibnamefont {Liu}}, \bibinfo {author}
  {\bibfnamefont {R.}~\bibnamefont {Gupta}}, \bibinfo {author} {\bibfnamefont
  {Y.-X.}\ \bibnamefont {Jiang}}, \bibinfo {author} {\bibfnamefont
  {M.}~\bibnamefont {Medarde}}, \bibinfo {author} {\bibfnamefont
  {X.}~\bibnamefont {Wu}}, \bibinfo {author} {\bibfnamefont {H.~C.}\
  \bibnamefont {Lei}}, \bibinfo {author} {\bibfnamefont {J.}~\bibnamefont
  {Chang}}, \bibinfo {author} {\bibfnamefont {P.}~\bibnamefont {Dai}}, \bibinfo
  {author} {\bibfnamefont {Q.}~\bibnamefont {Si}}, \bibinfo {author}
  {\bibfnamefont {H.}~\bibnamefont {Miao}}, \bibinfo {author} {\bibfnamefont
  {R.}~\bibnamefont {Thomale}}, \bibinfo {author} {\bibfnamefont
  {T.}~\bibnamefont {Neupert}}, \bibinfo {author} {\bibfnamefont
  {Y.}~\bibnamefont {Shi}}, \bibinfo {author} {\bibfnamefont {R.}~\bibnamefont
  {Khasanov}}, \bibinfo {author} {\bibfnamefont {M.~Z.}\ \bibnamefont {Hasan}},
  \bibinfo {author} {\bibfnamefont {H.}~\bibnamefont {Luetkens}},\ and\
  \bibinfo {author} {\bibfnamefont {Z.}~\bibnamefont {Guguchia}},\ }\bibfield
  {title} {\bibinfo {title} {Time-reversal symmetry-breaking charge order in a
  kagome superconductor},\ }\href {https://doi.org/10.1038/s41586-021-04327-z}
  {\bibfield  {journal} {\bibinfo  {journal} {Nature}\ }\textbf {\bibinfo
  {volume} {602}},\ \bibinfo {pages} {245} (\bibinfo {year}
  {2022})}\BibitemShut {NoStop}%
\bibitem [{\citenamefont {Jiang}\ \emph {et~al.}(2021)\citenamefont {Jiang},
  \citenamefont {Yin}, \citenamefont {Denner}, \citenamefont {Shumiya},
  \citenamefont {Ortiz}, \citenamefont {Xu}, \citenamefont {Guguchia},
  \citenamefont {He}, \citenamefont {Hossain}, \citenamefont {Liu},
  \citenamefont {Ruff}, \citenamefont {Kautzsch}, \citenamefont {Zhang},
  \citenamefont {Chang}, \citenamefont {Belopolski}, \citenamefont {Zhang},
  \citenamefont {Cochran}, \citenamefont {Multer}, \citenamefont {Litskevich},
  \citenamefont {Cheng}, \citenamefont {Yang}, \citenamefont {Wang},
  \citenamefont {Thomale}, \citenamefont {Neupert}, \citenamefont {Wilson},\
  and\ \citenamefont {Hasan}}]{Jiang2021}%
  \BibitemOpen
  \bibfield  {author} {\bibinfo {author} {\bibfnamefont {Y.-X.}\ \bibnamefont
  {Jiang}}, \bibinfo {author} {\bibfnamefont {J.-X.}\ \bibnamefont {Yin}},
  \bibinfo {author} {\bibfnamefont {M.~M.}\ \bibnamefont {Denner}}, \bibinfo
  {author} {\bibfnamefont {N.}~\bibnamefont {Shumiya}}, \bibinfo {author}
  {\bibfnamefont {B.~R.}\ \bibnamefont {Ortiz}}, \bibinfo {author}
  {\bibfnamefont {G.}~\bibnamefont {Xu}}, \bibinfo {author} {\bibfnamefont
  {Z.}~\bibnamefont {Guguchia}}, \bibinfo {author} {\bibfnamefont
  {J.}~\bibnamefont {He}}, \bibinfo {author} {\bibfnamefont {M.~S.}\
  \bibnamefont {Hossain}}, \bibinfo {author} {\bibfnamefont {X.}~\bibnamefont
  {Liu}}, \bibinfo {author} {\bibfnamefont {J.}~\bibnamefont {Ruff}}, \bibinfo
  {author} {\bibfnamefont {L.}~\bibnamefont {Kautzsch}}, \bibinfo {author}
  {\bibfnamefont {S.~S.}\ \bibnamefont {Zhang}}, \bibinfo {author}
  {\bibfnamefont {G.}~\bibnamefont {Chang}}, \bibinfo {author} {\bibfnamefont
  {I.}~\bibnamefont {Belopolski}}, \bibinfo {author} {\bibfnamefont
  {Q.}~\bibnamefont {Zhang}}, \bibinfo {author} {\bibfnamefont {T.~A.}\
  \bibnamefont {Cochran}}, \bibinfo {author} {\bibfnamefont {D.}~\bibnamefont
  {Multer}}, \bibinfo {author} {\bibfnamefont {M.}~\bibnamefont {Litskevich}},
  \bibinfo {author} {\bibfnamefont {Z.-J.}\ \bibnamefont {Cheng}}, \bibinfo
  {author} {\bibfnamefont {X.~P.}\ \bibnamefont {Yang}}, \bibinfo {author}
  {\bibfnamefont {Z.}~\bibnamefont {Wang}}, \bibinfo {author} {\bibfnamefont
  {R.}~\bibnamefont {Thomale}}, \bibinfo {author} {\bibfnamefont
  {T.}~\bibnamefont {Neupert}}, \bibinfo {author} {\bibfnamefont {S.~D.}\
  \bibnamefont {Wilson}},\ and\ \bibinfo {author} {\bibfnamefont {M.~Z.}\
  \bibnamefont {Hasan}},\ }\bibfield  {title} {\bibinfo {title} {Unconventional
  chiral charge order in kagome superconductor $\mathrm{KV_3Sb_5}$},\ }\href
  {https://doi.org/10.1038/s41563-021-01034-y} {\bibfield  {journal} {\bibinfo
  {journal} {Nature Materials}\ }\textbf {\bibinfo {volume} {20}},\ \bibinfo
  {pages} {1353} (\bibinfo {year} {2021})}\BibitemShut {NoStop}%
\bibitem [{\citenamefont {Neupert}\ \emph {et~al.}(2022)\citenamefont
  {Neupert}, \citenamefont {Denner}, \citenamefont {Yin}, \citenamefont
  {Thomale},\ and\ \citenamefont {Hasan}}]{Neupert2022}%
  \BibitemOpen
  \bibfield  {author} {\bibinfo {author} {\bibfnamefont {T.}~\bibnamefont
  {Neupert}}, \bibinfo {author} {\bibfnamefont {M.~M.}\ \bibnamefont {Denner}},
  \bibinfo {author} {\bibfnamefont {J.-X.}\ \bibnamefont {Yin}}, \bibinfo
  {author} {\bibfnamefont {R.}~\bibnamefont {Thomale}},\ and\ \bibinfo {author}
  {\bibfnamefont {M.~Z.}\ \bibnamefont {Hasan}},\ }\bibfield  {title} {\bibinfo
  {title} {Charge order and superconductivity in kagome materials},\ }\href
  {https://doi.org/10.1038/s41567-021-01404-y} {\bibfield  {journal} {\bibinfo
  {journal} {Nature Physics}\ }\textbf {\bibinfo {volume} {18}},\ \bibinfo
  {pages} {137} (\bibinfo {year} {2022})}\BibitemShut {NoStop}%
\bibitem [{\citenamefont {Yin}\ \emph {et~al.}(2022{\natexlab{a}})\citenamefont
  {Yin}, \citenamefont {Lian},\ and\ \citenamefont {Hasan}}]{Yin2022}%
  \BibitemOpen
  \bibfield  {author} {\bibinfo {author} {\bibfnamefont {J.-X.}\ \bibnamefont
  {Yin}}, \bibinfo {author} {\bibfnamefont {B.}~\bibnamefont {Lian}},\ and\
  \bibinfo {author} {\bibfnamefont {M.~Z.}\ \bibnamefont {Hasan}},\ }\bibfield
  {title} {\bibinfo {title} {Topological kagome magnets and superconductors},\
  }\href {https://doi.org/10.1038/s41586-022-05516-0} {\bibfield  {journal}
  {\bibinfo  {journal} {Nature}\ }\textbf {\bibinfo {volume} {612}},\ \bibinfo
  {pages} {647} (\bibinfo {year} {2022}{\natexlab{a}})}\BibitemShut {NoStop}%
\bibitem [{\citenamefont {Denner}\ \emph {et~al.}(2021)\citenamefont {Denner},
  \citenamefont {Thomale},\ and\ \citenamefont {Neupert}}]{Denner_2021}%
  \BibitemOpen
  \bibfield  {author} {\bibinfo {author} {\bibfnamefont {M.~M.}\ \bibnamefont
  {Denner}}, \bibinfo {author} {\bibfnamefont {R.}~\bibnamefont {Thomale}},\
  and\ \bibinfo {author} {\bibfnamefont {T.}~\bibnamefont {Neupert}},\
  }\bibfield  {title} {\bibinfo {title} {Analysis of charge order in the kagome
  metal $\mathrm{AV}_{3}\mathrm{Sb}_{5}$
  ($\mathrm{A}=\mathrm{K},\mathrm{Rb},\mathrm{Cs}$)},\ }\href
  {https://doi.org/10.1103/PhysRevLett.127.217601} {\bibfield  {journal}
  {\bibinfo  {journal} {Phys. Rev. Lett.}\ }\textbf {\bibinfo {volume} {127}},\
  \bibinfo {pages} {217601} (\bibinfo {year} {2021})}\BibitemShut {NoStop}%
\bibitem [{\citenamefont {Li}\ \emph {et~al.}(2024)\citenamefont {Li},
  \citenamefont {Kim},\ and\ \citenamefont {Kee}}]{Li2024}%
  \BibitemOpen
  \bibfield  {author} {\bibinfo {author} {\bibfnamefont {H.}~\bibnamefont
  {Li}}, \bibinfo {author} {\bibfnamefont {Y.~B.}\ \bibnamefont {Kim}},\ and\
  \bibinfo {author} {\bibfnamefont {H.-Y.}\ \bibnamefont {Kee}},\ }\bibfield
  {title} {\bibinfo {title} {Intertwined van hove singularities as a mechanism
  for loop current order in kagome metals},\ }\href
  {https://doi.org/10.1103/PhysRevLett.132.146501} {\bibfield  {journal}
  {\bibinfo  {journal} {Phys. Rev. Lett.}\ }\textbf {\bibinfo {volume} {132}},\
  \bibinfo {pages} {146501} (\bibinfo {year} {2024})}\BibitemShut {NoStop}%
\bibitem [{\citenamefont {Dong}\ \emph {et~al.}(2023)\citenamefont {Dong},
  \citenamefont {Wang},\ and\ \citenamefont {Zhou}}]{Dong2023}%
  \BibitemOpen
  \bibfield  {author} {\bibinfo {author} {\bibfnamefont {J.-W.}\ \bibnamefont
  {Dong}}, \bibinfo {author} {\bibfnamefont {Z.}~\bibnamefont {Wang}},\ and\
  \bibinfo {author} {\bibfnamefont {S.}~\bibnamefont {Zhou}},\ }\bibfield
  {title} {\bibinfo {title} {Loop-current charge density wave driven by
  long-range coulomb repulsion on the kagom\'e lattice},\ }\href
  {https://doi.org/10.1103/PhysRevB.107.045127} {\bibfield  {journal} {\bibinfo
   {journal} {Phys. Rev. B}\ }\textbf {\bibinfo {volume} {107}},\ \bibinfo
  {pages} {045127} (\bibinfo {year} {2023})}\BibitemShut {NoStop}%
\bibitem [{\citenamefont {Christensen}\ \emph {et~al.}(2022)\citenamefont
  {Christensen}, \citenamefont {Birol}, \citenamefont {Andersen},\ and\
  \citenamefont {Fernandes}}]{Christensen2022}%
  \BibitemOpen
  \bibfield  {author} {\bibinfo {author} {\bibfnamefont {M.~H.}\ \bibnamefont
  {Christensen}}, \bibinfo {author} {\bibfnamefont {T.}~\bibnamefont {Birol}},
  \bibinfo {author} {\bibfnamefont {B.~M.}\ \bibnamefont {Andersen}},\ and\
  \bibinfo {author} {\bibfnamefont {R.~M.}\ \bibnamefont {Fernandes}},\
  }\bibfield  {title} {\bibinfo {title} {Loop currents in
  $a{\mathrm{v}}_{3}{\mathrm{sb}}_{5}$ kagome metals: Multipolar and toroidal
  magnetic orders},\ }\href {https://doi.org/10.1103/PhysRevB.106.144504}
  {\bibfield  {journal} {\bibinfo  {journal} {Phys. Rev. B}\ }\textbf {\bibinfo
  {volume} {106}},\ \bibinfo {pages} {144504} (\bibinfo {year}
  {2022})}\BibitemShut {NoStop}%
\bibitem [{\citenamefont {Tazai}\ \emph {et~al.}(2023)\citenamefont {Tazai},
  \citenamefont {Yamakawa},\ and\ \citenamefont {Kontani}}]{Tazai2023}%
  \BibitemOpen
  \bibfield  {author} {\bibinfo {author} {\bibfnamefont {R.}~\bibnamefont
  {Tazai}}, \bibinfo {author} {\bibfnamefont {Y.}~\bibnamefont {Yamakawa}},\
  and\ \bibinfo {author} {\bibfnamefont {H.}~\bibnamefont {Kontani}},\
  }\bibfield  {title} {\bibinfo {title} {Charge-loop current order and z$_3$
  nematicity mediated by bond order fluctuations in kagome metals},\ }\href
  {https://doi.org/10.1038/s41467-023-42952-6} {\bibfield  {journal} {\bibinfo
  {journal} {Nature Communications}\ }\textbf {\bibinfo {volume} {14}},\
  \bibinfo {pages} {7845} (\bibinfo {year} {2023})}\BibitemShut {NoStop}%
\bibitem [{\citenamefont {Kiesel}\ and\ \citenamefont
  {Thomale}(2012)}]{Ronny_PRB_2012}%
  \BibitemOpen
  \bibfield  {author} {\bibinfo {author} {\bibfnamefont {M.~L.}\ \bibnamefont
  {Kiesel}}\ and\ \bibinfo {author} {\bibfnamefont {R.}~\bibnamefont
  {Thomale}},\ }\bibfield  {title} {\bibinfo {title} {Sublattice interference
  in the kagome hubbard model},\ }\href
  {https://doi.org/10.1103/PhysRevB.86.121105} {\bibfield  {journal} {\bibinfo
  {journal} {Phys. Rev. B}\ }\textbf {\bibinfo {volume} {86}},\ \bibinfo
  {pages} {121105} (\bibinfo {year} {2012})}\BibitemShut {NoStop}%
\bibitem [{\citenamefont {Kiesel}\ \emph {et~al.}(2013)\citenamefont {Kiesel},
  \citenamefont {Platt},\ and\ \citenamefont {Thomale}}]{Ronny_PRL_2013}%
  \BibitemOpen
  \bibfield  {author} {\bibinfo {author} {\bibfnamefont {M.~L.}\ \bibnamefont
  {Kiesel}}, \bibinfo {author} {\bibfnamefont {C.}~\bibnamefont {Platt}},\ and\
  \bibinfo {author} {\bibfnamefont {R.}~\bibnamefont {Thomale}},\ }\bibfield
  {title} {\bibinfo {title} {Unconventional fermi surface instabilities in the
  kagome hubbard model},\ }\href
  {https://doi.org/10.1103/PhysRevLett.110.126405} {\bibfield  {journal}
  {\bibinfo  {journal} {Phys. Rev. Lett.}\ }\textbf {\bibinfo {volume} {110}},\
  \bibinfo {pages} {126405} (\bibinfo {year} {2013})}\BibitemShut {NoStop}%
\bibitem [{\citenamefont {Huang}\ and\ \citenamefont {Lu}(2020)}]{Huang2020}%
  \BibitemOpen
  \bibfield  {author} {\bibinfo {author} {\bibfnamefont {L.}~\bibnamefont
  {Huang}}\ and\ \bibinfo {author} {\bibfnamefont {H.}~\bibnamefont {Lu}},\
  }\bibfield  {title} {\bibinfo {title} {Protracted kondo screening and kagome
  bands in the heavy-fermion metal ${\mathrm{ce}}_{3}\mathrm{Al}$},\ }\href
  {https://doi.org/10.1103/PhysRevB.102.155140} {\bibfield  {journal} {\bibinfo
   {journal} {Phys. Rev. B}\ }\textbf {\bibinfo {volume} {102}},\ \bibinfo
  {pages} {155140} (\bibinfo {year} {2020})}\BibitemShut {NoStop}%
\bibitem [{\citenamefont {Kourris}\ and\ \citenamefont
  {Vojta}(2023)}]{Kourris2023}%
  \BibitemOpen
  \bibfield  {author} {\bibinfo {author} {\bibfnamefont {C.}~\bibnamefont
  {Kourris}}\ and\ \bibinfo {author} {\bibfnamefont {M.}~\bibnamefont
  {Vojta}},\ }\bibfield  {title} {\bibinfo {title} {Kondo screening and
  coherence in kagome local-moment metals: Energy scales of heavy fermions in
  the presence of flat bands},\ }\href
  {https://doi.org/10.1103/PhysRevB.108.235106} {\bibfield  {journal} {\bibinfo
   {journal} {Phys. Rev. B}\ }\textbf {\bibinfo {volume} {108}},\ \bibinfo
  {pages} {235106} (\bibinfo {year} {2023})}\BibitemShut {NoStop}%
\bibitem [{\citenamefont {Ortiz}\ \emph {et~al.}(2019)\citenamefont {Ortiz},
  \citenamefont {Gomes}, \citenamefont {Morey}, \citenamefont {Winiarski},
  \citenamefont {Bordelon}, \citenamefont {Mangum}, \citenamefont {Oswald},
  \citenamefont {Rodriguez-Rivera}, \citenamefont {Neilson}, \citenamefont
  {Wilson}, \citenamefont {Ertekin}, \citenamefont {McQueen},\ and\
  \citenamefont {Toberer}}]{Ortiz_2019}%
  \BibitemOpen
  \bibfield  {author} {\bibinfo {author} {\bibfnamefont {B.~R.}\ \bibnamefont
  {Ortiz}}, \bibinfo {author} {\bibfnamefont {L.~C.}\ \bibnamefont {Gomes}},
  \bibinfo {author} {\bibfnamefont {J.~R.}\ \bibnamefont {Morey}}, \bibinfo
  {author} {\bibfnamefont {M.}~\bibnamefont {Winiarski}}, \bibinfo {author}
  {\bibfnamefont {M.}~\bibnamefont {Bordelon}}, \bibinfo {author}
  {\bibfnamefont {J.~S.}\ \bibnamefont {Mangum}}, \bibinfo {author}
  {\bibfnamefont {I.~W.~H.}\ \bibnamefont {Oswald}}, \bibinfo {author}
  {\bibfnamefont {J.~A.}\ \bibnamefont {Rodriguez-Rivera}}, \bibinfo {author}
  {\bibfnamefont {J.~R.}\ \bibnamefont {Neilson}}, \bibinfo {author}
  {\bibfnamefont {S.~D.}\ \bibnamefont {Wilson}}, \bibinfo {author}
  {\bibfnamefont {E.}~\bibnamefont {Ertekin}}, \bibinfo {author} {\bibfnamefont
  {T.~M.}\ \bibnamefont {McQueen}},\ and\ \bibinfo {author} {\bibfnamefont
  {E.~S.}\ \bibnamefont {Toberer}},\ }\bibfield  {title} {\bibinfo {title} {New
  kagome prototype materials: discovery of $\mathrm{KV_3Sb_5}$,
  $\mathrm{RbV_3Sb_5}$, and $\mathrm{CsV_3Sb_5}$},\ }\href
  {https://doi.org/10.1103/PhysRevMaterials.3.094407} {\bibfield  {journal}
  {\bibinfo  {journal} {Phys. Rev. Mater.}\ }\textbf {\bibinfo {volume} {3}},\
  \bibinfo {pages} {094407} (\bibinfo {year} {2019})}\BibitemShut {NoStop}%
\bibitem [{\citenamefont {Ortiz}\ \emph {et~al.}(2020)\citenamefont {Ortiz},
  \citenamefont {Teicher}, \citenamefont {Hu}, \citenamefont {Zuo},
  \citenamefont {Sarte}, \citenamefont {Schueller}, \citenamefont {Abeykoon},
  \citenamefont {Krogstad}, \citenamefont {Rosenkranz}, \citenamefont {Osborn},
  \citenamefont {Seshadri}, \citenamefont {Balents}, \citenamefont {He},\ and\
  \citenamefont {Wilson}}]{Ortiz_2020}%
  \BibitemOpen
  \bibfield  {author} {\bibinfo {author} {\bibfnamefont {B.~R.}\ \bibnamefont
  {Ortiz}}, \bibinfo {author} {\bibfnamefont {S.~M.~L.}\ \bibnamefont
  {Teicher}}, \bibinfo {author} {\bibfnamefont {Y.}~\bibnamefont {Hu}},
  \bibinfo {author} {\bibfnamefont {J.~L.}\ \bibnamefont {Zuo}}, \bibinfo
  {author} {\bibfnamefont {P.~M.}\ \bibnamefont {Sarte}}, \bibinfo {author}
  {\bibfnamefont {E.~C.}\ \bibnamefont {Schueller}}, \bibinfo {author}
  {\bibfnamefont {A.~M.~M.}\ \bibnamefont {Abeykoon}}, \bibinfo {author}
  {\bibfnamefont {M.~J.}\ \bibnamefont {Krogstad}}, \bibinfo {author}
  {\bibfnamefont {S.}~\bibnamefont {Rosenkranz}}, \bibinfo {author}
  {\bibfnamefont {R.}~\bibnamefont {Osborn}}, \bibinfo {author} {\bibfnamefont
  {R.}~\bibnamefont {Seshadri}}, \bibinfo {author} {\bibfnamefont
  {L.}~\bibnamefont {Balents}}, \bibinfo {author} {\bibfnamefont
  {J.}~\bibnamefont {He}},\ and\ \bibinfo {author} {\bibfnamefont {S.~D.}\
  \bibnamefont {Wilson}},\ }\bibfield  {title} {\bibinfo {title}
  {$\mathrm{CsV_3Sb_5}$: A ${\mathbb{z}}_{2}$ topological kagome metal with a
  superconducting ground state},\ }\href
  {https://doi.org/10.1103/PhysRevLett.125.247002} {\bibfield  {journal}
  {\bibinfo  {journal} {Phys. Rev. Lett.}\ }\textbf {\bibinfo {volume} {125}},\
  \bibinfo {pages} {247002} (\bibinfo {year} {2020})}\BibitemShut {NoStop}%
\bibitem [{\citenamefont {Arachchige}\ \emph {et~al.}(2022)\citenamefont
  {Arachchige}, \citenamefont {Meier}, \citenamefont {Marshall}, \citenamefont
  {Matsuoka}, \citenamefont {Xue}, \citenamefont {McGuire}, \citenamefont
  {Hermann}, \citenamefont {Cao},\ and\ \citenamefont
  {Mandrus}}]{Arachchige_2022}%
  \BibitemOpen
  \bibfield  {author} {\bibinfo {author} {\bibfnamefont {H.~W.~S.}\
  \bibnamefont {Arachchige}}, \bibinfo {author} {\bibfnamefont {W.~R.}\
  \bibnamefont {Meier}}, \bibinfo {author} {\bibfnamefont {M.}~\bibnamefont
  {Marshall}}, \bibinfo {author} {\bibfnamefont {T.}~\bibnamefont {Matsuoka}},
  \bibinfo {author} {\bibfnamefont {R.}~\bibnamefont {Xue}}, \bibinfo {author}
  {\bibfnamefont {M.~A.}\ \bibnamefont {McGuire}}, \bibinfo {author}
  {\bibfnamefont {R.~P.}\ \bibnamefont {Hermann}}, \bibinfo {author}
  {\bibfnamefont {H.}~\bibnamefont {Cao}},\ and\ \bibinfo {author}
  {\bibfnamefont {D.}~\bibnamefont {Mandrus}},\ }\bibfield  {title} {\bibinfo
  {title} {Charge density wave in kagome lattice intermetallic
  $\mathrm{ScV_6Sn_6}$},\ }\href
  {https://doi.org/10.1103/PhysRevLett.129.216402} {\bibfield  {journal}
  {\bibinfo  {journal} {Phys. Rev. Lett.}\ }\textbf {\bibinfo {volume} {129}},\
  \bibinfo {pages} {216402} (\bibinfo {year} {2022})}\BibitemShut {NoStop}%
\bibitem [{\citenamefont {Tan}\ \emph {et~al.}(2021)\citenamefont {Tan},
  \citenamefont {Liu}, \citenamefont {Wang},\ and\ \citenamefont
  {Yan}}]{Tan_2021}%
  \BibitemOpen
  \bibfield  {author} {\bibinfo {author} {\bibfnamefont {H.}~\bibnamefont
  {Tan}}, \bibinfo {author} {\bibfnamefont {Y.}~\bibnamefont {Liu}}, \bibinfo
  {author} {\bibfnamefont {Z.}~\bibnamefont {Wang}},\ and\ \bibinfo {author}
  {\bibfnamefont {B.}~\bibnamefont {Yan}},\ }\bibfield  {title} {\bibinfo
  {title} {Charge density waves and electronic properties of superconducting
  kagome metals},\ }\href {https://doi.org/10.1103/PhysRevLett.127.046401}
  {\bibfield  {journal} {\bibinfo  {journal} {Phys. Rev. Lett.}\ }\textbf
  {\bibinfo {volume} {127}},\ \bibinfo {pages} {046401} (\bibinfo {year}
  {2021})}\BibitemShut {NoStop}%
\bibitem [{\citenamefont {Subedi}(2022)}]{Subedi_2022}%
  \BibitemOpen
  \bibfield  {author} {\bibinfo {author} {\bibfnamefont {A.}~\bibnamefont
  {Subedi}},\ }\bibfield  {title} {\bibinfo {title}
  {Hexagonal-to-base-centered-orthorhombic $4q$ charge density wave order in
  kagome metals $\mathrm{KV_3Sb_5}$, $\mathrm{RbV_3Sb_5}$ and
  $\mathrm{CsV_3Sb_5}$},\ }\href
  {https://doi.org/10.1103/PhysRevMaterials.6.015001} {\bibfield  {journal}
  {\bibinfo  {journal} {Phys. Rev. Mater.}\ }\textbf {\bibinfo {volume} {6}},\
  \bibinfo {pages} {015001} (\bibinfo {year} {2022})}\BibitemShut {NoStop}%
\bibitem [{\citenamefont {Li}\ \emph {et~al.}(2021)\citenamefont {Li},
  \citenamefont {Zhang}, \citenamefont {Yilmaz}, \citenamefont {Pai},
  \citenamefont {Marvinney}, \citenamefont {Said}, \citenamefont {Yin},
  \citenamefont {Gong}, \citenamefont {Tu}, \citenamefont {Vescovo},
  \citenamefont {Nelson}, \citenamefont {Moore}, \citenamefont {Murakami},
  \citenamefont {Lei}, \citenamefont {Lee}, \citenamefont {Lawrie},\ and\
  \citenamefont {Miao}}]{Miao_2021}%
  \BibitemOpen
  \bibfield  {author} {\bibinfo {author} {\bibfnamefont {H.}~\bibnamefont
  {Li}}, \bibinfo {author} {\bibfnamefont {T.~T.}\ \bibnamefont {Zhang}},
  \bibinfo {author} {\bibfnamefont {T.}~\bibnamefont {Yilmaz}}, \bibinfo
  {author} {\bibfnamefont {Y.~Y.}\ \bibnamefont {Pai}}, \bibinfo {author}
  {\bibfnamefont {C.~E.}\ \bibnamefont {Marvinney}}, \bibinfo {author}
  {\bibfnamefont {A.}~\bibnamefont {Said}}, \bibinfo {author} {\bibfnamefont
  {Q.~W.}\ \bibnamefont {Yin}}, \bibinfo {author} {\bibfnamefont {C.~S.}\
  \bibnamefont {Gong}}, \bibinfo {author} {\bibfnamefont {Z.~J.}\ \bibnamefont
  {Tu}}, \bibinfo {author} {\bibfnamefont {E.}~\bibnamefont {Vescovo}},
  \bibinfo {author} {\bibfnamefont {C.~S.}\ \bibnamefont {Nelson}}, \bibinfo
  {author} {\bibfnamefont {R.~G.}\ \bibnamefont {Moore}}, \bibinfo {author}
  {\bibfnamefont {S.}~\bibnamefont {Murakami}}, \bibinfo {author}
  {\bibfnamefont {H.~C.}\ \bibnamefont {Lei}}, \bibinfo {author} {\bibfnamefont
  {H.~N.}\ \bibnamefont {Lee}}, \bibinfo {author} {\bibfnamefont {B.~J.}\
  \bibnamefont {Lawrie}},\ and\ \bibinfo {author} {\bibfnamefont
  {H.}~\bibnamefont {Miao}},\ }\bibfield  {title} {\bibinfo {title}
  {Observation of unconventional charge density wave without acoustic phonon
  anomaly in kagome superconductors $\mathrm{AV_3Sb_5}$ ($\mathrm{A=Rb,
  Cs}$)},\ }\href {https://doi.org/10.1103/PhysRevX.11.031050} {\bibfield
  {journal} {\bibinfo  {journal} {Phys. Rev. X}\ }\textbf {\bibinfo {volume}
  {11}},\ \bibinfo {pages} {031050} (\bibinfo {year} {2021})}\BibitemShut
  {NoStop}%
\bibitem [{\citenamefont {Ratcliff}\ \emph {et~al.}(2021)\citenamefont
  {Ratcliff}, \citenamefont {Hallett}, \citenamefont {Ortiz}, \citenamefont
  {Wilson},\ and\ \citenamefont {Harter}}]{Ratcliff_2021}%
  \BibitemOpen
  \bibfield  {author} {\bibinfo {author} {\bibfnamefont {N.}~\bibnamefont
  {Ratcliff}}, \bibinfo {author} {\bibfnamefont {L.}~\bibnamefont {Hallett}},
  \bibinfo {author} {\bibfnamefont {B.~R.}\ \bibnamefont {Ortiz}}, \bibinfo
  {author} {\bibfnamefont {S.~D.}\ \bibnamefont {Wilson}},\ and\ \bibinfo
  {author} {\bibfnamefont {J.~W.}\ \bibnamefont {Harter}},\ }\bibfield  {title}
  {\bibinfo {title} {Coherent phonon spectroscopy and interlayer modulation of
  charge density wave order in the kagome metal $\mathrm{CsV_3Sb_5}$},\ }\href
  {https://doi.org/10.1103/PhysRevMaterials.5.L111801} {\bibfield  {journal}
  {\bibinfo  {journal} {Phys. Rev. Mater.}\ }\textbf {\bibinfo {volume} {5}},\
  \bibinfo {pages} {L111801} (\bibinfo {year} {2021})}\BibitemShut {NoStop}%
\bibitem [{\citenamefont {Subires}\ \emph {et~al.}(2023)\citenamefont
  {Subires}, \citenamefont {Korshunov}, \citenamefont {Said}, \citenamefont
  {S{\'a}nchez}, \citenamefont {Ortiz}, \citenamefont {Wilson}, \citenamefont
  {Bosak},\ and\ \citenamefont {Blanco-Canosa}}]{Subires_2023}%
  \BibitemOpen
  \bibfield  {author} {\bibinfo {author} {\bibfnamefont {D.}~\bibnamefont
  {Subires}}, \bibinfo {author} {\bibfnamefont {A.}~\bibnamefont {Korshunov}},
  \bibinfo {author} {\bibfnamefont {A.~H.}\ \bibnamefont {Said}}, \bibinfo
  {author} {\bibfnamefont {L.}~\bibnamefont {S{\'a}nchez}}, \bibinfo {author}
  {\bibfnamefont {B.~R.}\ \bibnamefont {Ortiz}}, \bibinfo {author}
  {\bibfnamefont {S.~D.}\ \bibnamefont {Wilson}}, \bibinfo {author}
  {\bibfnamefont {A.}~\bibnamefont {Bosak}},\ and\ \bibinfo {author}
  {\bibfnamefont {S.}~\bibnamefont {Blanco-Canosa}},\ }\bibfield  {title}
  {\bibinfo {title} {Order-disorder charge density wave instability in the
  kagome metal $\mathrm{(Cs,Rb)V_3Sb_5}$},\ }\href
  {https://doi.org/10.1038/s41467-023-36668-w} {\bibfield  {journal} {\bibinfo
  {journal} {Nature Communications}\ }\textbf {\bibinfo {volume} {14}},\
  \bibinfo {pages} {1015} (\bibinfo {year} {2023})}\BibitemShut {NoStop}%
\bibitem [{\citenamefont {Korshunov}\ \emph {et~al.}(2023)\citenamefont
  {Korshunov}, \citenamefont {Hu}, \citenamefont {Subires}, \citenamefont
  {Jiang}, \citenamefont {C{\u{a}}lug{\u{a}}ru}, \citenamefont {Feng},
  \citenamefont {Rajapitamahuni}, \citenamefont {Yi}, \citenamefont
  {Roychowdhury}, \citenamefont {Vergniory}, \citenamefont {Strempfer},
  \citenamefont {Shekhar}, \citenamefont {Vescovo}, \citenamefont {Chernyshov},
  \citenamefont {Said}, \citenamefont {Bosak}, \citenamefont {Felser},
  \citenamefont {Bernevig},\ and\ \citenamefont
  {Blanco-Canosa}}]{Korshunov_2023}%
  \BibitemOpen
  \bibfield  {author} {\bibinfo {author} {\bibfnamefont {A.}~\bibnamefont
  {Korshunov}}, \bibinfo {author} {\bibfnamefont {H.}~\bibnamefont {Hu}},
  \bibinfo {author} {\bibfnamefont {D.}~\bibnamefont {Subires}}, \bibinfo
  {author} {\bibfnamefont {Y.}~\bibnamefont {Jiang}}, \bibinfo {author}
  {\bibfnamefont {D.}~\bibnamefont {C{\u{a}}lug{\u{a}}ru}}, \bibinfo {author}
  {\bibfnamefont {X.}~\bibnamefont {Feng}}, \bibinfo {author} {\bibfnamefont
  {A.}~\bibnamefont {Rajapitamahuni}}, \bibinfo {author} {\bibfnamefont
  {C.}~\bibnamefont {Yi}}, \bibinfo {author} {\bibfnamefont {S.}~\bibnamefont
  {Roychowdhury}}, \bibinfo {author} {\bibfnamefont {M.~G.}\ \bibnamefont
  {Vergniory}}, \bibinfo {author} {\bibfnamefont {J.}~\bibnamefont
  {Strempfer}}, \bibinfo {author} {\bibfnamefont {C.}~\bibnamefont {Shekhar}},
  \bibinfo {author} {\bibfnamefont {E.}~\bibnamefont {Vescovo}}, \bibinfo
  {author} {\bibfnamefont {D.}~\bibnamefont {Chernyshov}}, \bibinfo {author}
  {\bibfnamefont {A.~H.}\ \bibnamefont {Said}}, \bibinfo {author}
  {\bibfnamefont {A.}~\bibnamefont {Bosak}}, \bibinfo {author} {\bibfnamefont
  {C.}~\bibnamefont {Felser}}, \bibinfo {author} {\bibfnamefont {B.~A.}\
  \bibnamefont {Bernevig}},\ and\ \bibinfo {author} {\bibfnamefont
  {S.}~\bibnamefont {Blanco-Canosa}},\ }\bibfield  {title} {\bibinfo {title}
  {Softening of a flat phonon mode in the kagome $\mathrm{ScV_6Sn_6}$},\ }\href
  {https://doi.org/10.1038/s41467-023-42186-6} {\bibfield  {journal} {\bibinfo
  {journal} {Nature Communications}\ }\textbf {\bibinfo {volume} {14}},\
  \bibinfo {pages} {6646} (\bibinfo {year} {2023})}\BibitemShut {NoStop}%
\bibitem [{\citenamefont {Hu}\ \emph {et~al.}(2023)\citenamefont {Hu},
  \citenamefont {Jiang}, \citenamefont {Călugăru}, \citenamefont {Feng},
  \citenamefont {Subires}, \citenamefont {Vergniory}, \citenamefont {Felser},
  \citenamefont {Blanco-Canosa},\ and\ \citenamefont {Bernevig}}]{Hu_2023}%
  \BibitemOpen
  \bibfield  {author} {\bibinfo {author} {\bibfnamefont {H.}~\bibnamefont
  {Hu}}, \bibinfo {author} {\bibfnamefont {Y.}~\bibnamefont {Jiang}}, \bibinfo
  {author} {\bibfnamefont {D.}~\bibnamefont {Călugăru}}, \bibinfo {author}
  {\bibfnamefont {X.}~\bibnamefont {Feng}}, \bibinfo {author} {\bibfnamefont
  {D.}~\bibnamefont {Subires}}, \bibinfo {author} {\bibfnamefont {M.~G.}\
  \bibnamefont {Vergniory}}, \bibinfo {author} {\bibfnamefont {C.}~\bibnamefont
  {Felser}}, \bibinfo {author} {\bibfnamefont {S.}~\bibnamefont
  {Blanco-Canosa}},\ and\ \bibinfo {author} {\bibfnamefont {B.~A.}\
  \bibnamefont {Bernevig}},\ }\href@noop {} {\bibinfo {title} {Kagome materials
  i: Sg 191, scv$_6$sn$_6$. flat phonon soft modes and unconventional cdw
  formation: Microscopic and effective theory}} (\bibinfo {year} {2023}),\
  \Eprint {https://arxiv.org/abs/2305.15469} {arXiv:2305.15469
  [cond-mat.str-el]} \BibitemShut {NoStop}%
\bibitem [{\citenamefont {Cao}\ \emph {et~al.}(2023)\citenamefont {Cao},
  \citenamefont {Xu}, \citenamefont {Fukui}, \citenamefont {Manjo},
  \citenamefont {Dong}, \citenamefont {Shi}, \citenamefont {Liu}, \citenamefont
  {Cao},\ and\ \citenamefont {Song}}]{Cao_2023}%
  \BibitemOpen
  \bibfield  {author} {\bibinfo {author} {\bibfnamefont {S.}~\bibnamefont
  {Cao}}, \bibinfo {author} {\bibfnamefont {C.}~\bibnamefont {Xu}}, \bibinfo
  {author} {\bibfnamefont {H.}~\bibnamefont {Fukui}}, \bibinfo {author}
  {\bibfnamefont {T.}~\bibnamefont {Manjo}}, \bibinfo {author} {\bibfnamefont
  {Y.}~\bibnamefont {Dong}}, \bibinfo {author} {\bibfnamefont {M.}~\bibnamefont
  {Shi}}, \bibinfo {author} {\bibfnamefont {Y.}~\bibnamefont {Liu}}, \bibinfo
  {author} {\bibfnamefont {C.}~\bibnamefont {Cao}},\ and\ \bibinfo {author}
  {\bibfnamefont {Y.}~\bibnamefont {Song}},\ }\bibfield  {title} {\bibinfo
  {title} {Competing charge-density wave instabilities in the kagome metal
  scv6sn6},\ }\href {https://doi.org/10.1038/s41467-023-43454-1} {\bibfield
  {journal} {\bibinfo  {journal} {Nature Communications}\ }\textbf {\bibinfo
  {volume} {14}},\ \bibinfo {pages} {7671} (\bibinfo {year}
  {2023})}\BibitemShut {NoStop}%
\bibitem [{\citenamefont {Tan}\ and\ \citenamefont {Yan}(2023)}]{Tan_2023}%
  \BibitemOpen
  \bibfield  {author} {\bibinfo {author} {\bibfnamefont {H.}~\bibnamefont
  {Tan}}\ and\ \bibinfo {author} {\bibfnamefont {B.}~\bibnamefont {Yan}},\
  }\bibfield  {title} {\bibinfo {title} {Abundant lattice instability in kagome
  metal $\mathrm{ScV_6Sn_6}$},\ }\href
  {https://doi.org/10.1103/PhysRevLett.130.266402} {\bibfield  {journal}
  {\bibinfo  {journal} {Phys. Rev. Lett.}\ }\textbf {\bibinfo {volume} {130}},\
  \bibinfo {pages} {266402} (\bibinfo {year} {2023})}\BibitemShut {NoStop}%
\bibitem [{\citenamefont {Lee}\ \emph {et~al.}(2023)\citenamefont {Lee},
  \citenamefont {Won}, \citenamefont {Kim}, \citenamefont {Yoo}, \citenamefont
  {Park}, \citenamefont {Denlinger}, \citenamefont {Jozwiak}, \citenamefont
  {Bostwick}, \citenamefont {Rotenberg}, \citenamefont {Comin}, \citenamefont
  {Kang},\ and\ \citenamefont {Park}}]{Lee_2023}%
  \BibitemOpen
  \bibfield  {author} {\bibinfo {author} {\bibfnamefont {S.}~\bibnamefont
  {Lee}}, \bibinfo {author} {\bibfnamefont {C.}~\bibnamefont {Won}}, \bibinfo
  {author} {\bibfnamefont {J.}~\bibnamefont {Kim}}, \bibinfo {author}
  {\bibfnamefont {J.}~\bibnamefont {Yoo}}, \bibinfo {author} {\bibfnamefont
  {S.}~\bibnamefont {Park}}, \bibinfo {author} {\bibfnamefont {J.}~\bibnamefont
  {Denlinger}}, \bibinfo {author} {\bibfnamefont {C.}~\bibnamefont {Jozwiak}},
  \bibinfo {author} {\bibfnamefont {A.}~\bibnamefont {Bostwick}}, \bibinfo
  {author} {\bibfnamefont {E.}~\bibnamefont {Rotenberg}}, \bibinfo {author}
  {\bibfnamefont {R.}~\bibnamefont {Comin}}, \bibinfo {author} {\bibfnamefont
  {M.}~\bibnamefont {Kang}},\ and\ \bibinfo {author} {\bibfnamefont {J.-H.}\
  \bibnamefont {Park}},\ }\href@noop {} {\bibinfo {title} {Nature of charge
  density wave in kagome metal $\mathrm{ScV_6Sn_6}$}} (\bibinfo {year}
  {2023}),\ \Eprint {https://arxiv.org/abs/2304.11820} {arXiv:2304.11820
  [cond-mat.str-el]} \BibitemShut {NoStop}%
\bibitem [{\citenamefont {Kang}\ \emph {et~al.}(2023)\citenamefont {Kang},
  \citenamefont {Li}, \citenamefont {Meier}, \citenamefont {Villanova},
  \citenamefont {Hus}, \citenamefont {Jeon}, \citenamefont {Arachchige},
  \citenamefont {Lu}, \citenamefont {Gai}, \citenamefont {Denlinger},
  \citenamefont {Moore}, \citenamefont {Yoon},\ and\ \citenamefont
  {Mandrus}}]{Kang_2023}%
  \BibitemOpen
  \bibfield  {author} {\bibinfo {author} {\bibfnamefont {S.-H.}\ \bibnamefont
  {Kang}}, \bibinfo {author} {\bibfnamefont {H.}~\bibnamefont {Li}}, \bibinfo
  {author} {\bibfnamefont {W.~R.}\ \bibnamefont {Meier}}, \bibinfo {author}
  {\bibfnamefont {J.~W.}\ \bibnamefont {Villanova}}, \bibinfo {author}
  {\bibfnamefont {S.}~\bibnamefont {Hus}}, \bibinfo {author} {\bibfnamefont
  {H.}~\bibnamefont {Jeon}}, \bibinfo {author} {\bibfnamefont {H.~W.~S.}\
  \bibnamefont {Arachchige}}, \bibinfo {author} {\bibfnamefont
  {Q.}~\bibnamefont {Lu}}, \bibinfo {author} {\bibfnamefont {Z.}~\bibnamefont
  {Gai}}, \bibinfo {author} {\bibfnamefont {J.}~\bibnamefont {Denlinger}},
  \bibinfo {author} {\bibfnamefont {R.}~\bibnamefont {Moore}}, \bibinfo
  {author} {\bibfnamefont {M.}~\bibnamefont {Yoon}},\ and\ \bibinfo {author}
  {\bibfnamefont {D.}~\bibnamefont {Mandrus}},\ }\href@noop {} {\bibinfo
  {title} {Emergence of a new band and the lifshitz transition in kagome metal
  scv$_6$sn$_6$ with charge density wave}} (\bibinfo {year} {2023}),\ \Eprint
  {https://arxiv.org/abs/2302.14041} {arXiv:2302.14041 [cond-mat.str-el]}
  \BibitemShut {NoStop}%
\bibitem [{\citenamefont {Cheng}\ \emph {et~al.}(2023)\citenamefont {Cheng},
  \citenamefont {Ren}, \citenamefont {Li}, \citenamefont {Oh}, \citenamefont
  {Tan}, \citenamefont {Pokharel}, \citenamefont {DeStefano}, \citenamefont
  {Rosenberg}, \citenamefont {Guo}, \citenamefont {Zhang}, \citenamefont {Yue},
  \citenamefont {Lee}, \citenamefont {Gorovikov}, \citenamefont {Zonno},
  \citenamefont {Hashimoto}, \citenamefont {Lu}, \citenamefont {Ke},
  \citenamefont {Mazzola}, \citenamefont {Kono}, \citenamefont {Birgeneau},
  \citenamefont {Chu}, \citenamefont {Wilson}, \citenamefont {Wang},
  \citenamefont {Yan}, \citenamefont {Yi},\ and\ \citenamefont
  {Zeljkovic}}]{Cheng_2023}%
  \BibitemOpen
  \bibfield  {author} {\bibinfo {author} {\bibfnamefont {S.}~\bibnamefont
  {Cheng}}, \bibinfo {author} {\bibfnamefont {Z.}~\bibnamefont {Ren}}, \bibinfo
  {author} {\bibfnamefont {H.}~\bibnamefont {Li}}, \bibinfo {author}
  {\bibfnamefont {J.}~\bibnamefont {Oh}}, \bibinfo {author} {\bibfnamefont
  {H.}~\bibnamefont {Tan}}, \bibinfo {author} {\bibfnamefont {G.}~\bibnamefont
  {Pokharel}}, \bibinfo {author} {\bibfnamefont {J.~M.}\ \bibnamefont
  {DeStefano}}, \bibinfo {author} {\bibfnamefont {E.}~\bibnamefont
  {Rosenberg}}, \bibinfo {author} {\bibfnamefont {Y.}~\bibnamefont {Guo}},
  \bibinfo {author} {\bibfnamefont {Y.}~\bibnamefont {Zhang}}, \bibinfo
  {author} {\bibfnamefont {Z.}~\bibnamefont {Yue}}, \bibinfo {author}
  {\bibfnamefont {Y.}~\bibnamefont {Lee}}, \bibinfo {author} {\bibfnamefont
  {S.}~\bibnamefont {Gorovikov}}, \bibinfo {author} {\bibfnamefont
  {M.}~\bibnamefont {Zonno}}, \bibinfo {author} {\bibfnamefont
  {M.}~\bibnamefont {Hashimoto}}, \bibinfo {author} {\bibfnamefont
  {D.}~\bibnamefont {Lu}}, \bibinfo {author} {\bibfnamefont {L.}~\bibnamefont
  {Ke}}, \bibinfo {author} {\bibfnamefont {F.}~\bibnamefont {Mazzola}},
  \bibinfo {author} {\bibfnamefont {J.}~\bibnamefont {Kono}}, \bibinfo {author}
  {\bibfnamefont {R.~J.}\ \bibnamefont {Birgeneau}}, \bibinfo {author}
  {\bibfnamefont {J.-H.}\ \bibnamefont {Chu}}, \bibinfo {author} {\bibfnamefont
  {S.~D.}\ \bibnamefont {Wilson}}, \bibinfo {author} {\bibfnamefont
  {Z.}~\bibnamefont {Wang}}, \bibinfo {author} {\bibfnamefont {B.}~\bibnamefont
  {Yan}}, \bibinfo {author} {\bibfnamefont {M.}~\bibnamefont {Yi}},\ and\
  \bibinfo {author} {\bibfnamefont {I.}~\bibnamefont {Zeljkovic}},\ }\href@noop
  {} {\bibinfo {title} {Nanoscale visualization and spectral fingerprints of
  the charge order in $\mathrm{ScV_6Sn_6}$ distinct from other kagome metals}}
  (\bibinfo {year} {2023}),\ \Eprint {https://arxiv.org/abs/2302.12227}
  {arXiv:2302.12227 [cond-mat.str-el]} \BibitemShut {NoStop}%
\bibitem [{\citenamefont {Tuniz}\ \emph {et~al.}(2023)\citenamefont {Tuniz},
  \citenamefont {Consiglio}, \citenamefont {Puntel}, \citenamefont {Bigi},
  \citenamefont {Enzner}, \citenamefont {Pokharel}, \citenamefont {Orgiani},
  \citenamefont {Bronsch}, \citenamefont {Parmigiani}, \citenamefont
  {Polewczyk}, \citenamefont {King}, \citenamefont {Wells}, \citenamefont
  {Zeljkovic}, \citenamefont {Carrara}, \citenamefont {Rossi}, \citenamefont
  {Fujii}, \citenamefont {Vobornik}, \citenamefont {Wilson}, \citenamefont
  {Thomale}, \citenamefont {Wehling}, \citenamefont {Sangiovanni},
  \citenamefont {Panaccione}, \citenamefont {Cilento}, \citenamefont {Sante},\
  and\ \citenamefont {Mazzola}}]{Tuniz_2023}%
  \BibitemOpen
  \bibfield  {author} {\bibinfo {author} {\bibfnamefont {M.}~\bibnamefont
  {Tuniz}}, \bibinfo {author} {\bibfnamefont {A.}~\bibnamefont {Consiglio}},
  \bibinfo {author} {\bibfnamefont {D.}~\bibnamefont {Puntel}}, \bibinfo
  {author} {\bibfnamefont {C.}~\bibnamefont {Bigi}}, \bibinfo {author}
  {\bibfnamefont {S.}~\bibnamefont {Enzner}}, \bibinfo {author} {\bibfnamefont
  {G.}~\bibnamefont {Pokharel}}, \bibinfo {author} {\bibfnamefont
  {P.}~\bibnamefont {Orgiani}}, \bibinfo {author} {\bibfnamefont
  {W.}~\bibnamefont {Bronsch}}, \bibinfo {author} {\bibfnamefont
  {F.}~\bibnamefont {Parmigiani}}, \bibinfo {author} {\bibfnamefont
  {V.}~\bibnamefont {Polewczyk}}, \bibinfo {author} {\bibfnamefont {P.~D.~C.}\
  \bibnamefont {King}}, \bibinfo {author} {\bibfnamefont {J.~W.}\ \bibnamefont
  {Wells}}, \bibinfo {author} {\bibfnamefont {I.}~\bibnamefont {Zeljkovic}},
  \bibinfo {author} {\bibfnamefont {P.}~\bibnamefont {Carrara}}, \bibinfo
  {author} {\bibfnamefont {G.}~\bibnamefont {Rossi}}, \bibinfo {author}
  {\bibfnamefont {J.}~\bibnamefont {Fujii}}, \bibinfo {author} {\bibfnamefont
  {I.}~\bibnamefont {Vobornik}}, \bibinfo {author} {\bibfnamefont {S.~D.}\
  \bibnamefont {Wilson}}, \bibinfo {author} {\bibfnamefont {R.}~\bibnamefont
  {Thomale}}, \bibinfo {author} {\bibfnamefont {T.}~\bibnamefont {Wehling}},
  \bibinfo {author} {\bibfnamefont {G.}~\bibnamefont {Sangiovanni}}, \bibinfo
  {author} {\bibfnamefont {G.}~\bibnamefont {Panaccione}}, \bibinfo {author}
  {\bibfnamefont {F.}~\bibnamefont {Cilento}}, \bibinfo {author} {\bibfnamefont
  {D.~D.}\ \bibnamefont {Sante}},\ and\ \bibinfo {author} {\bibfnamefont
  {F.}~\bibnamefont {Mazzola}},\ }\href@noop {} {\bibinfo {title} {Dynamics and
  resilience of the charge density wave in a bilayer kagome metal}} (\bibinfo
  {year} {2023}),\ \Eprint {https://arxiv.org/abs/2302.10699} {arXiv:2302.10699
  [cond-mat.str-el]} \BibitemShut {NoStop}%
\bibitem [{\citenamefont {Kenney}\ \emph {et~al.}(2021)\citenamefont {Kenney},
  \citenamefont {Ortiz}, \citenamefont {Wang}, \citenamefont {Wilson},\ and\
  \citenamefont {Graf}}]{Kenney_2021}%
  \BibitemOpen
  \bibfield  {author} {\bibinfo {author} {\bibfnamefont {E.~M.}\ \bibnamefont
  {Kenney}}, \bibinfo {author} {\bibfnamefont {B.~R.}\ \bibnamefont {Ortiz}},
  \bibinfo {author} {\bibfnamefont {C.}~\bibnamefont {Wang}}, \bibinfo {author}
  {\bibfnamefont {S.~D.}\ \bibnamefont {Wilson}},\ and\ \bibinfo {author}
  {\bibfnamefont {M.~J.}\ \bibnamefont {Graf}},\ }\bibfield  {title} {\bibinfo
  {title} {Absence of local moments in the kagome metal $\mathrm{KV_3Sb_5}$ as
  determined by muon spin spectroscopy},\ }\href
  {https://doi.org/10.1088/1361-648X/abe8f9} {\bibfield  {journal} {\bibinfo
  {journal} {Journal of Physics: Condensed Matter}\ }\textbf {\bibinfo {volume}
  {33}},\ \bibinfo {pages} {235801} (\bibinfo {year} {2021})}\BibitemShut
  {NoStop}%
\bibitem [{\citenamefont {Teng}\ \emph {et~al.}(2022)\citenamefont {Teng},
  \citenamefont {Chen}, \citenamefont {Ye}, \citenamefont {Rosenberg},
  \citenamefont {Liu}, \citenamefont {Yin}, \citenamefont {Jiang},
  \citenamefont {Oh}, \citenamefont {Hasan}, \citenamefont {Neubauer},
  \citenamefont {Gao}, \citenamefont {Xie}, \citenamefont {Hashimoto},
  \citenamefont {Lu}, \citenamefont {Jozwiak}, \citenamefont {Bostwick},
  \citenamefont {Rotenberg}, \citenamefont {Birgeneau}, \citenamefont {Chu},
  \citenamefont {Yi},\ and\ \citenamefont {Dai}}]{Teng_2022}%
  \BibitemOpen
  \bibfield  {author} {\bibinfo {author} {\bibfnamefont {X.}~\bibnamefont
  {Teng}}, \bibinfo {author} {\bibfnamefont {L.}~\bibnamefont {Chen}}, \bibinfo
  {author} {\bibfnamefont {F.}~\bibnamefont {Ye}}, \bibinfo {author}
  {\bibfnamefont {E.}~\bibnamefont {Rosenberg}}, \bibinfo {author}
  {\bibfnamefont {Z.}~\bibnamefont {Liu}}, \bibinfo {author} {\bibfnamefont
  {J.-X.}\ \bibnamefont {Yin}}, \bibinfo {author} {\bibfnamefont {Y.-X.}\
  \bibnamefont {Jiang}}, \bibinfo {author} {\bibfnamefont {J.~S.}\ \bibnamefont
  {Oh}}, \bibinfo {author} {\bibfnamefont {M.~Z.}\ \bibnamefont {Hasan}},
  \bibinfo {author} {\bibfnamefont {K.~J.}\ \bibnamefont {Neubauer}}, \bibinfo
  {author} {\bibfnamefont {B.}~\bibnamefont {Gao}}, \bibinfo {author}
  {\bibfnamefont {Y.}~\bibnamefont {Xie}}, \bibinfo {author} {\bibfnamefont
  {M.}~\bibnamefont {Hashimoto}}, \bibinfo {author} {\bibfnamefont
  {D.}~\bibnamefont {Lu}}, \bibinfo {author} {\bibfnamefont {C.}~\bibnamefont
  {Jozwiak}}, \bibinfo {author} {\bibfnamefont {A.}~\bibnamefont {Bostwick}},
  \bibinfo {author} {\bibfnamefont {E.}~\bibnamefont {Rotenberg}}, \bibinfo
  {author} {\bibfnamefont {R.~J.}\ \bibnamefont {Birgeneau}}, \bibinfo {author}
  {\bibfnamefont {J.-H.}\ \bibnamefont {Chu}}, \bibinfo {author} {\bibfnamefont
  {M.}~\bibnamefont {Yi}},\ and\ \bibinfo {author} {\bibfnamefont
  {P.}~\bibnamefont {Dai}},\ }\bibfield  {title} {\bibinfo {title} {Discovery
  of charge density wave in a kagome lattice antiferromagnet},\ }\href
  {https://doi.org/10.1038/s41586-022-05034-z} {\bibfield  {journal} {\bibinfo
  {journal} {Nature}\ }\textbf {\bibinfo {volume} {609}},\ \bibinfo {pages}
  {490} (\bibinfo {year} {2022})}\BibitemShut {NoStop}%
\bibitem [{\citenamefont {Yin}\ \emph {et~al.}(2022{\natexlab{b}})\citenamefont
  {Yin}, \citenamefont {Jiang}, \citenamefont {Teng}, \citenamefont {Hossain},
  \citenamefont {Mardanya}, \citenamefont {Chang}, \citenamefont {Ye},
  \citenamefont {Xu}, \citenamefont {Denner}, \citenamefont {Neupert},
  \citenamefont {Lienhard}, \citenamefont {Deng}, \citenamefont {Setty},
  \citenamefont {Si}, \citenamefont {Chang}, \citenamefont {Guguchia},
  \citenamefont {Gao}, \citenamefont {Shumiya}, \citenamefont {Zhang},
  \citenamefont {Cochran}, \citenamefont {Multer}, \citenamefont {Yi},
  \citenamefont {Dai},\ and\ \citenamefont {Hasan}}]{Yin_2022}%
  \BibitemOpen
  \bibfield  {author} {\bibinfo {author} {\bibfnamefont {J.-X.}\ \bibnamefont
  {Yin}}, \bibinfo {author} {\bibfnamefont {Y.-X.}\ \bibnamefont {Jiang}},
  \bibinfo {author} {\bibfnamefont {X.}~\bibnamefont {Teng}}, \bibinfo {author}
  {\bibfnamefont {M.~S.}\ \bibnamefont {Hossain}}, \bibinfo {author}
  {\bibfnamefont {S.}~\bibnamefont {Mardanya}}, \bibinfo {author}
  {\bibfnamefont {T.-R.}\ \bibnamefont {Chang}}, \bibinfo {author}
  {\bibfnamefont {Z.}~\bibnamefont {Ye}}, \bibinfo {author} {\bibfnamefont
  {G.}~\bibnamefont {Xu}}, \bibinfo {author} {\bibfnamefont {M.~M.}\
  \bibnamefont {Denner}}, \bibinfo {author} {\bibfnamefont {T.}~\bibnamefont
  {Neupert}}, \bibinfo {author} {\bibfnamefont {B.}~\bibnamefont {Lienhard}},
  \bibinfo {author} {\bibfnamefont {H.-B.}\ \bibnamefont {Deng}}, \bibinfo
  {author} {\bibfnamefont {C.}~\bibnamefont {Setty}}, \bibinfo {author}
  {\bibfnamefont {Q.}~\bibnamefont {Si}}, \bibinfo {author} {\bibfnamefont
  {G.}~\bibnamefont {Chang}}, \bibinfo {author} {\bibfnamefont
  {Z.}~\bibnamefont {Guguchia}}, \bibinfo {author} {\bibfnamefont
  {B.}~\bibnamefont {Gao}}, \bibinfo {author} {\bibfnamefont {N.}~\bibnamefont
  {Shumiya}}, \bibinfo {author} {\bibfnamefont {Q.}~\bibnamefont {Zhang}},
  \bibinfo {author} {\bibfnamefont {T.~A.}\ \bibnamefont {Cochran}}, \bibinfo
  {author} {\bibfnamefont {D.}~\bibnamefont {Multer}}, \bibinfo {author}
  {\bibfnamefont {M.}~\bibnamefont {Yi}}, \bibinfo {author} {\bibfnamefont
  {P.}~\bibnamefont {Dai}},\ and\ \bibinfo {author} {\bibfnamefont {M.~Z.}\
  \bibnamefont {Hasan}},\ }\bibfield  {title} {\bibinfo {title} {Discovery of
  charge order and corresponding edge state in kagome magnet $\mathrm{FeGe}$},\
  }\href {https://doi.org/10.1103/PhysRevLett.129.166401} {\bibfield  {journal}
  {\bibinfo  {journal} {Phys. Rev. Lett.}\ }\textbf {\bibinfo {volume} {129}},\
  \bibinfo {pages} {166401} (\bibinfo {year} {2022}{\natexlab{b}})}\BibitemShut
  {NoStop}%
\bibitem [{\citenamefont {Chen}\ \emph
  {et~al.}(2023{\natexlab{a}})\citenamefont {Chen}, \citenamefont {Teng},
  \citenamefont {Tan}, \citenamefont {Winn}, \citenamefont {Granorth},
  \citenamefont {Ye}, \citenamefont {Yu}, \citenamefont {Mole}, \citenamefont
  {Gao}, \citenamefont {Yan}, \citenamefont {Yi},\ and\ \citenamefont
  {Dai}}]{Chen_Yi_2023}%
  \BibitemOpen
  \bibfield  {author} {\bibinfo {author} {\bibfnamefont {L.}~\bibnamefont
  {Chen}}, \bibinfo {author} {\bibfnamefont {X.}~\bibnamefont {Teng}}, \bibinfo
  {author} {\bibfnamefont {H.}~\bibnamefont {Tan}}, \bibinfo {author}
  {\bibfnamefont {B.~L.}\ \bibnamefont {Winn}}, \bibinfo {author}
  {\bibfnamefont {G.~E.}\ \bibnamefont {Granorth}}, \bibinfo {author}
  {\bibfnamefont {F.}~\bibnamefont {Ye}}, \bibinfo {author} {\bibfnamefont
  {D.~H.}\ \bibnamefont {Yu}}, \bibinfo {author} {\bibfnamefont {R.~A.}\
  \bibnamefont {Mole}}, \bibinfo {author} {\bibfnamefont {B.}~\bibnamefont
  {Gao}}, \bibinfo {author} {\bibfnamefont {B.}~\bibnamefont {Yan}}, \bibinfo
  {author} {\bibfnamefont {M.}~\bibnamefont {Yi}},\ and\ \bibinfo {author}
  {\bibfnamefont {P.}~\bibnamefont {Dai}},\ }\href@noop {} {\bibinfo {title}
  {Competing itinerant and local spin interactions in kagome metal
  $\mathrm{FeGe}$}} (\bibinfo {year} {2023}{\natexlab{a}}),\ \Eprint
  {https://arxiv.org/abs/2308.04815} {arXiv:2308.04815 [cond-mat.str-el]}
  \BibitemShut {NoStop}%
\bibitem [{\citenamefont {Teng}\ \emph
  {et~al.}(2024{\natexlab{a}})\citenamefont {Teng}, \citenamefont {Tam},
  \citenamefont {Chen}, \citenamefont {Tan}, \citenamefont {Xie}, \citenamefont
  {Gao}, \citenamefont {Granroth}, \citenamefont {Ivanov}, \citenamefont
  {Bourges}, \citenamefont {Yan}, \citenamefont {Yi},\ and\ \citenamefont
  {Dai}}]{Teng_2024}%
  \BibitemOpen
  \bibfield  {author} {\bibinfo {author} {\bibfnamefont {X.}~\bibnamefont
  {Teng}}, \bibinfo {author} {\bibfnamefont {D.~W.}\ \bibnamefont {Tam}},
  \bibinfo {author} {\bibfnamefont {L.}~\bibnamefont {Chen}}, \bibinfo {author}
  {\bibfnamefont {H.}~\bibnamefont {Tan}}, \bibinfo {author} {\bibfnamefont
  {Y.}~\bibnamefont {Xie}}, \bibinfo {author} {\bibfnamefont {B.}~\bibnamefont
  {Gao}}, \bibinfo {author} {\bibfnamefont {G.~E.}\ \bibnamefont {Granroth}},
  \bibinfo {author} {\bibfnamefont {A.}~\bibnamefont {Ivanov}}, \bibinfo
  {author} {\bibfnamefont {P.}~\bibnamefont {Bourges}}, \bibinfo {author}
  {\bibfnamefont {B.}~\bibnamefont {Yan}}, \bibinfo {author} {\bibfnamefont
  {M.}~\bibnamefont {Yi}},\ and\ \bibinfo {author} {\bibfnamefont
  {P.}~\bibnamefont {Dai}},\ }\href@noop {} {\bibinfo {title}
  {Spin-charge-lattice coupling across the charge density wave transition in a
  kagome lattice antiferromagnet}} (\bibinfo {year} {2024}{\natexlab{a}}),\
  \Eprint {https://arxiv.org/abs/2404.04459} {arXiv:2404.04459
  [cond-mat.str-el]} \BibitemShut {NoStop}%
\bibitem [{\citenamefont {Jiang}\ \emph
  {et~al.}(2023{\natexlab{a}})\citenamefont {Jiang}, \citenamefont {Hu},
  \citenamefont {C{\u{a}}lug{\u{a}}ru}, \citenamefont {Felser}, \citenamefont
  {Blanco-Canosa}, \citenamefont {Weng}, \citenamefont {Xu},\ and\
  \citenamefont {Bernevig}}]{jiang2023kagome}%
  \BibitemOpen
  \bibfield  {author} {\bibinfo {author} {\bibfnamefont {Y.}~\bibnamefont
  {Jiang}}, \bibinfo {author} {\bibfnamefont {H.}~\bibnamefont {Hu}}, \bibinfo
  {author} {\bibfnamefont {D.}~\bibnamefont {C{\u{a}}lug{\u{a}}ru}}, \bibinfo
  {author} {\bibfnamefont {C.}~\bibnamefont {Felser}}, \bibinfo {author}
  {\bibfnamefont {S.}~\bibnamefont {Blanco-Canosa}}, \bibinfo {author}
  {\bibfnamefont {H.}~\bibnamefont {Weng}}, \bibinfo {author} {\bibfnamefont
  {Y.}~\bibnamefont {Xu}},\ and\ \bibinfo {author} {\bibfnamefont {B.~A.}\
  \bibnamefont {Bernevig}},\ }\bibfield  {title} {\bibinfo {title} {Kagome
  materials ii: Sg 191: $\mathrm{FeGe}$ as a lego building block for the entire
  1: 6: 6 series: hidden d-orbital decoupling of flat band sectors, effective
  models and interaction hamiltonians},\ }\href@noop {} {\bibfield  {journal}
  {\bibinfo  {journal} {arXiv preprint arXiv:2311.09290}\ } (\bibinfo {year}
  {2023}{\natexlab{a}})}\BibitemShut {NoStop}%
\bibitem [{\citenamefont {Miao}\ \emph {et~al.}(2023)\citenamefont {Miao},
  \citenamefont {Zhang}, \citenamefont {Li}, \citenamefont {Fabbris},
  \citenamefont {Said}, \citenamefont {Tartaglia}, \citenamefont {Yilmaz},
  \citenamefont {Vescovo}, \citenamefont {Yin}, \citenamefont {Murakami},
  \citenamefont {Feng}, \citenamefont {Jiang}, \citenamefont {Wu},
  \citenamefont {Wang}, \citenamefont {Okamoto}, \citenamefont {Wang},\ and\
  \citenamefont {Lee}}]{Miao_2023}%
  \BibitemOpen
  \bibfield  {author} {\bibinfo {author} {\bibfnamefont {H.}~\bibnamefont
  {Miao}}, \bibinfo {author} {\bibfnamefont {T.~T.}\ \bibnamefont {Zhang}},
  \bibinfo {author} {\bibfnamefont {H.~X.}\ \bibnamefont {Li}}, \bibinfo
  {author} {\bibfnamefont {G.}~\bibnamefont {Fabbris}}, \bibinfo {author}
  {\bibfnamefont {A.~H.}\ \bibnamefont {Said}}, \bibinfo {author}
  {\bibfnamefont {R.}~\bibnamefont {Tartaglia}}, \bibinfo {author}
  {\bibfnamefont {T.}~\bibnamefont {Yilmaz}}, \bibinfo {author} {\bibfnamefont
  {E.}~\bibnamefont {Vescovo}}, \bibinfo {author} {\bibfnamefont {J.-X.}\
  \bibnamefont {Yin}}, \bibinfo {author} {\bibfnamefont {S.}~\bibnamefont
  {Murakami}}, \bibinfo {author} {\bibfnamefont {X.~L.}\ \bibnamefont {Feng}},
  \bibinfo {author} {\bibfnamefont {K.}~\bibnamefont {Jiang}}, \bibinfo
  {author} {\bibfnamefont {X.~L.}\ \bibnamefont {Wu}}, \bibinfo {author}
  {\bibfnamefont {A.~F.}\ \bibnamefont {Wang}}, \bibinfo {author}
  {\bibfnamefont {S.}~\bibnamefont {Okamoto}}, \bibinfo {author} {\bibfnamefont
  {Y.~L.}\ \bibnamefont {Wang}},\ and\ \bibinfo {author} {\bibfnamefont
  {H.~N.}\ \bibnamefont {Lee}},\ }\bibfield  {title} {\bibinfo {title}
  {Signature of spin-phonon coupling driven charge density wave in a kagome
  magnet},\ }\href {https://doi.org/10.1038/s41467-023-41957-5} {\bibfield
  {journal} {\bibinfo  {journal} {Nature Communications}\ }\textbf {\bibinfo
  {volume} {14}},\ \bibinfo {pages} {6183} (\bibinfo {year}
  {2023})}\BibitemShut {NoStop}%
\bibitem [{\citenamefont {Teng}\ \emph
  {et~al.}(2024{\natexlab{b}})\citenamefont {Teng}, \citenamefont {Tam},
  \citenamefont {Chen}, \citenamefont {Tan}, \citenamefont {Xie}, \citenamefont
  {Gao}, \citenamefont {Granroth}, \citenamefont {Ivanov}, \citenamefont
  {Bourges}, \citenamefont {Yan}, \citenamefont {Yi},\ and\ \citenamefont
  {Dai}}]{teng2024spinchargelattice}%
  \BibitemOpen
  \bibfield  {author} {\bibinfo {author} {\bibfnamefont {X.}~\bibnamefont
  {Teng}}, \bibinfo {author} {\bibfnamefont {D.~W.}\ \bibnamefont {Tam}},
  \bibinfo {author} {\bibfnamefont {L.}~\bibnamefont {Chen}}, \bibinfo {author}
  {\bibfnamefont {H.}~\bibnamefont {Tan}}, \bibinfo {author} {\bibfnamefont
  {Y.}~\bibnamefont {Xie}}, \bibinfo {author} {\bibfnamefont {B.}~\bibnamefont
  {Gao}}, \bibinfo {author} {\bibfnamefont {G.~E.}\ \bibnamefont {Granroth}},
  \bibinfo {author} {\bibfnamefont {A.}~\bibnamefont {Ivanov}}, \bibinfo
  {author} {\bibfnamefont {P.}~\bibnamefont {Bourges}}, \bibinfo {author}
  {\bibfnamefont {B.}~\bibnamefont {Yan}}, \bibinfo {author} {\bibfnamefont
  {M.}~\bibnamefont {Yi}},\ and\ \bibinfo {author} {\bibfnamefont
  {P.}~\bibnamefont {Dai}},\ }\href@noop {} {\bibinfo {title}
  {Spin-charge-lattice coupling across the charge density wave transition in a
  kagome lattice antiferromagnet}} (\bibinfo {year} {2024}{\natexlab{b}}),\
  \Eprint {https://arxiv.org/abs/2404.04459} {arXiv:2404.04459} \BibitemShut
  {NoStop}%
\bibitem [{\citenamefont {Teng}\ \emph {et~al.}(2023)\citenamefont {Teng},
  \citenamefont {Oh}, \citenamefont {Tan}, \citenamefont {Chen}, \citenamefont
  {Huang}, \citenamefont {Gao}, \citenamefont {Yin}, \citenamefont {Chu},
  \citenamefont {Hashimoto}, \citenamefont {Lu}, \citenamefont {Jozwiak},
  \citenamefont {Bostwick}, \citenamefont {Rotenberg}, \citenamefont
  {Granroth}, \citenamefont {Yan}, \citenamefont {Birgeneau}, \citenamefont
  {Dai},\ and\ \citenamefont {Yi}}]{Teng_2023}%
  \BibitemOpen
  \bibfield  {author} {\bibinfo {author} {\bibfnamefont {X.}~\bibnamefont
  {Teng}}, \bibinfo {author} {\bibfnamefont {J.~S.}\ \bibnamefont {Oh}},
  \bibinfo {author} {\bibfnamefont {H.}~\bibnamefont {Tan}}, \bibinfo {author}
  {\bibfnamefont {L.}~\bibnamefont {Chen}}, \bibinfo {author} {\bibfnamefont
  {J.}~\bibnamefont {Huang}}, \bibinfo {author} {\bibfnamefont
  {B.}~\bibnamefont {Gao}}, \bibinfo {author} {\bibfnamefont {J.-X.}\
  \bibnamefont {Yin}}, \bibinfo {author} {\bibfnamefont {J.-H.}\ \bibnamefont
  {Chu}}, \bibinfo {author} {\bibfnamefont {M.}~\bibnamefont {Hashimoto}},
  \bibinfo {author} {\bibfnamefont {D.}~\bibnamefont {Lu}}, \bibinfo {author}
  {\bibfnamefont {C.}~\bibnamefont {Jozwiak}}, \bibinfo {author} {\bibfnamefont
  {A.}~\bibnamefont {Bostwick}}, \bibinfo {author} {\bibfnamefont
  {E.}~\bibnamefont {Rotenberg}}, \bibinfo {author} {\bibfnamefont {G.~E.}\
  \bibnamefont {Granroth}}, \bibinfo {author} {\bibfnamefont {B.}~\bibnamefont
  {Yan}}, \bibinfo {author} {\bibfnamefont {R.~J.}\ \bibnamefont {Birgeneau}},
  \bibinfo {author} {\bibfnamefont {P.}~\bibnamefont {Dai}},\ and\ \bibinfo
  {author} {\bibfnamefont {M.}~\bibnamefont {Yi}},\ }\bibfield  {title}
  {\bibinfo {title} {Magnetism and charge density wave order in kagome
  $\mathrm{FeGe}$},\ }\href {https://doi.org/10.1038/s41567-023-01985-w}
  {\bibfield  {journal} {\bibinfo  {journal} {Nature Physics}\ }\textbf
  {\bibinfo {volume} {19}},\ \bibinfo {pages} {814} (\bibinfo {year}
  {2023})}\BibitemShut {NoStop}%
\bibitem [{\citenamefont {Oh}\ \emph {et~al.}(2024)\citenamefont {Oh},
  \citenamefont {Biswas}, \citenamefont {Klemm}, \citenamefont {Tan},
  \citenamefont {Hashimoto}, \citenamefont {Lu}, \citenamefont {Yan},
  \citenamefont {Dai}, \citenamefont {Birgeneau},\ and\ \citenamefont
  {Yi}}]{Oh2024}%
  \BibitemOpen
  \bibfield  {author} {\bibinfo {author} {\bibfnamefont {J.~S.}\ \bibnamefont
  {Oh}}, \bibinfo {author} {\bibfnamefont {A.}~\bibnamefont {Biswas}}, \bibinfo
  {author} {\bibfnamefont {M.}~\bibnamefont {Klemm}}, \bibinfo {author}
  {\bibfnamefont {H.}~\bibnamefont {Tan}}, \bibinfo {author} {\bibfnamefont
  {M.}~\bibnamefont {Hashimoto}}, \bibinfo {author} {\bibfnamefont
  {D.}~\bibnamefont {Lu}}, \bibinfo {author} {\bibfnamefont {B.}~\bibnamefont
  {Yan}}, \bibinfo {author} {\bibfnamefont {P.}~\bibnamefont {Dai}}, \bibinfo
  {author} {\bibfnamefont {R.~J.}\ \bibnamefont {Birgeneau}},\ and\ \bibinfo
  {author} {\bibfnamefont {M.}~\bibnamefont {Yi}},\ }\href@noop {} {\bibinfo
  {title} {Tunability of charge density wave in a magnetic kagome metal}}
  (\bibinfo {year} {2024}),\ \Eprint {https://arxiv.org/abs/2404.02231}
  {arXiv:2404.02231 [cond-mat.str-el]} \BibitemShut {NoStop}%
\bibitem [{\citenamefont {Wang}(2023)}]{Wang_2023}%
  \BibitemOpen
  \bibfield  {author} {\bibinfo {author} {\bibfnamefont {Y.}~\bibnamefont
  {Wang}},\ }\bibfield  {title} {\bibinfo {title} {Enhanced spin-polarization
  via partial ge-dimerization as the driving force of the charge density wave
  in $\mathrm{FeGe}$},\ }\href
  {https://doi.org/10.1103/PhysRevMaterials.7.104006} {\bibfield  {journal}
  {\bibinfo  {journal} {Phys. Rev. Mater.}\ }\textbf {\bibinfo {volume} {7}},\
  \bibinfo {pages} {104006} (\bibinfo {year} {2023})}\BibitemShut {NoStop}%
\bibitem [{\citenamefont {Zhao}\ \emph {et~al.}(2023)\citenamefont {Zhao},
  \citenamefont {Li}, \citenamefont {Li}, \citenamefont {Wu}, \citenamefont
  {Yao}, \citenamefont {Chen}, \citenamefont {Cui}, \citenamefont {Sun},
  \citenamefont {Yang}, \citenamefont {Jiang}, \citenamefont {Liu},
  \citenamefont {Louat}, \citenamefont {Kim}, \citenamefont {Cacho},
  \citenamefont {Wang}, \citenamefont {Wang}, \citenamefont {Shen},
  \citenamefont {Jiang},\ and\ \citenamefont {Feng}}]{Zhao_2023}%
  \BibitemOpen
  \bibfield  {author} {\bibinfo {author} {\bibfnamefont {Z.}~\bibnamefont
  {Zhao}}, \bibinfo {author} {\bibfnamefont {T.}~\bibnamefont {Li}}, \bibinfo
  {author} {\bibfnamefont {P.}~\bibnamefont {Li}}, \bibinfo {author}
  {\bibfnamefont {X.}~\bibnamefont {Wu}}, \bibinfo {author} {\bibfnamefont
  {J.}~\bibnamefont {Yao}}, \bibinfo {author} {\bibfnamefont {Z.}~\bibnamefont
  {Chen}}, \bibinfo {author} {\bibfnamefont {S.}~\bibnamefont {Cui}}, \bibinfo
  {author} {\bibfnamefont {Z.}~\bibnamefont {Sun}}, \bibinfo {author}
  {\bibfnamefont {Y.}~\bibnamefont {Yang}}, \bibinfo {author} {\bibfnamefont
  {Z.}~\bibnamefont {Jiang}}, \bibinfo {author} {\bibfnamefont
  {Z.}~\bibnamefont {Liu}}, \bibinfo {author} {\bibfnamefont {A.}~\bibnamefont
  {Louat}}, \bibinfo {author} {\bibfnamefont {T.}~\bibnamefont {Kim}}, \bibinfo
  {author} {\bibfnamefont {C.}~\bibnamefont {Cacho}}, \bibinfo {author}
  {\bibfnamefont {A.}~\bibnamefont {Wang}}, \bibinfo {author} {\bibfnamefont
  {Y.}~\bibnamefont {Wang}}, \bibinfo {author} {\bibfnamefont {D.}~\bibnamefont
  {Shen}}, \bibinfo {author} {\bibfnamefont {J.}~\bibnamefont {Jiang}},\ and\
  \bibinfo {author} {\bibfnamefont {D.}~\bibnamefont {Feng}},\ }\href@noop {}
  {\bibinfo {title} {Photoemission evidence of a novel charge order in kagome
  metal $\mathrm{FeGe}$}} (\bibinfo {year} {2023}),\ \Eprint
  {https://arxiv.org/abs/2308.08336} {arXiv:2308.08336 [cond-mat.str-el]}
  \BibitemShut {NoStop}%
\bibitem [{\citenamefont {Wu}\ \emph {et~al.}(2023)\citenamefont {Wu},
  \citenamefont {Hu}, \citenamefont {Fan}, \citenamefont {Wang},\ and\
  \citenamefont {Wan}}]{Wu2023}%
  \BibitemOpen
  \bibfield  {author} {\bibinfo {author} {\bibfnamefont {L.}~\bibnamefont
  {Wu}}, \bibinfo {author} {\bibfnamefont {Y.}~\bibnamefont {Hu}}, \bibinfo
  {author} {\bibfnamefont {D.}~\bibnamefont {Fan}}, \bibinfo {author}
  {\bibfnamefont {D.}~\bibnamefont {Wang}},\ and\ \bibinfo {author}
  {\bibfnamefont {X.}~\bibnamefont {Wan}},\ }\bibfield  {title} {\bibinfo
  {title} {Electron-correlation-induced charge density wave in
  $\mathrm{FeGe}$},\ }\href {https://doi.org/10.1088/0256-307X/40/11/117103}
  {\bibfield  {journal} {\bibinfo  {journal} {Chinese Physics Letters}\
  }\textbf {\bibinfo {volume} {40}},\ \bibinfo {pages} {117103} (\bibinfo
  {year} {2023})}\BibitemShut {NoStop}%
\bibitem [{\citenamefont {Shi}\ \emph {et~al.}(2023)\citenamefont {Shi},
  \citenamefont {Liu}, \citenamefont {Maity}, \citenamefont {Wang},
  \citenamefont {Kotla}, \citenamefont {Ramakrishnan}, \citenamefont {Eisele},
  \citenamefont {Agarwal}, \citenamefont {Noohinejad}, \citenamefont {Tao},
  \citenamefont {Kang}, \citenamefont {Lou}, \citenamefont {Yang},
  \citenamefont {Qi}, \citenamefont {Lin}, \citenamefont {Xu}, \citenamefont
  {Thamizhavel}, \citenamefont {Cao}, \citenamefont {van Smaalen},
  \citenamefont {Cao},\ and\ \citenamefont {Bao}}]{Shi_2023}%
  \BibitemOpen
  \bibfield  {author} {\bibinfo {author} {\bibfnamefont {C.}~\bibnamefont
  {Shi}}, \bibinfo {author} {\bibfnamefont {Y.}~\bibnamefont {Liu}}, \bibinfo
  {author} {\bibfnamefont {B.~B.}\ \bibnamefont {Maity}}, \bibinfo {author}
  {\bibfnamefont {Q.}~\bibnamefont {Wang}}, \bibinfo {author} {\bibfnamefont
  {S.~R.}\ \bibnamefont {Kotla}}, \bibinfo {author} {\bibfnamefont
  {S.}~\bibnamefont {Ramakrishnan}}, \bibinfo {author} {\bibfnamefont
  {C.}~\bibnamefont {Eisele}}, \bibinfo {author} {\bibfnamefont
  {H.}~\bibnamefont {Agarwal}}, \bibinfo {author} {\bibfnamefont
  {L.}~\bibnamefont {Noohinejad}}, \bibinfo {author} {\bibfnamefont
  {Q.}~\bibnamefont {Tao}}, \bibinfo {author} {\bibfnamefont {B.}~\bibnamefont
  {Kang}}, \bibinfo {author} {\bibfnamefont {Z.}~\bibnamefont {Lou}}, \bibinfo
  {author} {\bibfnamefont {X.}~\bibnamefont {Yang}}, \bibinfo {author}
  {\bibfnamefont {Y.}~\bibnamefont {Qi}}, \bibinfo {author} {\bibfnamefont
  {X.}~\bibnamefont {Lin}}, \bibinfo {author} {\bibfnamefont {Z.-A.}\
  \bibnamefont {Xu}}, \bibinfo {author} {\bibfnamefont {A.}~\bibnamefont
  {Thamizhavel}}, \bibinfo {author} {\bibfnamefont {G.-H.}\ \bibnamefont
  {Cao}}, \bibinfo {author} {\bibfnamefont {S.}~\bibnamefont {van Smaalen}},
  \bibinfo {author} {\bibfnamefont {S.}~\bibnamefont {Cao}},\ and\ \bibinfo
  {author} {\bibfnamefont {J.-K.}\ \bibnamefont {Bao}},\ }\href@noop {}
  {\bibinfo {title} {Disordered structure for long-range charge density wave
  order in annealed crystals of magnetic kagome $\mathrm{FeGe}$}} (\bibinfo
  {year} {2023}),\ \Eprint {https://arxiv.org/abs/2308.09034} {arXiv:2308.09034
  [cond-mat.str-el]} \BibitemShut {NoStop}%
\bibitem [{\citenamefont {Chen}\ \emph
  {et~al.}(2023{\natexlab{b}})\citenamefont {Chen}, \citenamefont {Wu},
  \citenamefont {Yin}, \citenamefont {Zhang}, \citenamefont {Wang},
  \citenamefont {Li}, \citenamefont {Li}, \citenamefont {Wang}, \citenamefont
  {Wang}, \citenamefont {Yan},\ and\ \citenamefont {Feng}}]{Chen_2023}%
  \BibitemOpen
  \bibfield  {author} {\bibinfo {author} {\bibfnamefont {Z.}~\bibnamefont
  {Chen}}, \bibinfo {author} {\bibfnamefont {X.}~\bibnamefont {Wu}}, \bibinfo
  {author} {\bibfnamefont {R.}~\bibnamefont {Yin}}, \bibinfo {author}
  {\bibfnamefont {J.}~\bibnamefont {Zhang}}, \bibinfo {author} {\bibfnamefont
  {S.}~\bibnamefont {Wang}}, \bibinfo {author} {\bibfnamefont {Y.}~\bibnamefont
  {Li}}, \bibinfo {author} {\bibfnamefont {M.}~\bibnamefont {Li}}, \bibinfo
  {author} {\bibfnamefont {A.}~\bibnamefont {Wang}}, \bibinfo {author}
  {\bibfnamefont {Y.}~\bibnamefont {Wang}}, \bibinfo {author} {\bibfnamefont
  {Y.-J.}\ \bibnamefont {Yan}},\ and\ \bibinfo {author} {\bibfnamefont {D.-L.}\
  \bibnamefont {Feng}},\ }\href@noop {} {\bibinfo {title} {Charge density wave
  with strong quantum phase fluctuations in kagome magnet $\mathrm{FeGe}$}}
  (\bibinfo {year} {2023}{\natexlab{b}}),\ \Eprint
  {https://arxiv.org/abs/2302.04490} {arXiv:2302.04490 [cond-mat.str-el]}
  \BibitemShut {NoStop}%
\bibitem [{\citenamefont {Wu}\ \emph {et~al.}(2024{\natexlab{a}})\citenamefont
  {Wu}, \citenamefont {Klemm}, \citenamefont {Shah}, \citenamefont {Ritz},
  \citenamefont {Duan}, \citenamefont {Teng}, \citenamefont {Gao},
  \citenamefont {Ye}, \citenamefont {Matsuda}, \citenamefont {Li},
  \citenamefont {Xu}, \citenamefont {Yi}, \citenamefont {Birol}, \citenamefont
  {Dai},\ and\ \citenamefont {Blumberg}}]{Wu_2023}%
  \BibitemOpen
  \bibfield  {author} {\bibinfo {author} {\bibfnamefont {S.}~\bibnamefont
  {Wu}}, \bibinfo {author} {\bibfnamefont {M.~L.}\ \bibnamefont {Klemm}},
  \bibinfo {author} {\bibfnamefont {J.}~\bibnamefont {Shah}}, \bibinfo {author}
  {\bibfnamefont {E.~T.}\ \bibnamefont {Ritz}}, \bibinfo {author}
  {\bibfnamefont {C.}~\bibnamefont {Duan}}, \bibinfo {author} {\bibfnamefont
  {X.}~\bibnamefont {Teng}}, \bibinfo {author} {\bibfnamefont {B.}~\bibnamefont
  {Gao}}, \bibinfo {author} {\bibfnamefont {F.}~\bibnamefont {Ye}}, \bibinfo
  {author} {\bibfnamefont {M.}~\bibnamefont {Matsuda}}, \bibinfo {author}
  {\bibfnamefont {F.}~\bibnamefont {Li}}, \bibinfo {author} {\bibfnamefont
  {X.}~\bibnamefont {Xu}}, \bibinfo {author} {\bibfnamefont {M.}~\bibnamefont
  {Yi}}, \bibinfo {author} {\bibfnamefont {T.}~\bibnamefont {Birol}}, \bibinfo
  {author} {\bibfnamefont {P.}~\bibnamefont {Dai}},\ and\ \bibinfo {author}
  {\bibfnamefont {G.}~\bibnamefont {Blumberg}},\ }\bibfield  {title} {\bibinfo
  {title} {Symmetry breaking and ascending in the magnetic kagome metal
  $\mathrm{FeGe}$},\ }\href {https://doi.org/10.1103/PhysRevX.14.011043}
  {\bibfield  {journal} {\bibinfo  {journal} {Phys. Rev. X}\ }\textbf {\bibinfo
  {volume} {14}},\ \bibinfo {pages} {011043} (\bibinfo {year}
  {2024}{\natexlab{a}})}\BibitemShut {NoStop}%
\bibitem [{\citenamefont {Huang}\ and\ \citenamefont
  {Stoebe}(1993)}]{Huang_2006}%
  \BibitemOpen
  \bibfield  {author} {\bibinfo {author} {\bibfnamefont {C.}~\bibnamefont
  {Huang}}\ and\ \bibinfo {author} {\bibfnamefont {T.}~\bibnamefont {Stoebe}},\
  }\bibfield  {title} {\bibinfo {title} {Thermal properties of ‘stacked
  hexatic phases’ in liquid crystals},\ }\href
  {https://doi.org/10.1080/00018739300101504} {\bibfield  {journal} {\bibinfo
  {journal} {Advances in Physics}\ }\textbf {\bibinfo {volume} {42}},\ \bibinfo
  {pages} {343} (\bibinfo {year} {1993})}\BibitemShut {NoStop}%
\bibitem [{\citenamefont {Wannier}(1950)}]{Wannier_1950}%
  \BibitemOpen
  \bibfield  {author} {\bibinfo {author} {\bibfnamefont {G.~H.}\ \bibnamefont
  {Wannier}},\ }\bibfield  {title} {\bibinfo {title} {Antiferromagnetism. the
  triangular ising net},\ }\href {https://doi.org/10.1103/PhysRev.79.357}
  {\bibfield  {journal} {\bibinfo  {journal} {Phys. Rev.}\ }\textbf {\bibinfo
  {volume} {79}},\ \bibinfo {pages} {357} (\bibinfo {year} {1950})}\BibitemShut
  {NoStop}%
\bibitem [{\citenamefont {Mermin}(1979)}]{Mermin_1979}%
  \BibitemOpen
  \bibfield  {author} {\bibinfo {author} {\bibfnamefont {N.~D.}\ \bibnamefont
  {Mermin}},\ }\bibfield  {title} {\bibinfo {title} {The topological theory of
  defects in ordered media},\ }\href
  {https://doi.org/10.1103/RevModPhys.51.591} {\bibfield  {journal} {\bibinfo
  {journal} {Rev. Mod. Phys.}\ }\textbf {\bibinfo {volume} {51}},\ \bibinfo
  {pages} {591} (\bibinfo {year} {1979})}\BibitemShut {NoStop}%
\bibitem [{\citenamefont {Brinkman}\ \emph {et~al.}(1982)\citenamefont
  {Brinkman}, \citenamefont {Fisher},\ and\ \citenamefont
  {Moncton}}]{Brinkman_1982}%
  \BibitemOpen
  \bibfield  {author} {\bibinfo {author} {\bibfnamefont {W.~F.}\ \bibnamefont
  {Brinkman}}, \bibinfo {author} {\bibfnamefont {D.~S.}\ \bibnamefont
  {Fisher}},\ and\ \bibinfo {author} {\bibfnamefont {D.~E.}\ \bibnamefont
  {Moncton}},\ }\bibfield  {title} {\bibinfo {title} {Melting of
  two-dimensional solids},\ }\href
  {https://doi.org/10.1126/science.217.4561.693} {\bibfield  {journal}
  {\bibinfo  {journal} {Science}\ }\textbf {\bibinfo {volume} {217}},\ \bibinfo
  {pages} {693} (\bibinfo {year} {1982})}\BibitemShut {NoStop}%
\bibitem [{\citenamefont {Kang}\ \emph {et~al.}(2020)\citenamefont {Kang},
  \citenamefont {Ye}, \citenamefont {Fang}, \citenamefont {You}, \citenamefont
  {Levitan}, \citenamefont {Han}, \citenamefont {Facio}, \citenamefont
  {Jozwiak}, \citenamefont {Bostwick}, \citenamefont {Rotenberg}, \citenamefont
  {Chan}, \citenamefont {McDonald}, \citenamefont {Graf}, \citenamefont
  {Kaznatcheev}, \citenamefont {Vescovo}, \citenamefont {Bell}, \citenamefont
  {Kaxiras}, \citenamefont {van~den Brink}, \citenamefont {Richter},
  \citenamefont {Prasad~Ghimire}, \citenamefont {Checkelsky},\ and\
  \citenamefont {Comin}}]{Kang_2020}%
  \BibitemOpen
  \bibfield  {author} {\bibinfo {author} {\bibfnamefont {M.}~\bibnamefont
  {Kang}}, \bibinfo {author} {\bibfnamefont {L.}~\bibnamefont {Ye}}, \bibinfo
  {author} {\bibfnamefont {S.}~\bibnamefont {Fang}}, \bibinfo {author}
  {\bibfnamefont {J.-S.}\ \bibnamefont {You}}, \bibinfo {author} {\bibfnamefont
  {A.}~\bibnamefont {Levitan}}, \bibinfo {author} {\bibfnamefont
  {M.}~\bibnamefont {Han}}, \bibinfo {author} {\bibfnamefont {J.~I.}\
  \bibnamefont {Facio}}, \bibinfo {author} {\bibfnamefont {C.}~\bibnamefont
  {Jozwiak}}, \bibinfo {author} {\bibfnamefont {A.}~\bibnamefont {Bostwick}},
  \bibinfo {author} {\bibfnamefont {E.}~\bibnamefont {Rotenberg}}, \bibinfo
  {author} {\bibfnamefont {M.~K.}\ \bibnamefont {Chan}}, \bibinfo {author}
  {\bibfnamefont {R.~D.}\ \bibnamefont {McDonald}}, \bibinfo {author}
  {\bibfnamefont {D.}~\bibnamefont {Graf}}, \bibinfo {author} {\bibfnamefont
  {K.}~\bibnamefont {Kaznatcheev}}, \bibinfo {author} {\bibfnamefont
  {E.}~\bibnamefont {Vescovo}}, \bibinfo {author} {\bibfnamefont {D.~C.}\
  \bibnamefont {Bell}}, \bibinfo {author} {\bibfnamefont {E.}~\bibnamefont
  {Kaxiras}}, \bibinfo {author} {\bibfnamefont {J.}~\bibnamefont {van~den
  Brink}}, \bibinfo {author} {\bibfnamefont {M.}~\bibnamefont {Richter}},
  \bibinfo {author} {\bibfnamefont {M.}~\bibnamefont {Prasad~Ghimire}},
  \bibinfo {author} {\bibfnamefont {J.~G.}\ \bibnamefont {Checkelsky}},\ and\
  \bibinfo {author} {\bibfnamefont {R.}~\bibnamefont {Comin}},\ }\bibfield
  {title} {\bibinfo {title} {Dirac fermions and flat bands in the ideal kagome
  metal fesn},\ }\href {https://doi.org/10.1038/s41563-019-0531-0} {\bibfield
  {journal} {\bibinfo  {journal} {Nature Materials}\ }\textbf {\bibinfo
  {volume} {19}},\ \bibinfo {pages} {163} (\bibinfo {year} {2020})}\BibitemShut
  {NoStop}%
\bibitem [{\citenamefont {Zhou}\ \emph {et~al.}(2023)\citenamefont {Zhou},
  \citenamefont {Yan}, \citenamefont {Fan}, \citenamefont {Wang},\ and\
  \citenamefont {Wan}}]{Zhou2023}%
  \BibitemOpen
  \bibfield  {author} {\bibinfo {author} {\bibfnamefont {H.}~\bibnamefont
  {Zhou}}, \bibinfo {author} {\bibfnamefont {S.}~\bibnamefont {Yan}}, \bibinfo
  {author} {\bibfnamefont {D.}~\bibnamefont {Fan}}, \bibinfo {author}
  {\bibfnamefont {D.}~\bibnamefont {Wang}},\ and\ \bibinfo {author}
  {\bibfnamefont {X.}~\bibnamefont {Wan}},\ }\bibfield  {title} {\bibinfo
  {title} {Magnetic interactions and possible structural distortion in kagome
  $\mathrm{FeGe}$ from first-principles calculations and symmetry analysis},\
  }\href {https://doi.org/10.1103/PhysRevB.108.035138} {\bibfield  {journal}
  {\bibinfo  {journal} {Phys. Rev. B}\ }\textbf {\bibinfo {volume} {108}},\
  \bibinfo {pages} {035138} (\bibinfo {year} {2023})}\BibitemShut {NoStop}%
\bibitem [{\citenamefont {Weschke}\ \emph {et~al.}(2004)\citenamefont
  {Weschke}, \citenamefont {Ott}, \citenamefont {Schierle}, \citenamefont
  {Sch\'u\ss{}ler-Langeheine}, \citenamefont {Vyalikh}, \citenamefont {Kaindl},
  \citenamefont {Leiner}, \citenamefont {Ay}, \citenamefont {Schmitte},
  \citenamefont {Zabel},\ and\ \citenamefont {Jensen}}]{Ott_2004}%
  \BibitemOpen
  \bibfield  {author} {\bibinfo {author} {\bibfnamefont {E.}~\bibnamefont
  {Weschke}}, \bibinfo {author} {\bibfnamefont {H.}~\bibnamefont {Ott}},
  \bibinfo {author} {\bibfnamefont {E.}~\bibnamefont {Schierle}}, \bibinfo
  {author} {\bibfnamefont {C.}~\bibnamefont {Sch\'u\ss{}ler-Langeheine}},
  \bibinfo {author} {\bibfnamefont {D.~V.}\ \bibnamefont {Vyalikh}}, \bibinfo
  {author} {\bibfnamefont {G.}~\bibnamefont {Kaindl}}, \bibinfo {author}
  {\bibfnamefont {V.}~\bibnamefont {Leiner}}, \bibinfo {author} {\bibfnamefont
  {M.}~\bibnamefont {Ay}}, \bibinfo {author} {\bibfnamefont {T.}~\bibnamefont
  {Schmitte}}, \bibinfo {author} {\bibfnamefont {H.}~\bibnamefont {Zabel}},\
  and\ \bibinfo {author} {\bibfnamefont {P.~J.}\ \bibnamefont {Jensen}},\
  }\bibfield  {title} {\bibinfo {title} {Finite-size effect on magnetic
  ordering temperatures in long-period antiferromagnets: Holmium thin films},\
  }\href {https://doi.org/10.1103/PhysRevLett.93.157204} {\bibfield  {journal}
  {\bibinfo  {journal} {Phys. Rev. Lett.}\ }\textbf {\bibinfo {volume} {93}},\
  \bibinfo {pages} {157204} (\bibinfo {year} {2004})}\BibitemShut {NoStop}%
\bibitem [{\citenamefont {Kang}\ \emph {et~al.}(2022)\citenamefont {Kang},
  \citenamefont {Fang}, \citenamefont {Kim}, \citenamefont {Ortiz},
  \citenamefont {Ryu}, \citenamefont {Kim}, \citenamefont {Yoo}, \citenamefont
  {Sangiovanni}, \citenamefont {Di~Sante}, \citenamefont {Park}, \citenamefont
  {Jozwiak}, \citenamefont {Bostwick}, \citenamefont {Rotenberg}, \citenamefont
  {Kaxiras}, \citenamefont {Wilson}, \citenamefont {Park},\ and\ \citenamefont
  {Comin}}]{Kang_2022}%
  \BibitemOpen
  \bibfield  {author} {\bibinfo {author} {\bibfnamefont {M.}~\bibnamefont
  {Kang}}, \bibinfo {author} {\bibfnamefont {S.}~\bibnamefont {Fang}}, \bibinfo
  {author} {\bibfnamefont {J.-K.}\ \bibnamefont {Kim}}, \bibinfo {author}
  {\bibfnamefont {B.~R.}\ \bibnamefont {Ortiz}}, \bibinfo {author}
  {\bibfnamefont {S.~H.}\ \bibnamefont {Ryu}}, \bibinfo {author} {\bibfnamefont
  {J.}~\bibnamefont {Kim}}, \bibinfo {author} {\bibfnamefont {J.}~\bibnamefont
  {Yoo}}, \bibinfo {author} {\bibfnamefont {G.}~\bibnamefont {Sangiovanni}},
  \bibinfo {author} {\bibfnamefont {D.}~\bibnamefont {Di~Sante}}, \bibinfo
  {author} {\bibfnamefont {B.-G.}\ \bibnamefont {Park}}, \bibinfo {author}
  {\bibfnamefont {C.}~\bibnamefont {Jozwiak}}, \bibinfo {author} {\bibfnamefont
  {A.}~\bibnamefont {Bostwick}}, \bibinfo {author} {\bibfnamefont
  {E.}~\bibnamefont {Rotenberg}}, \bibinfo {author} {\bibfnamefont
  {E.}~\bibnamefont {Kaxiras}}, \bibinfo {author} {\bibfnamefont {S.~D.}\
  \bibnamefont {Wilson}}, \bibinfo {author} {\bibfnamefont {J.-H.}\
  \bibnamefont {Park}},\ and\ \bibinfo {author} {\bibfnamefont
  {R.}~\bibnamefont {Comin}},\ }\bibfield  {title} {\bibinfo {title} {Twofold
  van hove singularity and origin of charge order in topological kagome
  superconductor csv3sb5},\ }\href {https://doi.org/10.1038/s41567-021-01451-5}
  {\bibfield  {journal} {\bibinfo  {journal} {Nature Physics}\ }\textbf
  {\bibinfo {volume} {18}},\ \bibinfo {pages} {301} (\bibinfo {year}
  {2022})}\BibitemShut {NoStop}%
\bibitem [{\citenamefont {Jiang}\ \emph
  {et~al.}(2023{\natexlab{b}})\citenamefont {Jiang}, \citenamefont {Hu},
  \citenamefont {Călugăru}, \citenamefont {Felser}, \citenamefont
  {Blanco-Canosa}, \citenamefont {Weng}, \citenamefont {Xu},\ and\
  \citenamefont {Bernevig}}]{Yi_2023}%
  \BibitemOpen
  \bibfield  {author} {\bibinfo {author} {\bibfnamefont {Y.}~\bibnamefont
  {Jiang}}, \bibinfo {author} {\bibfnamefont {H.}~\bibnamefont {Hu}}, \bibinfo
  {author} {\bibfnamefont {D.}~\bibnamefont {Călugăru}}, \bibinfo {author}
  {\bibfnamefont {C.}~\bibnamefont {Felser}}, \bibinfo {author} {\bibfnamefont
  {S.}~\bibnamefont {Blanco-Canosa}}, \bibinfo {author} {\bibfnamefont
  {H.}~\bibnamefont {Weng}}, \bibinfo {author} {\bibfnamefont {Y.}~\bibnamefont
  {Xu}},\ and\ \bibinfo {author} {\bibfnamefont {B.~A.}\ \bibnamefont
  {Bernevig}},\ }\href@noop {} {\bibinfo {title} {Kagome materials
  $\mathrm{II}$: Sg 191: $\mathrm{FeGe}$ as a lego building block for the
  entire 1:6:6 series: hidden d-orbital decoupling of flat band sectors,
  effective models and interaction hamiltonians}} (\bibinfo {year}
  {2023}{\natexlab{b}}),\ \Eprint {https://arxiv.org/abs/2311.09290}
  {arXiv:2311.09290 [cond-mat.str-el]} \BibitemShut {NoStop}%
\bibitem [{\citenamefont {Kang}\ \emph {et~al.}(2019)\citenamefont {Kang},
  \citenamefont {Pelliciari}, \citenamefont {Frano}, \citenamefont {Breznay},
  \citenamefont {Schierle}, \citenamefont {Weschke}, \citenamefont {Sutarto},
  \citenamefont {He}, \citenamefont {Shafer}, \citenamefont {Arenholz},
  \citenamefont {Chen}, \citenamefont {Zhang}, \citenamefont {Ruiz},
  \citenamefont {Hao}, \citenamefont {Lewin}, \citenamefont {Analytis},
  \citenamefont {Krockenberger}, \citenamefont {Yamamoto}, \citenamefont
  {Das},\ and\ \citenamefont {Comin}}]{Kang_2019}%
  \BibitemOpen
  \bibfield  {author} {\bibinfo {author} {\bibfnamefont {M.}~\bibnamefont
  {Kang}}, \bibinfo {author} {\bibfnamefont {J.}~\bibnamefont {Pelliciari}},
  \bibinfo {author} {\bibfnamefont {A.}~\bibnamefont {Frano}}, \bibinfo
  {author} {\bibfnamefont {N.}~\bibnamefont {Breznay}}, \bibinfo {author}
  {\bibfnamefont {E.}~\bibnamefont {Schierle}}, \bibinfo {author}
  {\bibfnamefont {E.}~\bibnamefont {Weschke}}, \bibinfo {author} {\bibfnamefont
  {R.}~\bibnamefont {Sutarto}}, \bibinfo {author} {\bibfnamefont
  {F.}~\bibnamefont {He}}, \bibinfo {author} {\bibfnamefont {P.}~\bibnamefont
  {Shafer}}, \bibinfo {author} {\bibfnamefont {E.}~\bibnamefont {Arenholz}},
  \bibinfo {author} {\bibfnamefont {M.}~\bibnamefont {Chen}}, \bibinfo {author}
  {\bibfnamefont {K.}~\bibnamefont {Zhang}}, \bibinfo {author} {\bibfnamefont
  {A.}~\bibnamefont {Ruiz}}, \bibinfo {author} {\bibfnamefont {Z.}~\bibnamefont
  {Hao}}, \bibinfo {author} {\bibfnamefont {S.}~\bibnamefont {Lewin}}, \bibinfo
  {author} {\bibfnamefont {J.}~\bibnamefont {Analytis}}, \bibinfo {author}
  {\bibfnamefont {Y.}~\bibnamefont {Krockenberger}}, \bibinfo {author}
  {\bibfnamefont {H.}~\bibnamefont {Yamamoto}}, \bibinfo {author}
  {\bibfnamefont {T.}~\bibnamefont {Das}},\ and\ \bibinfo {author}
  {\bibfnamefont {R.}~\bibnamefont {Comin}},\ }\bibfield  {title} {\bibinfo
  {title} {Evolution of charge order topology across a magnetic phase
  transition in cuprate superconductors},\ }\href
  {https://doi.org/10.1038/s41567-018-0401-8} {\bibfield  {journal} {\bibinfo
  {journal} {Nature Physics}\ }\textbf {\bibinfo {volume} {15}},\ \bibinfo
  {pages} {335} (\bibinfo {year} {2019})}\BibitemShut {NoStop}%
\bibitem [{\citenamefont {Schoeffel}\ \emph {et~al.}(1996)\citenamefont
  {Schoeffel}, \citenamefont {Pouget}, \citenamefont {Dhalenne},\ and\
  \citenamefont {Revcolevschi}}]{Schoeffel1996}%
  \BibitemOpen
  \bibfield  {author} {\bibinfo {author} {\bibfnamefont {J.~P.}\ \bibnamefont
  {Schoeffel}}, \bibinfo {author} {\bibfnamefont {J.~P.}\ \bibnamefont
  {Pouget}}, \bibinfo {author} {\bibfnamefont {G.}~\bibnamefont {Dhalenne}},\
  and\ \bibinfo {author} {\bibfnamefont {A.}~\bibnamefont {Revcolevschi}},\
  }\bibfield  {title} {\bibinfo {title} {Spin-peierls lattice fluctuations of
  pure and si- and zn-substituted ${\mathrm{cugeo}}_{3}$},\ }\href
  {https://doi.org/10.1103/PhysRevB.53.14971} {\bibfield  {journal} {\bibinfo
  {journal} {Phys. Rev. B}\ }\textbf {\bibinfo {volume} {53}},\ \bibinfo
  {pages} {14971} (\bibinfo {year} {1996})}\BibitemShut {NoStop}%
\bibitem [{\citenamefont {Gama}\ \emph {et~al.}(1993)\citenamefont {Gama},
  \citenamefont {Henriques}, \citenamefont {Almeida},\ and\ \citenamefont
  {Pouget}}]{Gama1993}%
  \BibitemOpen
  \bibfield  {author} {\bibinfo {author} {\bibfnamefont {V.}~\bibnamefont
  {Gama}}, \bibinfo {author} {\bibfnamefont {R.}~\bibnamefont {Henriques}},
  \bibinfo {author} {\bibfnamefont {M.}~\bibnamefont {Almeida}},\ and\ \bibinfo
  {author} {\bibfnamefont {J.}~\bibnamefont {Pouget}},\ }\bibfield  {title}
  {\bibinfo {title} {Diffuse x-ray scattering evidence for peierls and
  “spin-peierls” like transitions in the organic conductors
  $\mathrm{(Perylene)_2[M(mnt)_2]}$ ($\mathrm{M = Cu, Ni, Co and Fe}$)},\
  }\href {https://doi.org/https://doi.org/10.1016/0379-6779(93)90305-G}
  {\bibfield  {journal} {\bibinfo  {journal} {Synthetic Metals}\ }\textbf
  {\bibinfo {volume} {56}},\ \bibinfo {pages} {1677} (\bibinfo {year}
  {1993})}\BibitemShut {NoStop}%
\bibitem [{\citenamefont {Shi}\ \emph {et~al.}(2024)\citenamefont {Shi},
  \citenamefont {Deng}, \citenamefont {Kotla}, \citenamefont {Liu},
  \citenamefont {Ramakrishnan}, \citenamefont {Eisele}, \citenamefont
  {Agarwal}, \citenamefont {Noohinejad}, \citenamefont {Liu}, \citenamefont
  {Yang}, \citenamefont {Liu}, \citenamefont {Maity}, \citenamefont {Wang},
  \citenamefont {Lin}, \citenamefont {Kang}, \citenamefont {Yang},
  \citenamefont {Li}, \citenamefont {Yang}, \citenamefont {Li}, \citenamefont
  {Qi}, \citenamefont {Thamizhavel}, \citenamefont {Ren}, \citenamefont {Cao},
  \citenamefont {Yin}, \citenamefont {van Smaalen}, \citenamefont {Cao},\ and\
  \citenamefont {Bao}}]{shi2024}%
  \BibitemOpen
  \bibfield  {author} {\bibinfo {author} {\bibfnamefont {C.}~\bibnamefont
  {Shi}}, \bibinfo {author} {\bibfnamefont {H.}~\bibnamefont {Deng}}, \bibinfo
  {author} {\bibfnamefont {S.~R.}\ \bibnamefont {Kotla}}, \bibinfo {author}
  {\bibfnamefont {Y.}~\bibnamefont {Liu}}, \bibinfo {author} {\bibfnamefont
  {S.}~\bibnamefont {Ramakrishnan}}, \bibinfo {author} {\bibfnamefont
  {C.}~\bibnamefont {Eisele}}, \bibinfo {author} {\bibfnamefont
  {H.}~\bibnamefont {Agarwal}}, \bibinfo {author} {\bibfnamefont
  {L.}~\bibnamefont {Noohinejad}}, \bibinfo {author} {\bibfnamefont {J.-Y.}\
  \bibnamefont {Liu}}, \bibinfo {author} {\bibfnamefont {T.}~\bibnamefont
  {Yang}}, \bibinfo {author} {\bibfnamefont {G.}~\bibnamefont {Liu}}, \bibinfo
  {author} {\bibfnamefont {B.~B.}\ \bibnamefont {Maity}}, \bibinfo {author}
  {\bibfnamefont {Q.}~\bibnamefont {Wang}}, \bibinfo {author} {\bibfnamefont
  {Z.}~\bibnamefont {Lin}}, \bibinfo {author} {\bibfnamefont {B.}~\bibnamefont
  {Kang}}, \bibinfo {author} {\bibfnamefont {W.}~\bibnamefont {Yang}}, \bibinfo
  {author} {\bibfnamefont {Y.}~\bibnamefont {Li}}, \bibinfo {author}
  {\bibfnamefont {Z.}~\bibnamefont {Yang}}, \bibinfo {author} {\bibfnamefont
  {Y.}~\bibnamefont {Li}}, \bibinfo {author} {\bibfnamefont {Y.}~\bibnamefont
  {Qi}}, \bibinfo {author} {\bibfnamefont {A.}~\bibnamefont {Thamizhavel}},
  \bibinfo {author} {\bibfnamefont {W.}~\bibnamefont {Ren}}, \bibinfo {author}
  {\bibfnamefont {G.-H.}\ \bibnamefont {Cao}}, \bibinfo {author} {\bibfnamefont
  {J.-X.}\ \bibnamefont {Yin}}, \bibinfo {author} {\bibfnamefont
  {S.}~\bibnamefont {van Smaalen}}, \bibinfo {author} {\bibfnamefont
  {S.}~\bibnamefont {Cao}},\ and\ \bibinfo {author} {\bibfnamefont {J.-K.}\
  \bibnamefont {Bao}},\ }\href@noop {} {\bibinfo {title} {Charge density wave
  without long-range structural modulation in canted antiferromagnetic kagome
  $\mathrm{FeGe}$}} (\bibinfo {year} {2024}),\ \Eprint
  {https://arxiv.org/abs/2404.00996} {arXiv:2404.00996 [cond-mat.str-el]}
  \BibitemShut {NoStop}%
\bibitem [{\citenamefont {Brazovskii}(1974)}]{Brazovskii}%
  \BibitemOpen
  \bibfield  {author} {\bibinfo {author} {\bibfnamefont {S.~A.}\ \bibnamefont
  {Brazovskii}},\ }\bibfield  {title} {\bibinfo {title} {Phase transition of an
  isotropic system to a nonuniform state.},\ }\href
  {https://doi.org/1975JETP...41...85B} {\bibfield  {journal} {\bibinfo
  {journal} {Sov. Phys. JETP 41, 85–89 (1974)}\ }\textbf {\bibinfo {volume}
  {41}},\ \bibinfo {pages} {85} (\bibinfo {year} {1974})}\BibitemShut {NoStop}%
\bibitem [{\citenamefont {Janoschek}\ \emph {et~al.}(2013)\citenamefont
  {Janoschek}, \citenamefont {Garst}, \citenamefont {Bauer}, \citenamefont
  {Krautscheid}, \citenamefont {Georgii}, \citenamefont {B\"oni},\ and\
  \citenamefont {Pfleiderer}}]{Jano2013}%
  \BibitemOpen
  \bibfield  {author} {\bibinfo {author} {\bibfnamefont {M.}~\bibnamefont
  {Janoschek}}, \bibinfo {author} {\bibfnamefont {M.}~\bibnamefont {Garst}},
  \bibinfo {author} {\bibfnamefont {A.}~\bibnamefont {Bauer}}, \bibinfo
  {author} {\bibfnamefont {P.}~\bibnamefont {Krautscheid}}, \bibinfo {author}
  {\bibfnamefont {R.}~\bibnamefont {Georgii}}, \bibinfo {author} {\bibfnamefont
  {P.}~\bibnamefont {B\"oni}},\ and\ \bibinfo {author} {\bibfnamefont
  {C.}~\bibnamefont {Pfleiderer}},\ }\bibfield  {title} {\bibinfo {title}
  {Fluctuation-induced first-order phase transition in dzyaloshinskii-moriya
  helimagnets},\ }\href {https://doi.org/10.1103/PhysRevB.87.134407} {\bibfield
   {journal} {\bibinfo  {journal} {Phys. Rev. B}\ }\textbf {\bibinfo {volume}
  {87}},\ \bibinfo {pages} {134407} (\bibinfo {year} {2013})}\BibitemShut
  {NoStop}%
\bibitem [{\citenamefont {Boschini}\ \emph {et~al.}(2021)\citenamefont
  {Boschini}, \citenamefont {Minola}, \citenamefont {Sutarto}, \citenamefont
  {Schierle}, \citenamefont {Bluschke}, \citenamefont {Das}, \citenamefont
  {Yang}, \citenamefont {Michiardi}, \citenamefont {Shao}, \citenamefont
  {Feng}, \citenamefont {Ono}, \citenamefont {Zhong}, \citenamefont
  {Schneeloch}, \citenamefont {Gu}, \citenamefont {Weschke}, \citenamefont
  {He}, \citenamefont {Chuang}, \citenamefont {Keimer}, \citenamefont
  {Damascelli}, \citenamefont {Frano},\ and\ \citenamefont
  {da~Silva~Neto}}]{Boschini_2021}%
  \BibitemOpen
  \bibfield  {author} {\bibinfo {author} {\bibfnamefont {F.}~\bibnamefont
  {Boschini}}, \bibinfo {author} {\bibfnamefont {M.}~\bibnamefont {Minola}},
  \bibinfo {author} {\bibfnamefont {R.}~\bibnamefont {Sutarto}}, \bibinfo
  {author} {\bibfnamefont {E.}~\bibnamefont {Schierle}}, \bibinfo {author}
  {\bibfnamefont {M.}~\bibnamefont {Bluschke}}, \bibinfo {author}
  {\bibfnamefont {S.}~\bibnamefont {Das}}, \bibinfo {author} {\bibfnamefont
  {Y.}~\bibnamefont {Yang}}, \bibinfo {author} {\bibfnamefont {M.}~\bibnamefont
  {Michiardi}}, \bibinfo {author} {\bibfnamefont {Y.~C.}\ \bibnamefont {Shao}},
  \bibinfo {author} {\bibfnamefont {X.}~\bibnamefont {Feng}}, \bibinfo {author}
  {\bibfnamefont {S.}~\bibnamefont {Ono}}, \bibinfo {author} {\bibfnamefont
  {R.~D.}\ \bibnamefont {Zhong}}, \bibinfo {author} {\bibfnamefont {J.~A.}\
  \bibnamefont {Schneeloch}}, \bibinfo {author} {\bibfnamefont {G.~D.}\
  \bibnamefont {Gu}}, \bibinfo {author} {\bibfnamefont {E.}~\bibnamefont
  {Weschke}}, \bibinfo {author} {\bibfnamefont {F.}~\bibnamefont {He}},
  \bibinfo {author} {\bibfnamefont {Y.~D.}\ \bibnamefont {Chuang}}, \bibinfo
  {author} {\bibfnamefont {B.}~\bibnamefont {Keimer}}, \bibinfo {author}
  {\bibfnamefont {A.}~\bibnamefont {Damascelli}}, \bibinfo {author}
  {\bibfnamefont {A.}~\bibnamefont {Frano}},\ and\ \bibinfo {author}
  {\bibfnamefont {E.~H.}\ \bibnamefont {da~Silva~Neto}},\ }\bibfield  {title}
  {\bibinfo {title} {Dynamic electron correlations with charge order wavelength
  along all directions in the copper oxide plane},\ }\href
  {https://doi.org/10.1038/s41467-020-20824-7} {\bibfield  {journal} {\bibinfo
  {journal} {Nature Communications}\ }\textbf {\bibinfo {volume} {12}},\
  \bibinfo {pages} {597} (\bibinfo {year} {2021})}\BibitemShut {NoStop}%
\bibitem [{\citenamefont {Aebischer}\ \emph {et~al.}(2006)\citenamefont
  {Aebischer}, \citenamefont {Hostettler}, \citenamefont {Hauser},
  \citenamefont {Krämer}, \citenamefont {Weber}, \citenamefont {Güdel},\ and\
  \citenamefont {Bürgi}}]{Aebischer2006}%
  \BibitemOpen
  \bibfield  {author} {\bibinfo {author} {\bibfnamefont {A.}~\bibnamefont
  {Aebischer}}, \bibinfo {author} {\bibfnamefont {M.}~\bibnamefont
  {Hostettler}}, \bibinfo {author} {\bibfnamefont {J.}~\bibnamefont {Hauser}},
  \bibinfo {author} {\bibfnamefont {K.}~\bibnamefont {Krämer}}, \bibinfo
  {author} {\bibfnamefont {T.}~\bibnamefont {Weber}}, \bibinfo {author}
  {\bibfnamefont {H.~U.}\ \bibnamefont {Güdel}},\ and\ \bibinfo {author}
  {\bibfnamefont {H.-B.}\ \bibnamefont {Bürgi}},\ }\bibfield  {title}
  {\bibinfo {title} {Structural and spectroscopic characterization of active
  sites in a family of light-emitting sodium lanthanide tetrafluorides},\
  }\href {https://doi.org/https://doi.org/10.1002/anie.200503966} {\bibfield
  {journal} {\bibinfo  {journal} {Angewandte Chemie International Edition}\
  }\textbf {\bibinfo {volume} {45}},\ \bibinfo {pages} {2802} (\bibinfo {year}
  {2006})}\BibitemShut {NoStop}%
\bibitem [{\citenamefont {Aeppli}\ and\ \citenamefont
  {Bruinsma}(1984)}]{Aeppli1984}%
  \BibitemOpen
  \bibfield  {author} {\bibinfo {author} {\bibfnamefont {G.}~\bibnamefont
  {Aeppli}}\ and\ \bibinfo {author} {\bibfnamefont {R.}~\bibnamefont
  {Bruinsma}},\ }\bibfield  {title} {\bibinfo {title} {Hexatic order and liquid
  density fluctuations},\ }\href {https://doi.org/10.1103/PhysRevLett.53.2133}
  {\bibfield  {journal} {\bibinfo  {journal} {Phys. Rev. Lett.}\ }\textbf
  {\bibinfo {volume} {53}},\ \bibinfo {pages} {2133} (\bibinfo {year}
  {1984})}\BibitemShut {NoStop}%
\bibitem [{\citenamefont {Zaluzhnyy}\ \emph {et~al.}(2017)\citenamefont
  {Zaluzhnyy}, \citenamefont {Kurta}, \citenamefont {Menushenkov},
  \citenamefont {I.},\ and\ \citenamefont {Vartanyants}}]{Zaluz2017}%
  \BibitemOpen
  \bibfield  {author} {\bibinfo {author} {\bibfnamefont {I.~A.}\ \bibnamefont
  {Zaluzhnyy}}, \bibinfo {author} {\bibfnamefont {R.~P.}\ \bibnamefont
  {Kurta}}, \bibinfo {author} {\bibfnamefont {A.~P.}\ \bibnamefont
  {Menushenkov}}, \bibinfo {author} {\bibfnamefont {O.~B.}\ \bibnamefont
  {I.}},\ and\ \bibinfo {author} {\bibfnamefont {I.~A.}\ \bibnamefont
  {Vartanyants}},\ }\bibfield  {title} {\bibinfo {title} {Analysis of the shape
  of x-ray diffraction peaks originating from the hexatic phase of liquid
  crystal films},\ }\href {https://doi.org/10.1080/15421406.2017.1289582}
  {\bibfield  {journal} {\bibinfo  {journal} {Molecular Crystals and Liquid
  Crystals}\ }\textbf {\bibinfo {volume} {647}},\ \bibinfo {pages} {169}
  (\bibinfo {year} {2017})}\BibitemShut {NoStop}%
\bibitem [{\citenamefont {Peterson}\ and\ \citenamefont
  {Kaganer}(1994)}]{Peterson1994}%
  \BibitemOpen
  \bibfield  {author} {\bibinfo {author} {\bibfnamefont {I.~R.}\ \bibnamefont
  {Peterson}}\ and\ \bibinfo {author} {\bibfnamefont {V.~M.}\ \bibnamefont
  {Kaganer}},\ }\bibfield  {title} {\bibinfo {title} {Diffraction line profile
  of a two-dimensional hexatic},\ }\href
  {https://doi.org/10.1103/PhysRevLett.73.102} {\bibfield  {journal} {\bibinfo
  {journal} {Phys. Rev. Lett.}\ }\textbf {\bibinfo {volume} {73}},\ \bibinfo
  {pages} {102} (\bibinfo {year} {1994})}\BibitemShut {NoStop}%
\bibitem [{\citenamefont {Birgeneau}\ and\ \citenamefont
  {Litster}(1978)}]{Birgeneau_1978}%
  \BibitemOpen
  \bibfield  {author} {\bibinfo {author} {\bibfnamefont {R.~J.}\ \bibnamefont
  {Birgeneau}}\ and\ \bibinfo {author} {\bibfnamefont {J.~D.}\ \bibnamefont
  {Litster}},\ }\bibfield  {title} {\bibinfo {title} {Bond orientational order
  model for smectic b liquid crystals},\ }\href
  {https://api.semanticscholar.org/CorpusID:43842886} {\bibfield  {journal}
  {\bibinfo  {journal} {Journal De Physique Lettres}\ }\textbf {\bibinfo
  {volume} {39}},\ \bibinfo {pages} {399} (\bibinfo {year} {1978})}\BibitemShut
  {NoStop}%
\bibitem [{\citenamefont {Pindak}\ \emph {et~al.}(1981)\citenamefont {Pindak},
  \citenamefont {Moncton}, \citenamefont {Davey},\ and\ \citenamefont
  {Goodby}}]{Pindak_1981}%
  \BibitemOpen
  \bibfield  {author} {\bibinfo {author} {\bibfnamefont {R.}~\bibnamefont
  {Pindak}}, \bibinfo {author} {\bibfnamefont {D.~E.}\ \bibnamefont {Moncton}},
  \bibinfo {author} {\bibfnamefont {S.~C.}\ \bibnamefont {Davey}},\ and\
  \bibinfo {author} {\bibfnamefont {J.~W.}\ \bibnamefont {Goodby}},\ }\bibfield
   {title} {\bibinfo {title} {X-ray observation of a stacked hexatic
  liquid-crystal $b$ phase},\ }\href
  {https://doi.org/10.1103/PhysRevLett.46.1135} {\bibfield  {journal} {\bibinfo
   {journal} {Phys. Rev. Lett.}\ }\textbf {\bibinfo {volume} {46}},\ \bibinfo
  {pages} {1135} (\bibinfo {year} {1981})}\BibitemShut {NoStop}%
\bibitem [{\citenamefont {Domr{\"o}se}\ \emph {et~al.}(2023)\citenamefont
  {Domr{\"o}se}, \citenamefont {Danz}, \citenamefont {Schaible}, \citenamefont
  {Rossnagel}, \citenamefont {Yalunin},\ and\ \citenamefont
  {Ropers}}]{Domrose_2023}%
  \BibitemOpen
  \bibfield  {author} {\bibinfo {author} {\bibfnamefont {T.}~\bibnamefont
  {Domr{\"o}se}}, \bibinfo {author} {\bibfnamefont {T.}~\bibnamefont {Danz}},
  \bibinfo {author} {\bibfnamefont {S.~F.}\ \bibnamefont {Schaible}}, \bibinfo
  {author} {\bibfnamefont {K.}~\bibnamefont {Rossnagel}}, \bibinfo {author}
  {\bibfnamefont {S.~V.}\ \bibnamefont {Yalunin}},\ and\ \bibinfo {author}
  {\bibfnamefont {C.}~\bibnamefont {Ropers}},\ }\bibfield  {title} {\bibinfo
  {title} {Light-induced hexatic state in a layered quantum material},\ }\href
  {https://doi.org/10.1038/s41563-023-01600-6} {\bibfield  {journal} {\bibinfo
  {journal} {Nature Materials}\ }\textbf {\bibinfo {volume} {22}},\ \bibinfo
  {pages} {1345} (\bibinfo {year} {2023})}\BibitemShut {NoStop}%
\bibitem [{\citenamefont {Aishwarya}\ \emph {et~al.}(2024)\citenamefont
  {Aishwarya}, \citenamefont {May-Mann}, \citenamefont {Almoalem},
  \citenamefont {Ran}, \citenamefont {Saha}, \citenamefont {Paglione},
  \citenamefont {Butch}, \citenamefont {Fradkin},\ and\ \citenamefont
  {Madhavan}}]{Aishwarya_2024}%
  \BibitemOpen
  \bibfield  {author} {\bibinfo {author} {\bibfnamefont {A.}~\bibnamefont
  {Aishwarya}}, \bibinfo {author} {\bibfnamefont {J.}~\bibnamefont {May-Mann}},
  \bibinfo {author} {\bibfnamefont {A.}~\bibnamefont {Almoalem}}, \bibinfo
  {author} {\bibfnamefont {S.}~\bibnamefont {Ran}}, \bibinfo {author}
  {\bibfnamefont {S.~R.}\ \bibnamefont {Saha}}, \bibinfo {author}
  {\bibfnamefont {J.}~\bibnamefont {Paglione}}, \bibinfo {author}
  {\bibfnamefont {N.~P.}\ \bibnamefont {Butch}}, \bibinfo {author}
  {\bibfnamefont {E.}~\bibnamefont {Fradkin}},\ and\ \bibinfo {author}
  {\bibfnamefont {V.}~\bibnamefont {Madhavan}},\ }\bibfield  {title} {\bibinfo
  {title} {Melting of the charge density wave by generation of pairs of
  topological defects in $\mathrm{UTe_2}$},\ }\bibfield  {journal} {\bibinfo
  {journal} {Nature Physics}\ }\href
  {https://doi.org/10.1038/s41567-024-02429-9} {10.1038/s41567-024-02429-9}
  (\bibinfo {year} {2024})\BibitemShut {NoStop}%
\bibitem [{\citenamefont {Guillam{\'o}n}\ \emph {et~al.}(2009)\citenamefont
  {Guillam{\'o}n}, \citenamefont {Suderow}, \citenamefont
  {Fern{\'a}ndez-Pacheco}, \citenamefont {Ses{\'e}}, \citenamefont
  {C{\'o}rdoba}, \citenamefont {De~Teresa}, \citenamefont {Ibarra},\ and\
  \citenamefont {Vieira}}]{Guillamon_2009}%
  \BibitemOpen
  \bibfield  {author} {\bibinfo {author} {\bibfnamefont {I.}~\bibnamefont
  {Guillam{\'o}n}}, \bibinfo {author} {\bibfnamefont {H.}~\bibnamefont
  {Suderow}}, \bibinfo {author} {\bibfnamefont {A.}~\bibnamefont
  {Fern{\'a}ndez-Pacheco}}, \bibinfo {author} {\bibfnamefont {J.}~\bibnamefont
  {Ses{\'e}}}, \bibinfo {author} {\bibfnamefont {R.}~\bibnamefont
  {C{\'o}rdoba}}, \bibinfo {author} {\bibfnamefont {J.~M.}\ \bibnamefont
  {De~Teresa}}, \bibinfo {author} {\bibfnamefont {M.~R.}\ \bibnamefont
  {Ibarra}},\ and\ \bibinfo {author} {\bibfnamefont {S.}~\bibnamefont
  {Vieira}},\ }\bibfield  {title} {\bibinfo {title} {Direct observation of
  melting in a two-dimensional superconducting vortex lattice},\ }\href
  {https://doi.org/10.1038/nphys1368} {\bibfield  {journal} {\bibinfo
  {journal} {Nature Physics}\ }\textbf {\bibinfo {volume} {5}},\ \bibinfo
  {pages} {651} (\bibinfo {year} {2009})}\BibitemShut {NoStop}%
\bibitem [{\citenamefont {Comin}\ \emph {et~al.}(2015)\citenamefont {Comin},
  \citenamefont {Sutarto}, \citenamefont {da~Silva~Neto}, \citenamefont
  {Chauviere}, \citenamefont {Liang}, \citenamefont {Hardy}, \citenamefont
  {Bonn}, \citenamefont {He}, \citenamefont {Sawatzky},\ and\ \citenamefont
  {Damascelli}}]{Rcomin_Science}%
  \BibitemOpen
  \bibfield  {author} {\bibinfo {author} {\bibfnamefont {R.}~\bibnamefont
  {Comin}}, \bibinfo {author} {\bibfnamefont {R.}~\bibnamefont {Sutarto}},
  \bibinfo {author} {\bibfnamefont {E.~H.}\ \bibnamefont {da~Silva~Neto}},
  \bibinfo {author} {\bibfnamefont {L.}~\bibnamefont {Chauviere}}, \bibinfo
  {author} {\bibfnamefont {R.}~\bibnamefont {Liang}}, \bibinfo {author}
  {\bibfnamefont {W.~N.}\ \bibnamefont {Hardy}}, \bibinfo {author}
  {\bibfnamefont {D.~A.}\ \bibnamefont {Bonn}}, \bibinfo {author}
  {\bibfnamefont {F.}~\bibnamefont {He}}, \bibinfo {author} {\bibfnamefont
  {G.~A.}\ \bibnamefont {Sawatzky}},\ and\ \bibinfo {author} {\bibfnamefont
  {A.}~\bibnamefont {Damascelli}},\ }\bibfield  {title} {\bibinfo {title}
  {Broken translational and rotational symmetry via charge stripe order in
  underdoped $\mathrm{YBa_2Cu_3O_{6+y}}$},\ }\href
  {https://doi.org/10.1126/science.1258399} {\bibfield  {journal} {\bibinfo
  {journal} {Science}\ }\textbf {\bibinfo {volume} {347}},\ \bibinfo {pages}
  {1335} (\bibinfo {year} {2015})}\BibitemShut {NoStop}%
\bibitem [{\citenamefont {Fine}(2016)}]{Rcomin_Comment}%
  \BibitemOpen
  \bibfield  {author} {\bibinfo {author} {\bibfnamefont {B.~V.}\ \bibnamefont
  {Fine}},\ }\bibfield  {title} {\bibinfo {title} {Comment on 'broken
  translational and rotational symmetry via charge stripe order in underdoped
  $\mathrm{YBa_2Cu_3O_{6+y}}$'},\ }\href
  {https://doi.org/10.1126/science.aac4454} {\bibfield  {journal} {\bibinfo
  {journal} {Science}\ }\textbf {\bibinfo {volume} {351}},\ \bibinfo {pages}
  {235} (\bibinfo {year} {2016})}\BibitemShut {NoStop}%
\bibitem [{\citenamefont {Comin}\ \emph {et~al.}(2016)\citenamefont {Comin},
  \citenamefont {Sutarto}, \citenamefont {da~Silva~Neto}, \citenamefont
  {Chauviere}, \citenamefont {Liang}, \citenamefont {Hardy}, \citenamefont
  {Bonn}, \citenamefont {He}, \citenamefont {Sawatzky},\ and\ \citenamefont
  {Damascelli}}]{Rcomin_Answer}%
  \BibitemOpen
  \bibfield  {author} {\bibinfo {author} {\bibfnamefont {R.}~\bibnamefont
  {Comin}}, \bibinfo {author} {\bibfnamefont {R.}~\bibnamefont {Sutarto}},
  \bibinfo {author} {\bibfnamefont {E.~H.}\ \bibnamefont {da~Silva~Neto}},
  \bibinfo {author} {\bibfnamefont {L.}~\bibnamefont {Chauviere}}, \bibinfo
  {author} {\bibfnamefont {R.}~\bibnamefont {Liang}}, \bibinfo {author}
  {\bibfnamefont {W.~N.}\ \bibnamefont {Hardy}}, \bibinfo {author}
  {\bibfnamefont {D.~A.}\ \bibnamefont {Bonn}}, \bibinfo {author}
  {\bibfnamefont {F.}~\bibnamefont {He}}, \bibinfo {author} {\bibfnamefont
  {G.~A.}\ \bibnamefont {Sawatzky}},\ and\ \bibinfo {author} {\bibfnamefont
  {A.}~\bibnamefont {Damascelli}},\ }\bibfield  {title} {\bibinfo {title}
  {Response to comment on 'broken translational and rotational symmetry via
  charge stripe order in underdoped $\mathrm{YBa_2Cu_3O_{6+y}}$'},\ }\href
  {https://doi.org/10.1126/science.aac4778} {\bibfield  {journal} {\bibinfo
  {journal} {Science}\ }\textbf {\bibinfo {volume} {351}},\ \bibinfo {pages}
  {235} (\bibinfo {year} {2016})}\BibitemShut {NoStop}%
\bibitem [{\citenamefont {Wenzel}\ \emph {et~al.}(2024)\citenamefont {Wenzel},
  \citenamefont {Uykur}, \citenamefont {Tsirlin}, \citenamefont {Pal},
  \citenamefont {Roy}, \citenamefont {Yi}, \citenamefont {Shekhar},
  \citenamefont {Felser}, \citenamefont {Pronin},\ and\ \citenamefont
  {Dressel}}]{Wenzel_2024}%
  \BibitemOpen
  \bibfield  {author} {\bibinfo {author} {\bibfnamefont {M.}~\bibnamefont
  {Wenzel}}, \bibinfo {author} {\bibfnamefont {E.}~\bibnamefont {Uykur}},
  \bibinfo {author} {\bibfnamefont {A.~A.}\ \bibnamefont {Tsirlin}}, \bibinfo
  {author} {\bibfnamefont {S.}~\bibnamefont {Pal}}, \bibinfo {author}
  {\bibfnamefont {R.~M.}\ \bibnamefont {Roy}}, \bibinfo {author} {\bibfnamefont
  {C.}~\bibnamefont {Yi}}, \bibinfo {author} {\bibfnamefont {C.}~\bibnamefont
  {Shekhar}}, \bibinfo {author} {\bibfnamefont {C.}~\bibnamefont {Felser}},
  \bibinfo {author} {\bibfnamefont {A.~V.}\ \bibnamefont {Pronin}},\ and\
  \bibinfo {author} {\bibfnamefont {M.}~\bibnamefont {Dressel}},\ }\href@noop
  {} {\bibinfo {title} {Intriguing low-temperature phase in the
  antiferromagnetic kagome metal $\mathrm{FeGe}$}} (\bibinfo {year} {2024}),\
  \Eprint {https://arxiv.org/abs/2401.13474} {arXiv:2401.13474
  [cond-mat.str-el]} \BibitemShut {NoStop}%
\bibitem [{\citenamefont {Song}\ \emph {et~al.}(2023)\citenamefont {Song},
  \citenamefont {Ying}, \citenamefont {Wu}, \citenamefont {Xia}, \citenamefont
  {Yin}, \citenamefont {Zhang}, \citenamefont {Song}, \citenamefont {Yang},
  \citenamefont {Guo}, \citenamefont {Gu}, \citenamefont {Chen}, \citenamefont
  {Hu}, \citenamefont {Schnyder}, \citenamefont {Lei}, \citenamefont {Guo},\
  and\ \citenamefont {Li}}]{Song2023}%
  \BibitemOpen
  \bibfield  {author} {\bibinfo {author} {\bibfnamefont {B.}~\bibnamefont
  {Song}}, \bibinfo {author} {\bibfnamefont {T.}~\bibnamefont {Ying}}, \bibinfo
  {author} {\bibfnamefont {X.}~\bibnamefont {Wu}}, \bibinfo {author}
  {\bibfnamefont {W.}~\bibnamefont {Xia}}, \bibinfo {author} {\bibfnamefont
  {Q.}~\bibnamefont {Yin}}, \bibinfo {author} {\bibfnamefont {Q.}~\bibnamefont
  {Zhang}}, \bibinfo {author} {\bibfnamefont {Y.}~\bibnamefont {Song}},
  \bibinfo {author} {\bibfnamefont {X.}~\bibnamefont {Yang}}, \bibinfo {author}
  {\bibfnamefont {J.}~\bibnamefont {Guo}}, \bibinfo {author} {\bibfnamefont
  {L.}~\bibnamefont {Gu}}, \bibinfo {author} {\bibfnamefont {X.}~\bibnamefont
  {Chen}}, \bibinfo {author} {\bibfnamefont {J.}~\bibnamefont {Hu}}, \bibinfo
  {author} {\bibfnamefont {A.~P.}\ \bibnamefont {Schnyder}}, \bibinfo {author}
  {\bibfnamefont {H.}~\bibnamefont {Lei}}, \bibinfo {author} {\bibfnamefont
  {Y.}~\bibnamefont {Guo}},\ and\ \bibinfo {author} {\bibfnamefont
  {S.}~\bibnamefont {Li}},\ }\bibfield  {title} {\bibinfo {title} {Anomalous
  enhancement of charge density wave in kagome superconductor csv3sb5
  approaching the 2d limit},\ }\href
  {https://doi.org/10.1038/s41467-023-38257-3} {\bibfield  {journal} {\bibinfo
  {journal} {Nature Communications}\ }\textbf {\bibinfo {volume} {14}},\
  \bibinfo {pages} {2492} (\bibinfo {year} {2023})}\BibitemShut {NoStop}%
\bibitem [{\citenamefont {Dyadkin}\ \emph {et~al.}(2016)\citenamefont
  {Dyadkin}, \citenamefont {Pattison}, \citenamefont {Dmitriev},\ and\
  \citenamefont {Chernyshov}}]{Dyadkin_2016}%
  \BibitemOpen
  \bibfield  {author} {\bibinfo {author} {\bibfnamefont {V.}~\bibnamefont
  {Dyadkin}}, \bibinfo {author} {\bibfnamefont {P.}~\bibnamefont {Pattison}},
  \bibinfo {author} {\bibfnamefont {V.}~\bibnamefont {Dmitriev}},\ and\
  \bibinfo {author} {\bibfnamefont {D.}~\bibnamefont {Chernyshov}},\ }\bibfield
   {title} {\bibinfo {title} {A new multipurpose diffractometer pilatus@snbl},\
  }\href {https://doi.org/10.1107/S1600577516002411} {\bibfield  {journal}
  {\bibinfo  {journal} {Journal of Synchrotron Radiation}\ }\textbf {\bibinfo
  {volume} {23}},\ \bibinfo {pages} {825} (\bibinfo {year} {2016})}\BibitemShut
  {NoStop}%
\bibitem [{\citenamefont {Kresse}\ and\ \citenamefont
  {Furthm{\"u}ller}(1996{\natexlab{a}})}]{kresse1996efficiency}%
  \BibitemOpen
  \bibfield  {author} {\bibinfo {author} {\bibfnamefont {G.}~\bibnamefont
  {Kresse}}\ and\ \bibinfo {author} {\bibfnamefont {J.}~\bibnamefont
  {Furthm{\"u}ller}},\ }\bibfield  {title} {\bibinfo {title} {Efficiency of
  ab-initio total energy calculations for metals and semiconductors using a
  plane-wave basis set},\ }\href@noop {} {\bibfield  {journal} {\bibinfo
  {journal} {Computational materials science}\ }\textbf {\bibinfo {volume}
  {6}},\ \bibinfo {pages} {15} (\bibinfo {year}
  {1996}{\natexlab{a}})}\BibitemShut {NoStop}%
\bibitem [{\citenamefont {Kresse}\ and\ \citenamefont
  {Hafner}(1993{\natexlab{a}})}]{kresse1993ab1}%
  \BibitemOpen
  \bibfield  {author} {\bibinfo {author} {\bibfnamefont {G.}~\bibnamefont
  {Kresse}}\ and\ \bibinfo {author} {\bibfnamefont {J.}~\bibnamefont
  {Hafner}},\ }\bibfield  {title} {\bibinfo {title} {Ab initio molecular
  dynamics for open-shell transition metals},\ }\href@noop {} {\bibfield
  {journal} {\bibinfo  {journal} {Physical Review B}\ }\textbf {\bibinfo
  {volume} {48}},\ \bibinfo {pages} {13115} (\bibinfo {year}
  {1993}{\natexlab{a}})}\BibitemShut {NoStop}%
\bibitem [{\citenamefont {Kresse}\ and\ \citenamefont
  {Hafner}(1993{\natexlab{b}})}]{kresse1993ab2}%
  \BibitemOpen
  \bibfield  {author} {\bibinfo {author} {\bibfnamefont {G.}~\bibnamefont
  {Kresse}}\ and\ \bibinfo {author} {\bibfnamefont {J.}~\bibnamefont
  {Hafner}},\ }\bibfield  {title} {\bibinfo {title} {Ab initio molecular
  dynamics for liquid metals},\ }\href@noop {} {\bibfield  {journal} {\bibinfo
  {journal} {Physical review B}\ }\textbf {\bibinfo {volume} {47}},\ \bibinfo
  {pages} {558} (\bibinfo {year} {1993}{\natexlab{b}})}\BibitemShut {NoStop}%
\bibitem [{\citenamefont {Kresse}\ and\ \citenamefont
  {Hafner}(1994)}]{kresse1994ab}%
  \BibitemOpen
  \bibfield  {author} {\bibinfo {author} {\bibfnamefont {G.}~\bibnamefont
  {Kresse}}\ and\ \bibinfo {author} {\bibfnamefont {J.}~\bibnamefont
  {Hafner}},\ }\bibfield  {title} {\bibinfo {title} {Ab initio
  molecular-dynamics simulation of the liquid-metal--amorphous-semiconductor
  transition in germanium},\ }\href@noop {} {\bibfield  {journal} {\bibinfo
  {journal} {Physical Review B}\ }\textbf {\bibinfo {volume} {49}},\ \bibinfo
  {pages} {14251} (\bibinfo {year} {1994})}\BibitemShut {NoStop}%
\bibitem [{\citenamefont {Kresse}\ and\ \citenamefont
  {Furthm{\"u}ller}(1996{\natexlab{b}})}]{kresse1996efficient}%
  \BibitemOpen
  \bibfield  {author} {\bibinfo {author} {\bibfnamefont {G.}~\bibnamefont
  {Kresse}}\ and\ \bibinfo {author} {\bibfnamefont {J.}~\bibnamefont
  {Furthm{\"u}ller}},\ }\bibfield  {title} {\bibinfo {title} {Efficient
  iterative schemes for ab initio total-energy calculations using a plane-wave
  basis set},\ }\href@noop {} {\bibfield  {journal} {\bibinfo  {journal}
  {Physical review B}\ }\textbf {\bibinfo {volume} {54}},\ \bibinfo {pages}
  {11169} (\bibinfo {year} {1996}{\natexlab{b}})}\BibitemShut {NoStop}%
\bibitem [{\citenamefont {Perdew}\ \emph {et~al.}(1996)\citenamefont {Perdew},
  \citenamefont {Burke},\ and\ \citenamefont
  {Ernzerhof}}]{perdew1996generalized}%
  \BibitemOpen
  \bibfield  {author} {\bibinfo {author} {\bibfnamefont {J.~P.}\ \bibnamefont
  {Perdew}}, \bibinfo {author} {\bibfnamefont {K.}~\bibnamefont {Burke}},\ and\
  \bibinfo {author} {\bibfnamefont {M.}~\bibnamefont {Ernzerhof}},\ }\bibfield
  {title} {\bibinfo {title} {Generalized gradient approximation made simple},\
  }\href@noop {} {\bibfield  {journal} {\bibinfo  {journal} {Physical review
  letters}\ }\textbf {\bibinfo {volume} {77}},\ \bibinfo {pages} {3865}
  (\bibinfo {year} {1996})}\BibitemShut {NoStop}%
\bibitem [{\citenamefont {Marzari}\ and\ \citenamefont
  {Vanderbilt}(1997)}]{marzari1997maximally}%
  \BibitemOpen
  \bibfield  {author} {\bibinfo {author} {\bibfnamefont {N.}~\bibnamefont
  {Marzari}}\ and\ \bibinfo {author} {\bibfnamefont {D.}~\bibnamefont
  {Vanderbilt}},\ }\bibfield  {title} {\bibinfo {title} {Maximally localized
  generalized wannier functions for composite energy bands},\ }\href@noop {}
  {\bibfield  {journal} {\bibinfo  {journal} {Physical review B}\ }\textbf
  {\bibinfo {volume} {56}},\ \bibinfo {pages} {12847} (\bibinfo {year}
  {1997})}\BibitemShut {NoStop}%
\bibitem [{\citenamefont {Souza}\ \emph {et~al.}(2001)\citenamefont {Souza},
  \citenamefont {Marzari},\ and\ \citenamefont
  {Vanderbilt}}]{souza2001maximally}%
  \BibitemOpen
  \bibfield  {author} {\bibinfo {author} {\bibfnamefont {I.}~\bibnamefont
  {Souza}}, \bibinfo {author} {\bibfnamefont {N.}~\bibnamefont {Marzari}},\
  and\ \bibinfo {author} {\bibfnamefont {D.}~\bibnamefont {Vanderbilt}},\
  }\bibfield  {title} {\bibinfo {title} {Maximally localized wannier functions
  for entangled energy bands},\ }\href@noop {} {\bibfield  {journal} {\bibinfo
  {journal} {Physical Review B}\ }\textbf {\bibinfo {volume} {65}},\ \bibinfo
  {pages} {035109} (\bibinfo {year} {2001})}\BibitemShut {NoStop}%
\bibitem [{\citenamefont {Marzari}\ \emph {et~al.}(2012)\citenamefont
  {Marzari}, \citenamefont {Mostofi}, \citenamefont {Yates}, \citenamefont
  {Souza},\ and\ \citenamefont {Vanderbilt}}]{marzari2012maximally}%
  \BibitemOpen
  \bibfield  {author} {\bibinfo {author} {\bibfnamefont {N.}~\bibnamefont
  {Marzari}}, \bibinfo {author} {\bibfnamefont {A.~A.}\ \bibnamefont
  {Mostofi}}, \bibinfo {author} {\bibfnamefont {J.~R.}\ \bibnamefont {Yates}},
  \bibinfo {author} {\bibfnamefont {I.}~\bibnamefont {Souza}},\ and\ \bibinfo
  {author} {\bibfnamefont {D.}~\bibnamefont {Vanderbilt}},\ }\bibfield  {title}
  {\bibinfo {title} {Maximally localized wannier functions: Theory and
  applications},\ }\href@noop {} {\bibfield  {journal} {\bibinfo  {journal}
  {Reviews of Modern Physics}\ }\textbf {\bibinfo {volume} {84}},\ \bibinfo
  {pages} {1419} (\bibinfo {year} {2012})}\BibitemShut {NoStop}%
\bibitem [{\citenamefont {Pizzi}\ \emph {et~al.}(2020)\citenamefont {Pizzi},
  \citenamefont {Vitale}, \citenamefont {Arita}, \citenamefont {Bl{\"u}gel},
  \citenamefont {Freimuth}, \citenamefont {G{\'e}ranton}, \citenamefont
  {Gibertini}, \citenamefont {Gresch}, \citenamefont {Johnson}, \citenamefont
  {Koretsune} \emph {et~al.}}]{pizzi2020wannier90}%
  \BibitemOpen
  \bibfield  {author} {\bibinfo {author} {\bibfnamefont {G.}~\bibnamefont
  {Pizzi}}, \bibinfo {author} {\bibfnamefont {V.}~\bibnamefont {Vitale}},
  \bibinfo {author} {\bibfnamefont {R.}~\bibnamefont {Arita}}, \bibinfo
  {author} {\bibfnamefont {S.}~\bibnamefont {Bl{\"u}gel}}, \bibinfo {author}
  {\bibfnamefont {F.}~\bibnamefont {Freimuth}}, \bibinfo {author}
  {\bibfnamefont {G.}~\bibnamefont {G{\'e}ranton}}, \bibinfo {author}
  {\bibfnamefont {M.}~\bibnamefont {Gibertini}}, \bibinfo {author}
  {\bibfnamefont {D.}~\bibnamefont {Gresch}}, \bibinfo {author} {\bibfnamefont
  {C.}~\bibnamefont {Johnson}}, \bibinfo {author} {\bibfnamefont
  {T.}~\bibnamefont {Koretsune}}, \emph {et~al.},\ }\bibfield  {title}
  {\bibinfo {title} {Wannier90 as a community code: new features and
  applications},\ }\href@noop {} {\bibfield  {journal} {\bibinfo  {journal}
  {Journal of Physics: Condensed Matter}\ }\textbf {\bibinfo {volume} {32}},\
  \bibinfo {pages} {165902} (\bibinfo {year} {2020})}\BibitemShut {NoStop}%
\bibitem [{\citenamefont {Wu}\ \emph {et~al.}(2018)\citenamefont {Wu},
  \citenamefont {Zhang}, \citenamefont {Song}, \citenamefont {Troyer},\ and\
  \citenamefont {Soluyanov}}]{wu2018wanniertools}%
  \BibitemOpen
  \bibfield  {author} {\bibinfo {author} {\bibfnamefont {Q.}~\bibnamefont
  {Wu}}, \bibinfo {author} {\bibfnamefont {S.}~\bibnamefont {Zhang}}, \bibinfo
  {author} {\bibfnamefont {H.-F.}\ \bibnamefont {Song}}, \bibinfo {author}
  {\bibfnamefont {M.}~\bibnamefont {Troyer}},\ and\ \bibinfo {author}
  {\bibfnamefont {A.~A.}\ \bibnamefont {Soluyanov}},\ }\bibfield  {title}
  {\bibinfo {title} {Wanniertools: An open-source software package for novel
  topological materials},\ }\href@noop {} {\bibfield  {journal} {\bibinfo
  {journal} {Computer Physics Communications}\ }\textbf {\bibinfo {volume}
  {224}},\ \bibinfo {pages} {405} (\bibinfo {year} {2018})}\BibitemShut
  {NoStop}%
\bibitem [{\citenamefont {Zheng}(2018)}]{zheng2018vaspbandunfolding}%
  \BibitemOpen
  \bibfield  {author} {\bibinfo {author} {\bibfnamefont {Q.}~\bibnamefont
  {Zheng}},\ }\bibfield  {title} {\bibinfo {title} {Vasp band unfolding},\
  }\href@noop {} {\bibfield  {journal} {\bibinfo  {journal} {URL
  https://github. com/QijingZheng/VaspBandUnfolding}\ } (\bibinfo {year}
  {2018})}\BibitemShut {NoStop}%
\bibitem [{\citenamefont {Popescu}\ and\ \citenamefont
  {Zunger}(2012)}]{popescu2012extracting}%
  \BibitemOpen
  \bibfield  {author} {\bibinfo {author} {\bibfnamefont {V.}~\bibnamefont
  {Popescu}}\ and\ \bibinfo {author} {\bibfnamefont {A.}~\bibnamefont
  {Zunger}},\ }\bibfield  {title} {\bibinfo {title} {Extracting e versus k
  effective band structure from supercell calculations on alloys and
  impurities},\ }\href@noop {} {\bibfield  {journal} {\bibinfo  {journal}
  {Physical Review B}\ }\textbf {\bibinfo {volume} {85}},\ \bibinfo {pages}
  {085201} (\bibinfo {year} {2012})}\BibitemShut {NoStop}%
\bibitem [{\citenamefont {Kawamura}(2019)}]{kawamura2019fermisurfer}%
  \BibitemOpen
  \bibfield  {author} {\bibinfo {author} {\bibfnamefont {M.}~\bibnamefont
  {Kawamura}},\ }\bibfield  {title} {\bibinfo {title} {Fermisurfer:
  Fermi-surface viewer providing multiple representation schemes},\ }\href@noop
  {} {\bibfield  {journal} {\bibinfo  {journal} {Computer Physics
  Communications}\ }\textbf {\bibinfo {volume} {239}},\ \bibinfo {pages} {197}
  (\bibinfo {year} {2019})}\BibitemShut {NoStop}%
\bibitem [{\citenamefont {Paddison}(2019)}]{Paddison:vk5031}%
  \BibitemOpen
  \bibfield  {author} {\bibinfo {author} {\bibfnamefont {J.~A.~M.}\
  \bibnamefont {Paddison}},\ }\bibfield  {title} {\bibinfo {title} {Ultrafast
  calculation of diffuse scattering from atomistic models},\ }\href
  {https://doi.org/10.1107/S2053273318015632} {\bibfield  {journal} {\bibinfo
  {journal} {Acta Crystallographica Section A}\ }\textbf {\bibinfo {volume}
  {75}},\ \bibinfo {pages} {14} (\bibinfo {year} {2019})}\BibitemShut {NoStop}%
\bibitem [{\citenamefont {Welberry}(2022)}]{Welberry2022diffuse}%
  \BibitemOpen
  \bibfield  {author} {\bibinfo {author} {\bibfnamefont {T.}~\bibnamefont
  {Welberry}},\ }\href {https://books.google.es/books?id=TKJnEAAAQBAJ} {\emph
  {\bibinfo {title} {Diffuse X-Ray Scattering and Models of Disorder}}},\ IUCr
  monographs on crystallography\ (\bibinfo  {publisher} {Oxford University
  Press},\ \bibinfo {year} {2022})\BibitemShut {NoStop}%
\bibitem [{\citenamefont {Wu}\ \emph {et~al.}(2024{\natexlab{b}})\citenamefont
  {Wu}, \citenamefont {Mi}, \citenamefont {Zhang}, \citenamefont {Wang},
  \citenamefont {Maraytta}, \citenamefont {Zhou}, \citenamefont {He},
  \citenamefont {Merz}, \citenamefont {Chai},\ and\ \citenamefont
  {Wang}}]{wu2024annealing}%
  \BibitemOpen
  \bibfield  {author} {\bibinfo {author} {\bibfnamefont {X.}~\bibnamefont
  {Wu}}, \bibinfo {author} {\bibfnamefont {X.}~\bibnamefont {Mi}}, \bibinfo
  {author} {\bibfnamefont {L.}~\bibnamefont {Zhang}}, \bibinfo {author}
  {\bibfnamefont {C.-W.}\ \bibnamefont {Wang}}, \bibinfo {author}
  {\bibfnamefont {N.}~\bibnamefont {Maraytta}}, \bibinfo {author}
  {\bibfnamefont {X.}~\bibnamefont {Zhou}}, \bibinfo {author} {\bibfnamefont
  {M.}~\bibnamefont {He}}, \bibinfo {author} {\bibfnamefont {M.}~\bibnamefont
  {Merz}}, \bibinfo {author} {\bibfnamefont {Y.}~\bibnamefont {Chai}},\ and\
  \bibinfo {author} {\bibfnamefont {A.}~\bibnamefont {Wang}},\ }\href@noop {}
  {\bibinfo {title} {Annealing-tunable charge density wave in the kagome
  antiferromagnet fege}} (\bibinfo {year} {2024}{\natexlab{b}}),\ \Eprint
  {https://arxiv.org/abs/2308.01291} {arXiv:2308.01291} \BibitemShut {NoStop}%
\bibitem [{\citenamefont {Ricci}\ \emph {et~al.}(2021)\citenamefont {Ricci},
  \citenamefont {Poccia}, \citenamefont {Campi}, \citenamefont {Mishra},
  \citenamefont {M\"uller}, \citenamefont {Joseph}, \citenamefont {Shi},
  \citenamefont {Zozulya}, \citenamefont {Buchholz}, \citenamefont {Trabant},
  \citenamefont {Lee}, \citenamefont {Viefhaus}, \citenamefont {Goedkoop},
  \citenamefont {Nugroho}, \citenamefont {Braden}, \citenamefont {Roy},
  \citenamefont {Sprung},\ and\ \citenamefont
  {Sch\"u\ss{}ler-Langeheine}}]{Ricci_2021}%
  \BibitemOpen
  \bibfield  {author} {\bibinfo {author} {\bibfnamefont {A.}~\bibnamefont
  {Ricci}}, \bibinfo {author} {\bibfnamefont {N.}~\bibnamefont {Poccia}},
  \bibinfo {author} {\bibfnamefont {G.}~\bibnamefont {Campi}}, \bibinfo
  {author} {\bibfnamefont {S.}~\bibnamefont {Mishra}}, \bibinfo {author}
  {\bibfnamefont {L.}~\bibnamefont {M\"uller}}, \bibinfo {author}
  {\bibfnamefont {B.}~\bibnamefont {Joseph}}, \bibinfo {author} {\bibfnamefont
  {B.}~\bibnamefont {Shi}}, \bibinfo {author} {\bibfnamefont {A.}~\bibnamefont
  {Zozulya}}, \bibinfo {author} {\bibfnamefont {M.}~\bibnamefont {Buchholz}},
  \bibinfo {author} {\bibfnamefont {C.}~\bibnamefont {Trabant}}, \bibinfo
  {author} {\bibfnamefont {J.~C.~T.}\ \bibnamefont {Lee}}, \bibinfo {author}
  {\bibfnamefont {J.}~\bibnamefont {Viefhaus}}, \bibinfo {author}
  {\bibfnamefont {J.~B.}\ \bibnamefont {Goedkoop}}, \bibinfo {author}
  {\bibfnamefont {A.~A.}\ \bibnamefont {Nugroho}}, \bibinfo {author}
  {\bibfnamefont {M.}~\bibnamefont {Braden}}, \bibinfo {author} {\bibfnamefont
  {S.}~\bibnamefont {Roy}}, \bibinfo {author} {\bibfnamefont {M.}~\bibnamefont
  {Sprung}},\ and\ \bibinfo {author} {\bibfnamefont {C.}~\bibnamefont
  {Sch\"u\ss{}ler-Langeheine}},\ }\bibfield  {title} {\bibinfo {title}
  {Measurement of spin dynamics in a layered nickelate using x-ray photon
  correlation spectroscopy: Evidence for intrinsic destabilization of
  incommensurate stripes at low temperatures},\ }\href
  {https://doi.org/10.1103/PhysRevLett.127.057001} {\bibfield  {journal}
  {\bibinfo  {journal} {Phys. Rev. Lett.}\ }\textbf {\bibinfo {volume} {127}},\
  \bibinfo {pages} {057001} (\bibinfo {year} {2021})}\BibitemShut {NoStop}%
\end{thebibliography}%

\section{Acknowledgments}
We acknowledge Ming Yi, Yu He, E. da Silva-Neto, A. Subedi, D. Efremov, C. Sch\"{u}ssler-Langeheine, A. Frano, Stephen Wilson, Philippe Bourges, Hu Miao, Riccardo Comin and Peter Keim for fruitful discussions and critical reading of the manuscript. D.S., A.Kar and S.B-C. acknowledge financial support from the MINECO of Spain through the project PID2021-122609NB-C21 and by MCIN and by the European Union Next Generation EU/PRTR-C17.I1, as well as by IKUR Strategy under the collaboration agreement between Ikerbasque Foundation and DIPC on behalf of the Department of Education of the Basque Government. A.K. thanks the Basque government for financial support through the project PIBA-2023-1-0051.   
Y.J. and H.H. were supported by the European Research Council (ERC) under the European Union’s Horizon 2020 research and innovation program (Grant Agreement No. 101020833) as well as by IKUR Strategy. 
D.C\u{a}l. acknowledges the hospitality of the Donostia International Physics Center, at which this work was carried out. 
D.C\u{a}l. was supported by the European Research Council (ERC) under the European Union’s Horizon 2020 research and innovation program (grant agreement no. 101020833) and by the Simons Investigator Grant No. 404513. B.A.B was supported by the Gordon and Betty Moore Foundation through Grant No.GBMF8685 towards the Princeton theory program, the Gordon and Betty Moore Foundation’s EPiQS Initiative (Grant No. GBMF11070), Office of Naval Research (ONR Grant No. N00014-20-1-2303), Global Collaborative Network Grant at Princeton University, BSF Israel US foundation No. 2018226, NSF-MERSEC (Grant No. MERSEC DMR 2011750), the Simons theory collaboration on frontiers of superconductivity, and 
Simons collaboration on mathematical sciences.
This work has been partly performed in the framework of the nanoscience foundry and fine analysis (NFFA-MUR Italy Progetti Internazionali) facility. This research used resources of the Advanced Photon Source, a U.S. Department of Energy (DOE) Office of Science User Facility operated for the DOE Office of Science by Argonne National Laboratory under Contract No. DE-AC02-06CH11357.

\section{Author Contribution}
S.B-C conceived and managed the project. C.Y, S.R, C.S and C.F synthesized the single crystals. D.S, A.Kar, A.J, I.V, J.D, M.T and S.B-C conducted the ARPES experiments. D.S and A.Kar analyzed the ARPES data. A.K and A.B measured the DS and C.Fuller performed the Monte Carlo and the DS simulations. J.S and S.B-C carried out the resonant x-ray scattering experiments. C.M, D.C, D.S, A. Kar and S.B-C took the hard x-ray diffraction and D.C. refined and solved the structure. H.H, Y.J, D.C. and B.A.B carried out the DFT calculations. D.S and S.B-C simulated the diffraction patterns and the correlation analysis. S.B-C wrote the manuscript with input from all coauthors.

\section{Competing Interests}
The authors declare no competing interests.

\clearpage
\onecolumngrid
\begin{center}
\textbf{Supplementary Information for 'Frustrated charge density wave and quasi-long-range bond-orientational order in the magnetic kagome FeGe'}
\end{center}

\renewcommand{\thefigure}{S\arabic{figure}}

\renewcommand{\thetable}{S\arabic{table}}

\renewcommand{\thesection}{S\arabic{section}}

\renewcommand{\theequation}{S\arabic{equation}}

\tableofcontents
\clearpage

\section{Sample growth and characterization}

\begin{figure*}[h]
    \centering
    \includegraphics[width=1.0\linewidth]{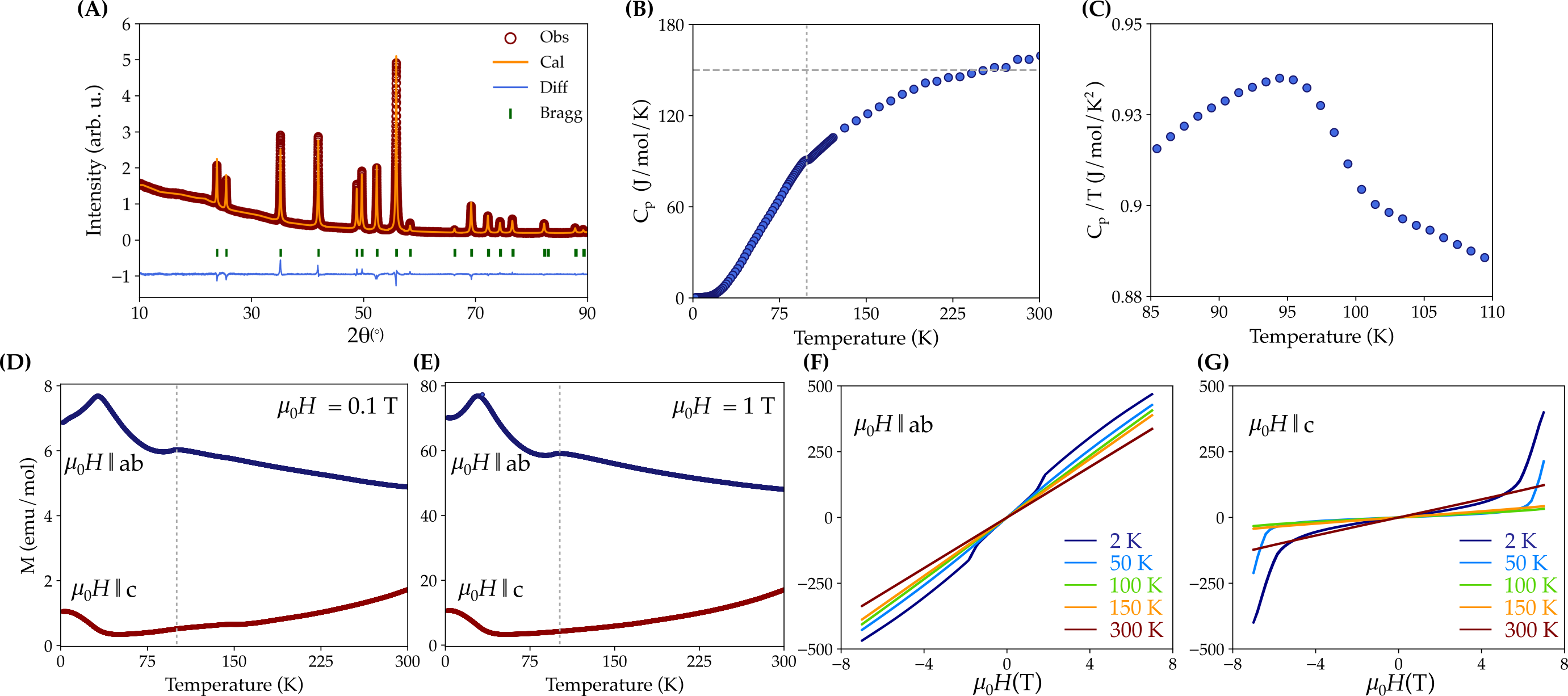}
    \caption{Characterization of FeGe. (A) Powder x-ray diffraction refinement and its  LeBail refinement. (B) Temperature dependence of the heat capacity. (C) Heat capacity divided by temperature, highlighting the CDW phase transition. (D-E) Temperature dependence of the magnetization for applied magnetic fields of 0.1 $\mathrm{T}$ and 1 $\mathrm{T}$, respectively. (F-G) Magnetization versus magnetic field parallel (F) and perpendicular (G) to the kagome plane.}
    \label{Fig:SI_characterization}
\end{figure*}

Single crystals of FeGe were grown by the chemical vapor transport method (see Methods) and characterized by means of energy dispersion x-ray (EDX) analysis, Laue diffraction, resistivity and magnetization (figure~\ref{Fig:SI_characterization}).

Figure \ref{Fig:SI_characterization} (A) shows the LeBail refinement of grounded FeGe single crystals. No impurity phases were observed, specially the presence of the cubic B20 phase. Both the specific heat (C$_p$), figure \ref{Fig:SI_characterization} (B-C) and the magnetization, figure \ref{Fig:SI_characterization} (D-E), identify the CDW transition at $\sim$ 105K. The transport properties are in nice agreement with the reports in the literature \cite{Teng_2022,Teng_2023}.

\section{x-ray diffraction}

This section contains the single crystal structural refinement of FeGe at 80 K. Both \textit{P6mm} (non-centrosymmetric), Table \ref{P6mm_sym}, and \textit{P6mm}
(centrosymmetric) Table \ref{P6mmm_sym} could be equally indexed with similar \textit{Goodness-of-Fit}. The x-ray refinements were carried out with SHELXL2018/1 code (see Methods).

\begin{figure*}[h]
    \centering
    \includegraphics[width=0.85\linewidth]{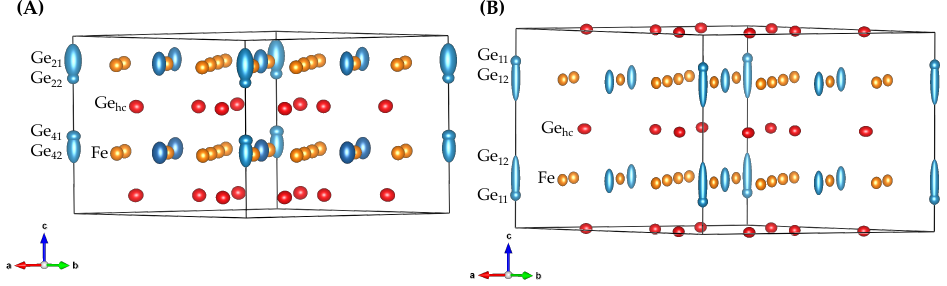}
    \caption{(A) CDW unit cell of the space group 183 and (B) space group 191.
    }
    \label{S2_space-groups}
\end{figure*}

 Figure \ref{S2_space-groups} displays the low temperature CDW unit cell within the space group 183 (non-centrosymmetric \textit{P6mm}) and 191 (centrosymmetric \textit{P6mmm}). The dimerized  trigonal Ge in the space group 183 creates 4 in-equivalent trigonal Ge in the kagome plane (Ge$_{21}$, Ge$_{22}$, Ge$_{41}$ and Ge$_{42}$), while the dimerization on the space group 191 splits the trigonal Ge into 2 Ge$_{11}$ and Ge$_{21}$. 
 
\begin{table*}[htbp]
\global\long\def\arraystretch{1.12}
\caption{Crystal data and structure refinement for 80 K: \textit{P6mm} symmetry, space group nº= 183. Unit cell dimensions \textit{a} =
\textit{b} = 9.97740(2) \r{A},  \textit{c} = 8.10070(10) \r{A}, $\alpha$=$\beta$= 90º, $\gamma$ = 120º. Volume =698.3738(20) \r{A}$^3$. $R_{1}$=0.056 }
\label{P6mm_sym}
\setlength{\tabcolsep}{0.6mm}{
\begin{tabular}{c|c|c|c|c|c|c|c|c}
\hline
\hline

  &    &     \textit{x}     &     \textit{y}    &      \textit{z}     &     Occ.  &   U &   Site  & Sym. \\
  \hline
  
    Ge & Ge$_1$     &    0.33333  &  0.66667  &  0.60251  &  1.000  &  0.012  &  2b    &   3m. \\
    Ge & Ge$_{22}$    &    1.00000  &  1.00000  &  0.76210  &  0.309  &  0.009  &  1a    &   6mm\\
    Ge & Ge$_{21}$  &      1.00000  &  1.00000  &  0.86285  &  0.691  &  0.036  &  1a    &   6mm\\
    Ge & Ge$_3$    &    0.67051  &  0.83526  &  0.10257  &  1.000  &  0.013  &  6e    &   .m.\\
    Ge & Ge$_{41}$    &    1.00000  &  1.00000  &  0.43604  &  0.378  &  0.011  &  1a    &   6mm\\
    Ge & Ge$_{42}$    &    1.00000  &  1.00000  &  0.36518  &  0.622  &  0.025  &  1a    &   6mm\\
    Ge & Ge$_5$     &    0.66412  &  0.83206  &  0.60263  &  1.000  &  0.013  &  6e    &   .m.\\
    Ge & Ge$_6$     &    0.33333  &  0.66667  &  0.10276  &  1.000  &  0.012  &  2b    &   3m.\\
    Ge & Ge$_7$     &    0.50000  &  1.00000  &  0.85698  &  1.000  &  0.020  &  3c    &   2mm\\
    Ge & Ge$_8$     &    0.50000  &  1.00000  &  0.35434  &  1.000  &  0.020  & 3c     &  2mm\\
   Fe & Fe$_1$      &   0.74954   & 1.00000   & 0.34468   & 1.000   & 0.012  &  6d     &  ..m\\
   Fe & Fe$_2$      &   0.50020   & 0.75010   & 0.84475   & 1.000   & 0.012  &  6e     &  .m.\\
   Fe & Fe$_3$      &   0.49999   & 0.75000   & 0.34241   & 1.000   & 0.012  &  6e     &  .m.\\
   Fe & Fe$_4$      &   0.74963   & 1.00000   & 0.84234   & 1.000   & 0.012  &  6d     &  ..m\\
\hline
\hline
\end{tabular}}
\end{table*}

\begin{table*}[htbp]
\global\long\def\arraystretch{1.12}
\caption{Crystal data and structure refinement for 80 K: \textit{P6/mmm} symmetry, space group nº= 191. Unit cell dimensions	\textit{a} = 9.97750(10)=\textit{b} = 9.97750(10) \r{A}, \textit{c} = 8.10030(10) \r{A}, $\alpha$=$\beta$= 90º, $\gamma$ = 120º. Volume =698.353(16) \r{A}$^3$, $R_{1}$=0.056}
\label{P6mmm_sym}
\setlength{\tabcolsep}{0.6mm}{
\begin{tabular}{c|c|c|c|c|c|c|c|c}
\hline
\hline

  &    &     \textit{x}     &     \textit{y}    &      \textit{z}     &     Occ.  &   U &   Site  & Sym. \\
  \hline
    Fe & Fe$_1$      &   0.25053  &  0.00000  &  0.75162  &  1.000  &  0.004  & 12n  &     ..m \\
    Fe & Fe$_2$      &   0.24996  &  0.49991  &  0.74843  &  1.000  &  0.004  & 12o  &     .m.\\
    Ge & Ge$_{11}$     &   0.00000  &  0.00000  &  0.83590  &  0.485  &  0.004  &  2e  &     6mm\\
    Ge & Ge$_{12}$     &   0.00000  &  0.00000  &  0.75170  &  0.515  &  0.028  &  2e  &     6mm\\
    Ge & Ge$_2$      &   0.50000  &  0.00000  &  0.74842  &  1.000  &  0.009  &  6i  &     2mm\\
    Ge & Ge$_3$      &   0.16844  &  0.33687  &  0.00000  &  1.000  &  0.005  &  6l  &    mm2\\
    Ge & Ge$_4$      &   0.83593  &  0.67186  &  0.50000  &  1.000  &  0.005  &  6m  &     mm2\\
    Ge & Ge$_5$      &   0.33333  &  0.66667  &  0.00000  &  1.000  &  0.005  &  2c  &    -6m2\\
    Ge & Ge$_6$      &   0.66667  &  0.33333  &  0.50000  &  1.000  &  0.005  &  2d  &   -6m2\\

\hline
\hline
\end{tabular}}
\end{table*}

\clearpage

\section{Angle Resolved Photoemission: ARPES}

We observe the following key features in the experimental band structure from ARPES:
\begin{enumerate}
\item A clear Dirac crossing appears at $K$ at about -0.65 eV. 
\item A V-shaped band centered at $\Gamma$ with the bottom at about -1.0 eV.  
\item A very broad spectrum weight connects $\Gamma$ and the Dirac crossing at $K$, which could be the weights from the Dirac bands and together with other bands. 
\item A U-shaped band centered at $\Gamma$ with the bottom at about -0.28 eV.   
\end{enumerate}

\begin{figure}[htbp]
    \centering
    \includegraphics[width=\textwidth]{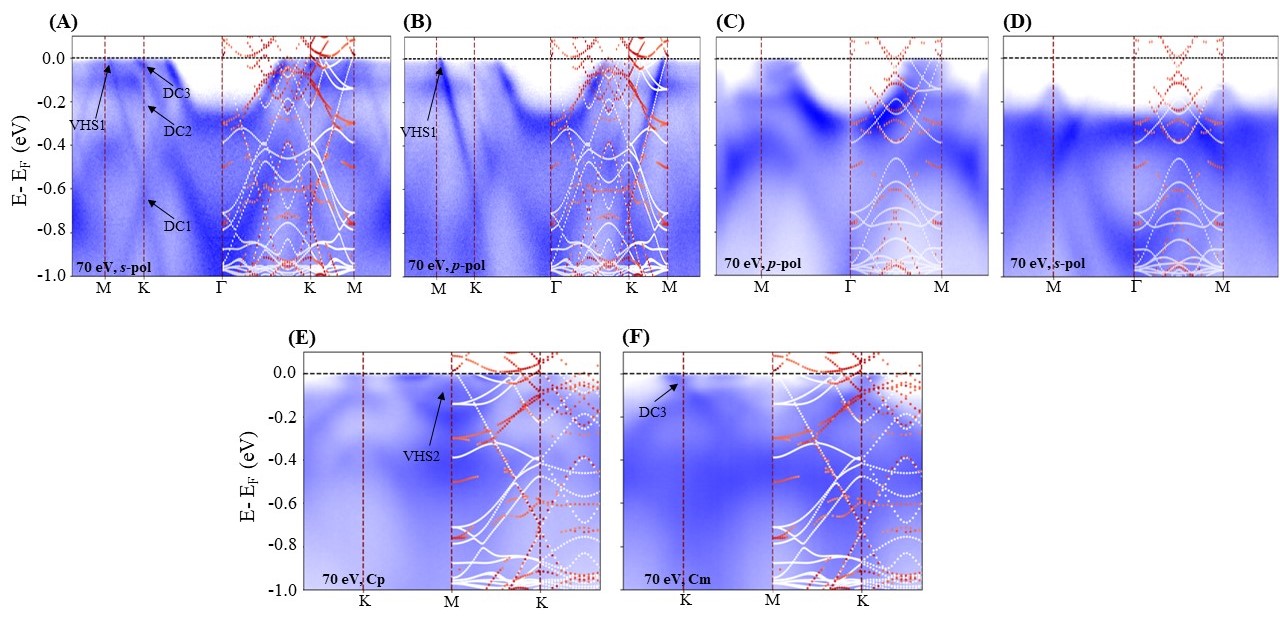}
    \caption{ A comparison between ARPES and DFT calculated energy dispersion spectra. Valance band electronic structure of FeGe obtained with 70 eV incident photon energy along (A)-(B) $\mathrm{M-K}-\Gamma-\mathrm{K-M}$ symmetry direction with $s$-pol and $p$-polarized incident light, respectively. (C)-(D) along $\mathrm{M}-\Gamma-\mathrm{M}$ symmetry direction with $p$-pol and $s$-polarized light, respectively. (E)-(F) Along $\mathrm{K-M-K}$ symmetry direction with circular positive (C$_p$) and circular negative (C$_m$) incident light, respectively. (`DC' and `VHS' represent the Dirac cone, and Van Hove singularity, respectively). On top of the VB spectra, DFT calculated bulk folded band structure in the CDW phase (white) and Ge terminated surface band structure (red) are overlapped.}
    \label{ARPES_fig1}
\end{figure}

The first two features in APRES, i.e., the Dirac crossing at $K$ and the $V$-shaped band at $\Gamma$ could be well-matched using the CDW bulk band structure in DFT, as shown in figure \ref{ARPES_fig1}. The broad spectrum weight connects $\Gamma$ and at $K$ is also seen in DFT, although the energy in DFT is slightly higher. However, the U-shape is not seen in the bulk bands. We argue that it can be matched using the surface bands in the honeycomb Ge termination, as shown in \ref{ARPES_fig1}. 

Besides the aforementioned features that have a good matching between DFT bands and ARPES, we also observe some features that cannot be well matched. They include: (i) In DFT, there exists another Dirac-like crossing at $K$ at about -0.4 eV, which is not seen in the APRES. (ii) The broad weight at -1 eV near $\Gamma$ in DFT is also not seen in the ARPES. (iii) In the folded surface bands, there exist many other bands that could not have a good match with the APRES, probably due to the inaccurate description of surface instructions in DFT. 

We conclude that the main features in the experimental ARPES bands could be well-matched by DFT. However, since the system has a large number of bands near the Fermi level, it shows a heavily broadened spectrum where the discrepancy in the detailed features between theory and experiments is expected. 

\begin{figure}[htbp]
    \centering
    \includegraphics[width=\textwidth]{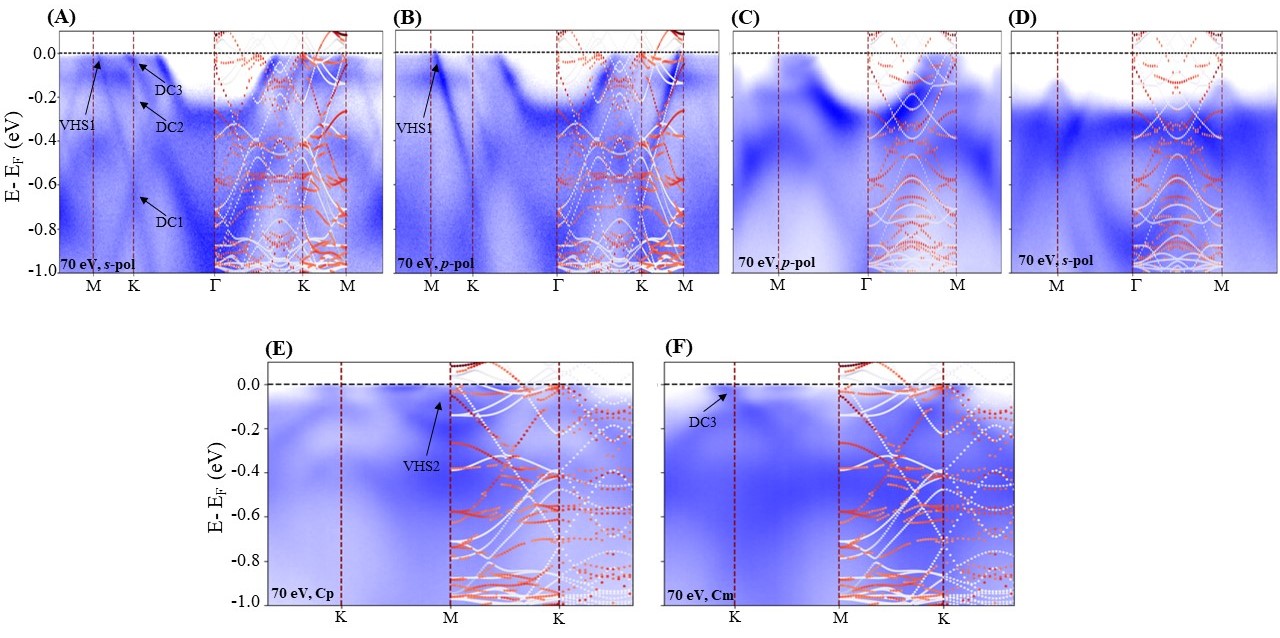}
    \caption{A comparison between ARPES and DFT calculated spectra. The valance band specters of FeGe were obtained with 70 eV incident photon energy along (A)-(B) $\mathrm{M-K}-\Gamma-\mathrm{K-M}$ symmetry direction with $s$-pol and $p$-polarized incident light, respectively. (C)-(D) along $\mathrm{M}-\Gamma-\mathrm{M}$ symmetry direction with $p$-pol and $s$-polarized light, respectively. (E)-(F) Along $\mathrm{K-M-K}$ symmetry direction with circular positive (C$_p$) and circular negative (C$_m$) incident light, respectively. (`DC' and `VHS' represent the Dirac cone, and Van Hove singularity, respectively).  Here, the `white' bands are folded bulk bands of FeGe calculated in CDW phase where as the `red' ones are the unfolded surface bands obtained with Kagome termination.}
    \label{ARPES_fig2}
\end{figure}

\clearpage

\section{First-principle calculation methods}

The first-principle calculations in this work use the Vienna ab-initio Simulation Package (VASP)\cite{kresse1996efficiency, kresse1993ab1, kresse1993ab2, kresse1994ab, kresse1996efficient} with generalized gradient approximation of Perdew-Burke-Ernzerhof (PBE) exchange-correlation potential\cite{perdew1996generalized}. A $8\times8\times8$ ($5\times5\times5$) $k$-mesh for non-CDW (CDW) phase and an energy cutoff of 500 eV are used. 
The maximally localized Wannier functions are obtained using WANNIER90\cite{marzari1997maximally, souza2001maximally, marzari2012maximally, pizzi2020wannier90}. A local coordinate system at the kagome site is adopted in order to decompose $d$ orbitals when construct MLWFs, the same as the one used in Ref.\cite{jiang2023kagome}.
The Wannier tight-binding models are symmetrized using \textit{Wannhr\_symm} in \textit{WannierTools}\cite{wu2018wanniertools}. The unfolding of CDW bands is performed using \textit{VaspBandUnfolding} package\cite{zheng2018vaspbandunfolding, popescu2012extracting}. The Fermi surface is computed using \textit{WannierTools}\cite{wu2018wanniertools} and visualized using \textit{Fermisurfer}\cite{kawamura2019fermisurfer}.

\section{Bulk band structure in non-CDW and CDW phases}

In this section, we discuss the bulk band structure in both non-CDW and CDW phases. 

\begin{figure}[htbp]
    \centering
    \includegraphics[width=\textwidth]{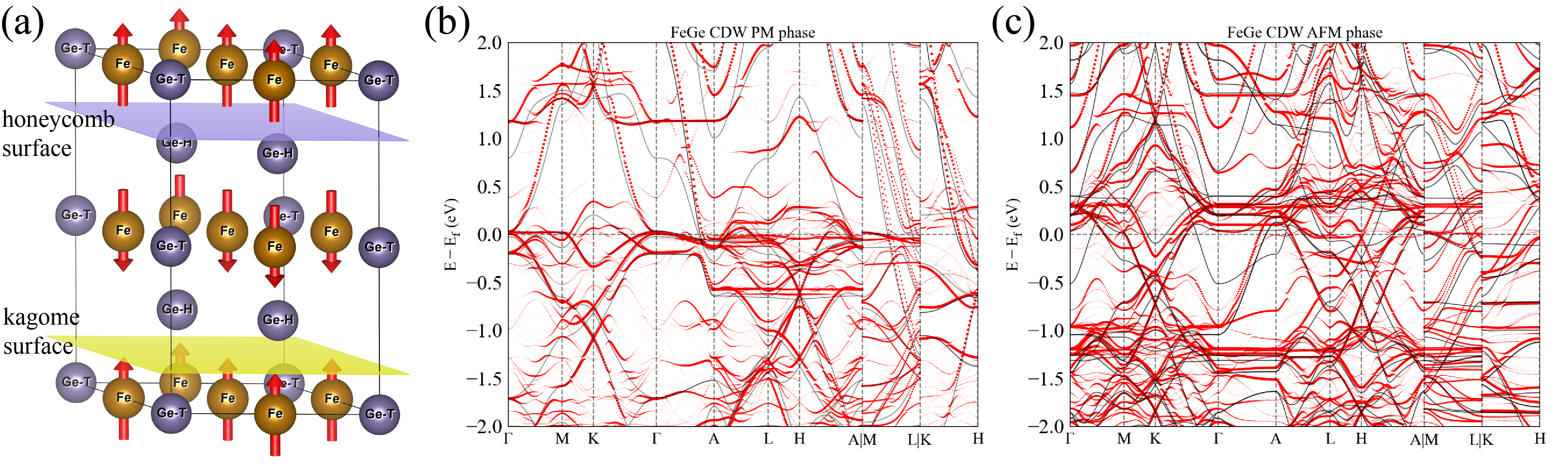}
    \caption{(a) The crystal structure of FeGe in the non-CDW phase, where Fe atoms form kagome lattices, while Ge atoms form triangular and honeycomb lattices (denoted as Ge-T and Ge-H, respectively). Two surface terminations, i.e., kagome and honeycomb surfaces are also marked, where the atoms below the plane define the surface. 
    (b) The band structures in non-CDW (black lines) and CDW (red lines) phases, where paramagnetic (PM) order is assumed. In the CDW phase, the bands are unfolded to the non-CDW Brillouin zone (BZ). 
    (c) Same as (b), but in the anti-ferromagnetic (AFM) phase. Spin-orbital coupling (SOC) is neglected for simplicity as SOC is weak in FeGe.  
    }
    \label{fig: structure-bands}
\end{figure}

In \cref{fig: structure-bands}(a), the crystal structure of FeGe in the non-CDW phase is shown together with two surface terminations, i.e., kagome and honeycomb surfaces. In the $2\times 2\times 1$ CDW phase, the main atomic displacements come from the dimerization of the triangular Ge. Experimentally, the dimerized Ge atoms have two possible positions as shown in the main text. In DFT, we fix the Ge atoms at the larger dimerized positions.

In \cref{fig: structure-bands}(b)(c), we show the comparison between the non-CDW and (unfolded) CDW bands in the PM and AFM phases. In \cref{fig: nonCDW-AFM-orbit-weight-full} and \cref{fig: nonCDW-AFM-orbit-weight}, the orbital weights in the non-CDW AFM phase are shown, while in \cref{fig: CDW-AFM-orbital-weight}, the orbital weights in the CDW phase are shown. We observe the following features in the CDW bulk bands:
\begin{itemize}
\item A Dirac crossing at $K$ centered at about -0.7 eV mainly comes from $d_{xy}$ orbitals, together with some $d_{x^2-y^2}$ weights. It exists in both CDW and non-CDW bands and has little changes, as shown in \cref{fig: nonCDW-AFM-orbit-weight}(a)(b) and \cref{fig: CDW-AFM-orbital-weight}(b).  

\item A V-shaped band centered at $\Gamma$ from about -0.5 to -1.0 eV mainly comes from the triangular Ge $p_z$ orbital, as shown in \cref{fig: CDW-AFM-orbital-weight}(d). This band accounts for the main reconstruction by CDW. In the non-CDW phase, it is located at high energy with the bottom at about -0.5 eV, as shown in \cref{fig: nonCDW-AFM-orbit-weight}(f). 

\item A quasi-flat bands at about -1 eV near $\Gamma$ mainly comes from $d_{z^2}$ orbital, as shown in \cref{fig: nonCDW-AFM-orbit-weight}(c) and \cref{fig: CDW-AFM-orbital-weight}(a).
\end{itemize}

In \cref{fig: compare-cdw-ncdw}, we also superimpose the CDW bulk bands with non-CDW bands for better comparison.

\begin{figure}[htbp]
    \centering
    \includegraphics[width=0.8\textwidth]{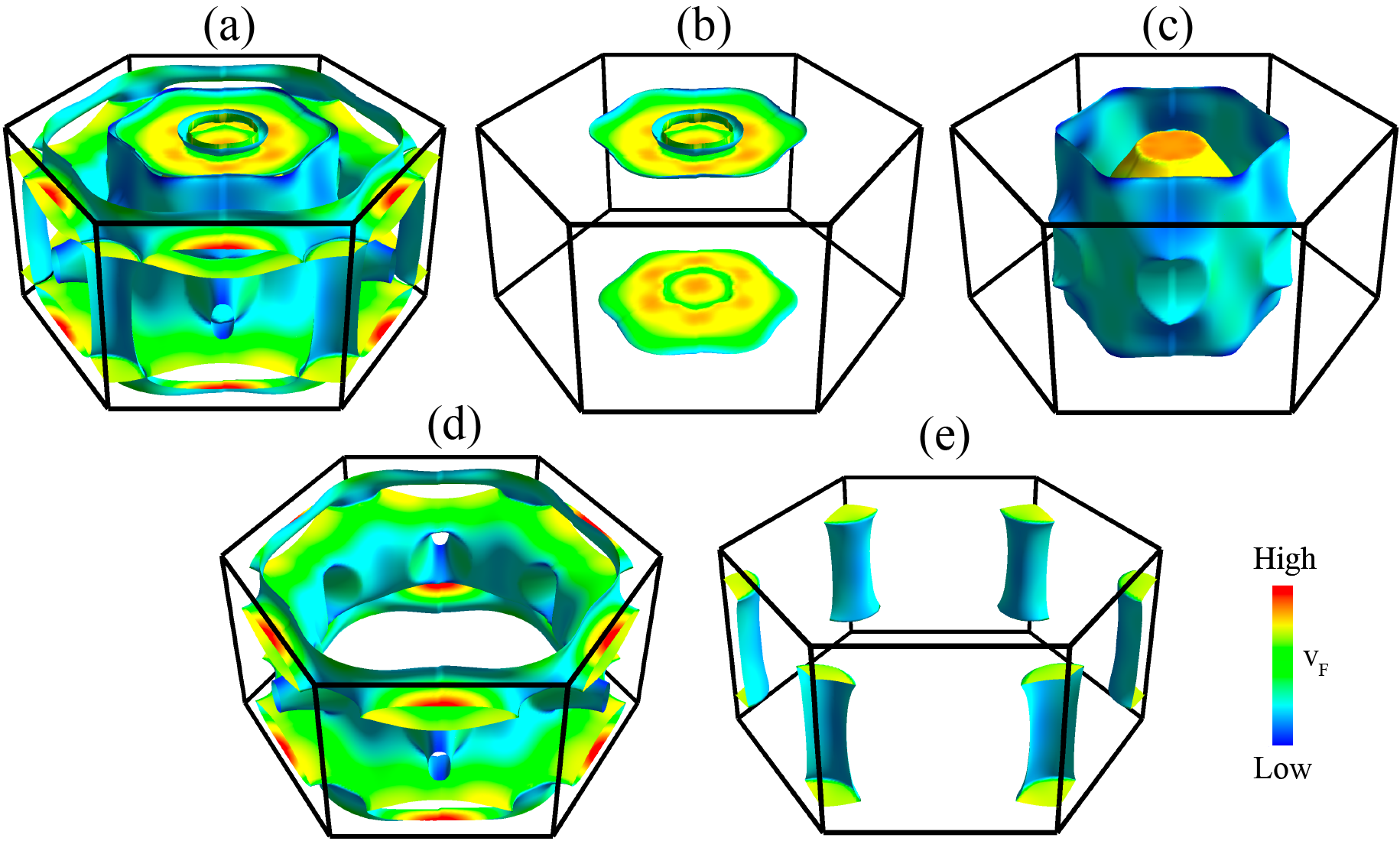}
    \caption{The Fermi surface (FS) of FeGe in the non-CDW AFM phase, where (a) is the full FS, and (b)-(e) are four parts of FSs contributed by different bands. The color on the FS denotes the Fermi velocity. A quasi-2D FS is shown in (c) with a weak $k_z$-dispersion. }
    \label{fig: FeGe-nonCDW-AFM-FS3D}
\end{figure}

\begin{figure}[htbp]
    \centering
    \includegraphics[width=0.8\textwidth]{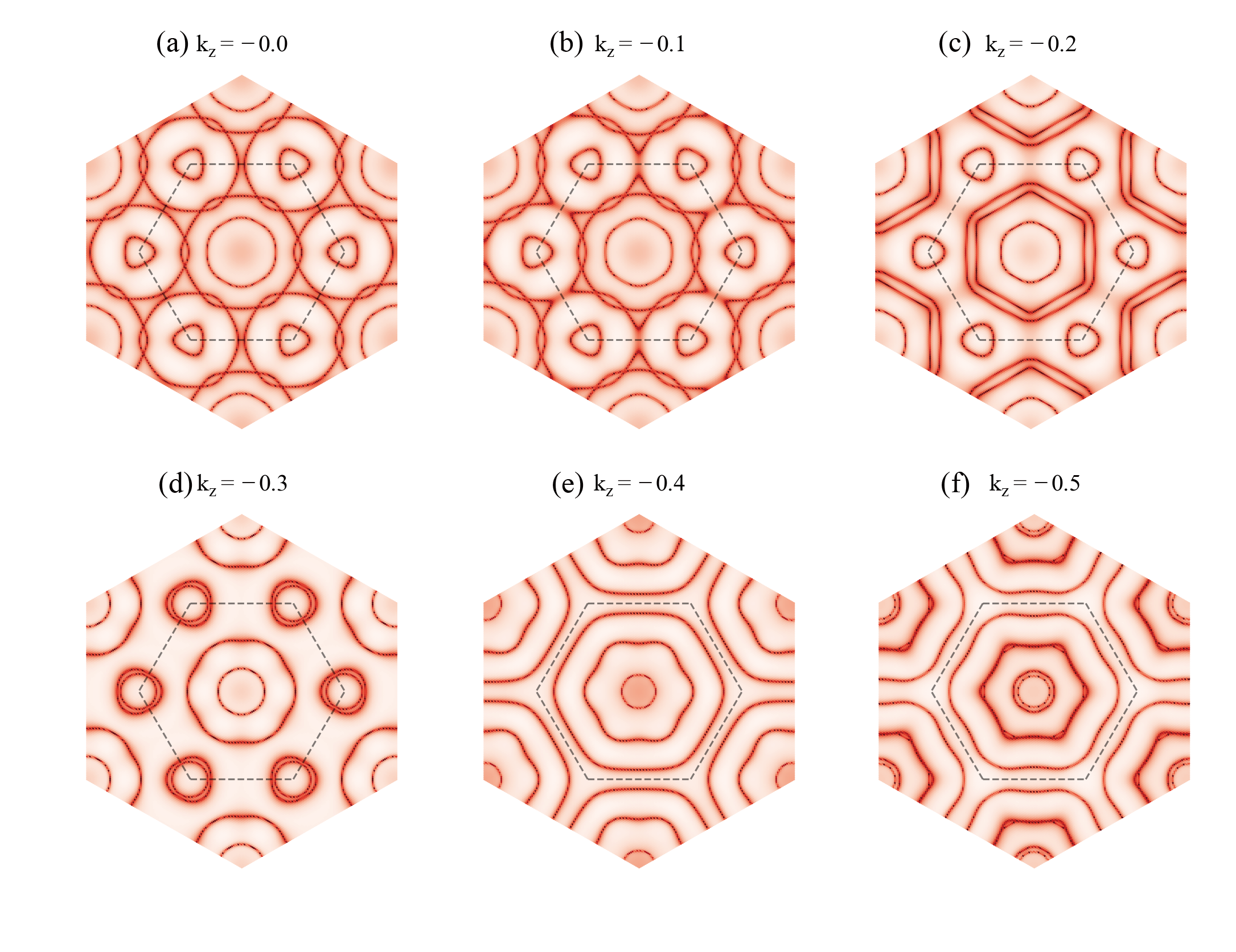}
    \caption{The 2D Fermi surface (FS) of FeGe in the non-CDW AFM phase on different $k_z$ planes. }
    \label{fig: FeGe-nonCDW-AFM-FS2D}
\end{figure}

\begin{figure}[htbp]
    \centering
    \includegraphics[width=\textwidth]{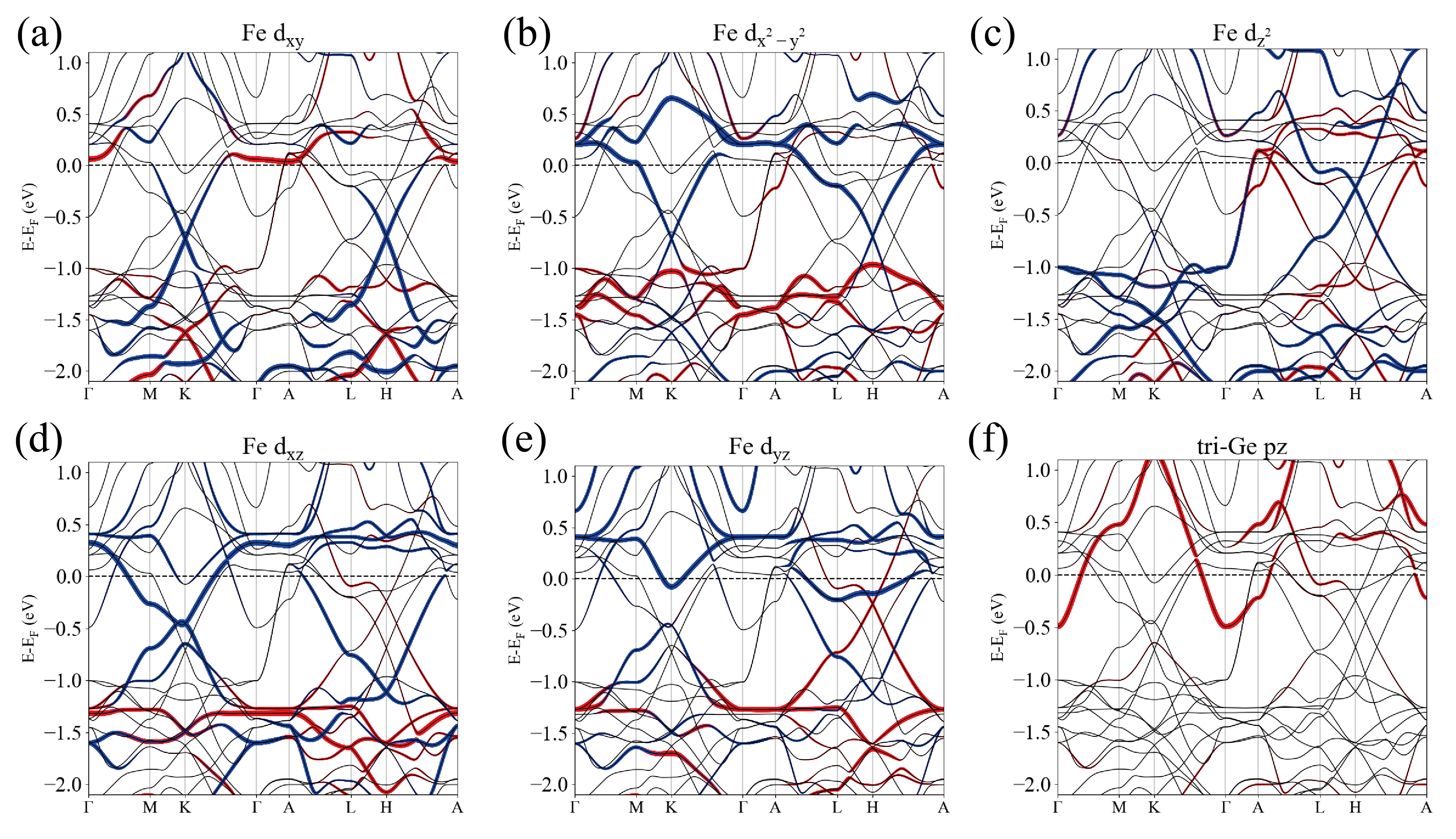}
    \caption{The orbital projections in the non-CDW AFM phase, for Fe (a) $d_{xy}$, (b) $d_{x^2-y^2}$, (c) $d_{z^2}$, (d) $d_{xz}$, (e) $d_{yz}$, and triangular Ge $p_z$ orbitals, respectively. Blue and red lines denote two spin-up and down bands from Fe one kagome layer. }
    \label{fig: nonCDW-AFM-orbit-weight-full}
\end{figure}

\begin{figure}[htbp]
    \centering
    \includegraphics[width=\textwidth]{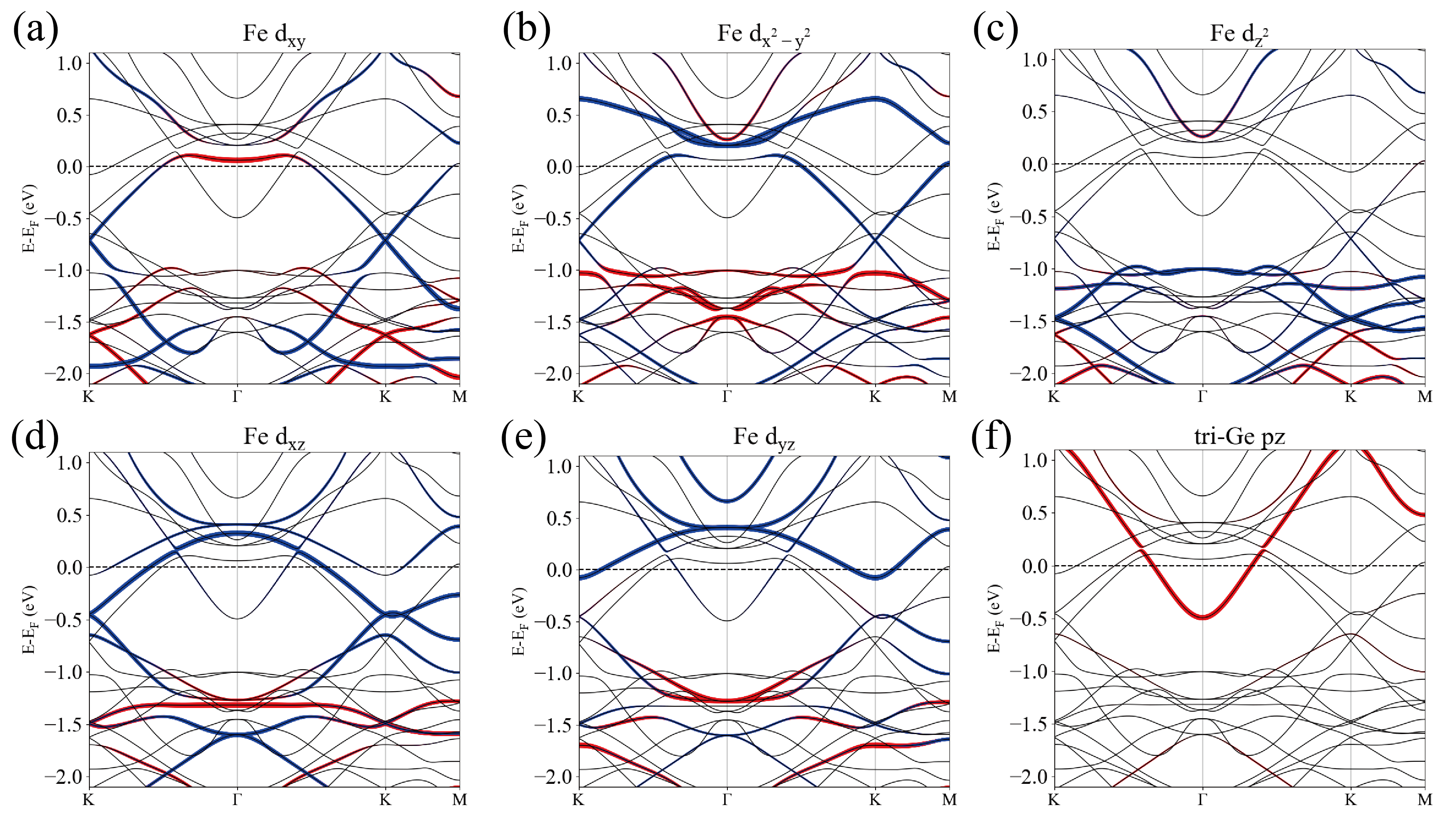}
    \caption{The orbital projections in the non-CDW AFM phase. It is the same as \cref{fig: nonCDW-AFM-orbit-weight-full} but on a different path in the BZ, in order to give a direct comparison with ARPES results.}
    \label{fig: nonCDW-AFM-orbit-weight}
\end{figure}

\begin{figure}[htbp]
    \centering
    \includegraphics[width=\textwidth]{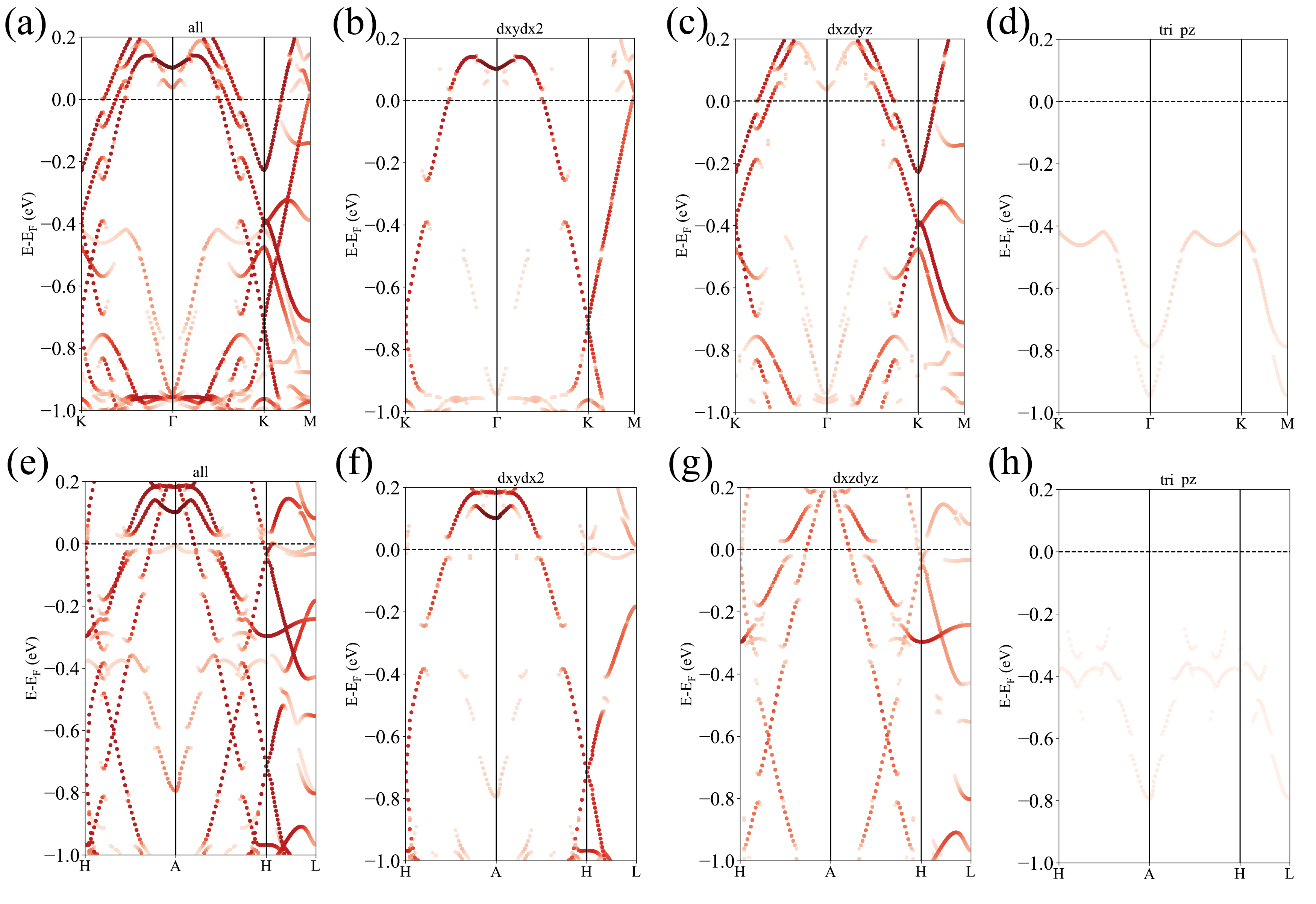}
    \caption{The unfolded bulk band structures and orbital projections in the CDW AFM phase. (a) The unfolded bulk bands. (b)-(d) The orbital weights for Fe $(d_{xy}, d_{x^2-y^2})$, $(d_{xz}, d_{yz})$, and triangular Ge $p_z$ orbitals, respectively. (e)-(h) in the second row is the same as the first row but on $k_z=\pi$ plane.}
    \label{fig: CDW-AFM-orbital-weight}
\end{figure}

\begin{figure}[htbp]
    \centering
    \includegraphics[width=\textwidth]{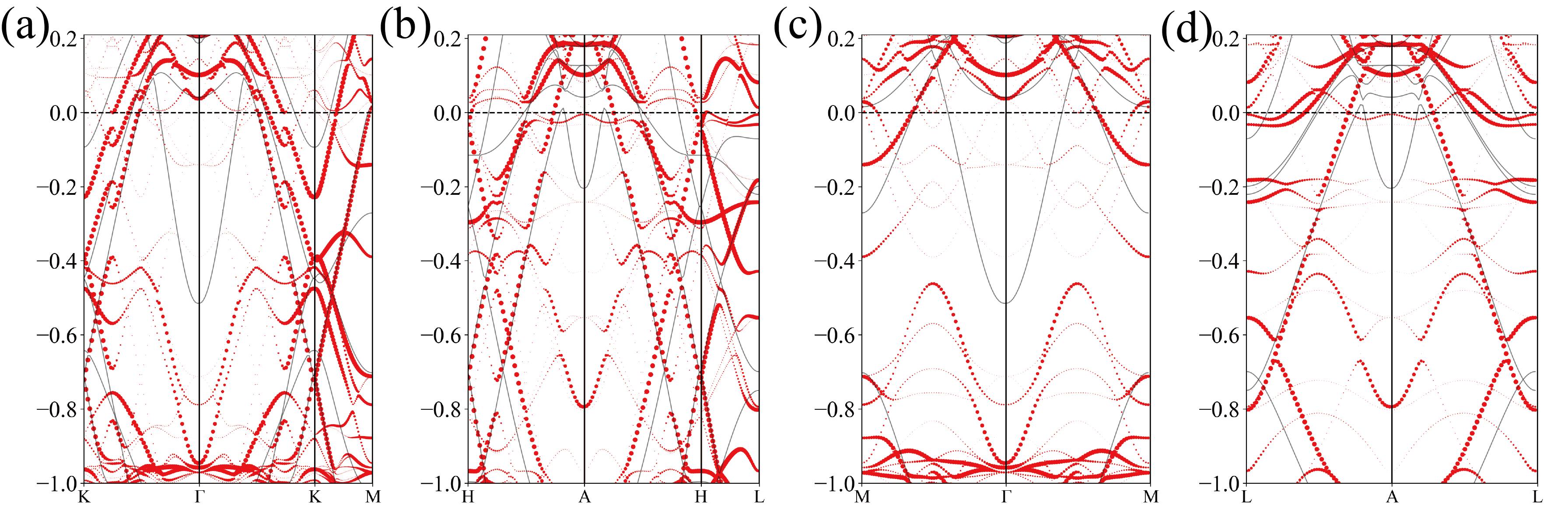}
    \caption{Comparsion of non-CDW (black) and unfolded CDW (red) bulk bands in the AFM phase, along four different paths in the BZ. The main difference between CDW and non-CDW bands is the V-shaped bands at $\Gamma$, which is mainly from triangular Ge $p_z$ orbitals (see \cref{fig: nonCDW-AFM-orbit-weight} and \cref{fig: CDW-AFM-orbital-weight} for orbital weights). This band moves down in the CDW phase.}
    \label{fig: compare-cdw-ncdw}
\end{figure}

In \cref{fig: FeGe-nonCDW-AFM-FS3D}, we show the Fermi surface (FS) of FeGe in the non-CDW AFM phase. It can be seen that there is quasi-2D FS with a weak $k_z$-dispersion in \cref{fig: FeGe-nonCDW-AFM-FS3D}(c), which is mainly contributed by the $d_{x^2-y^2}, d_{xy}$ orbitals of Fe (see \cref{fig: nonCDW-AFM-orbit-weight-full}). In \cref{fig: FeGe-nonCDW-AFM-FS2D}, we show the 2D slices of the FS on difference $k_z$ planes. On the $k_z=0$ plane, the smallest circular FS around $\Gamma$ is mainly given by the $p_z$ orbitals of the triangular Ge (see \cref{fig: nonCDW-AFM-orbit-weight-full}(f)). In the CDW phase, this band moves down and is far from $E_f$ (see \cref{fig: CDW-AFM-orbital-weight}(d)). However, there are some other bands close to $E_f$ near $\Gamma$ in the CDW phase, which could contribute to the FSs (see \cref{fig: compare-cdw-ncdw}).

\begin{figure}[htbp]
    \centering
    \includegraphics[width=\textwidth]{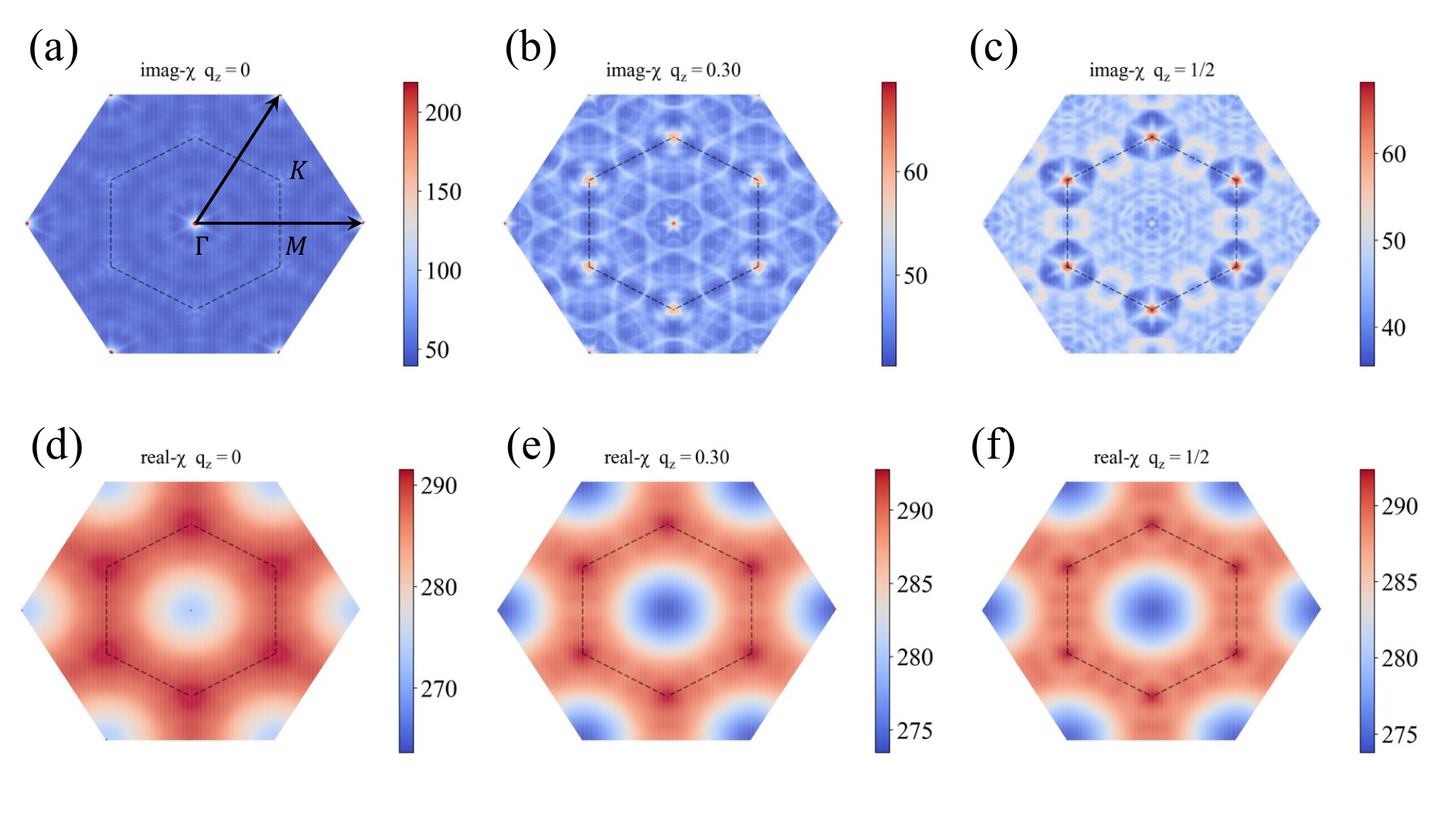}
    \caption{The nesting function (Im-$\chi$) and total susceptibility (Re-$\chi$) of FeGe in the AFM phase (non-CDW).  In the nesting function, the dominant peak appears at the $K$ point. In the total susceptibility, however, a broad peak appears along the boundary of the first BZ, with the highest point near $K$. }
    \label{fig:nesting-func}
\end{figure}

In \cref{fig:nesting-func}, we show the nesting function (Im-$\chi$) and total susceptibility (Re-$\chi$) of FeGe in the AFM phase (non-CDW). In the nesting function, the dominant peak appears at the $K$ point. In the total susceptibility, however, a broad peak appears along the boundary of the first BZ, with the highest point near $K$. Thus we conclude that the FS nesting cannot directly account for the CDW at $M$ point.

\section{Surface band structure in the CDW phase}

\begin{figure}[htbp]
    \centering
    \includegraphics[width=\textwidth]{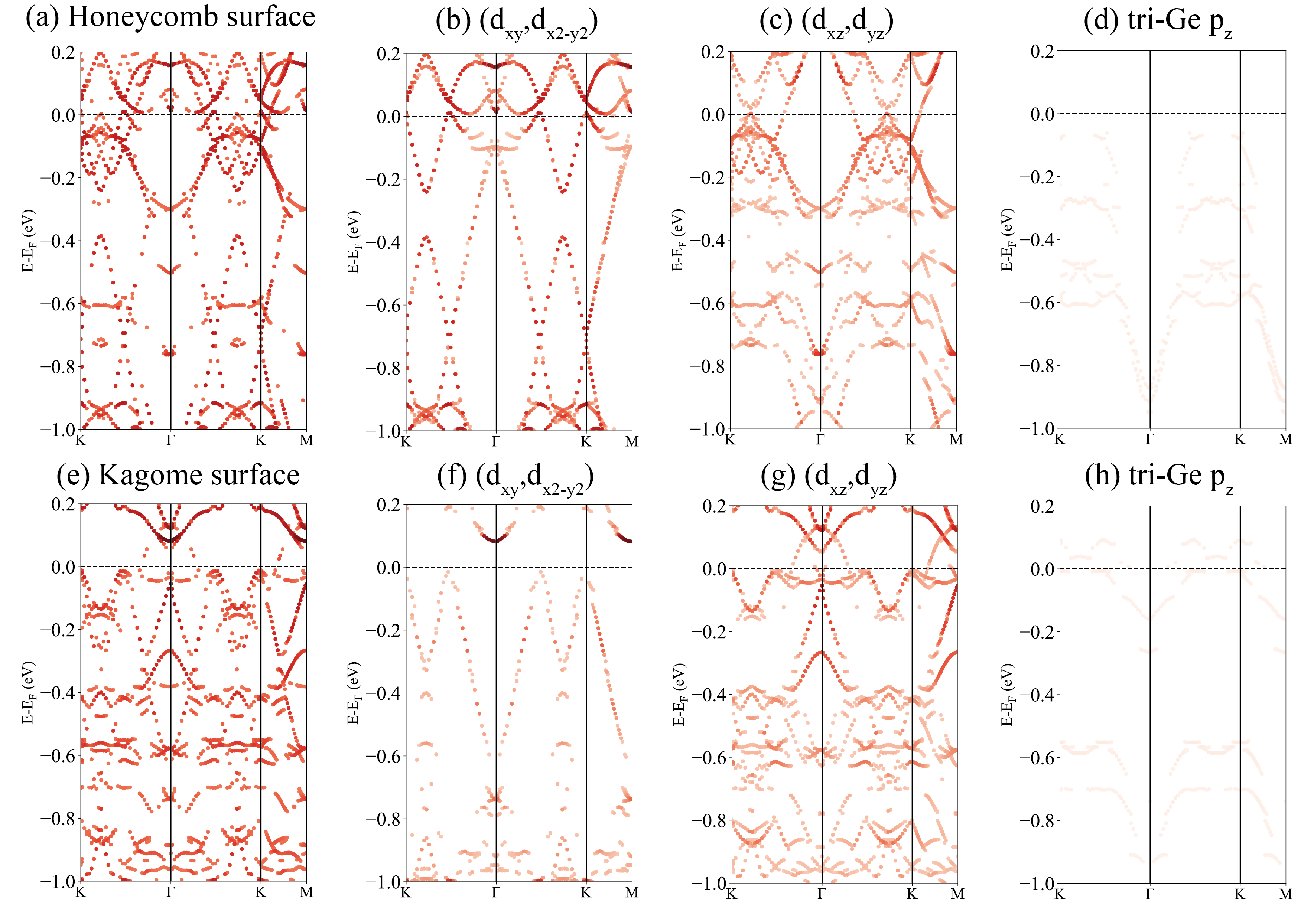}
    \caption{The folded surface bands in the CDW phase. (a) The honeycomb surface bands. (b)-(d) The orbital weights in the honeycomb surface bands, for Fe $(d_{xy}, d_{d^2-y^2})$, $(d_{xz}, d_{yz})$, and triangular Ge $p_z$ orbitals, respectively. A U-shaped band centered at $\Gamma$ at about -0.3 eV is observed at the honeycomb surface which mainly comes from $(d_{xz}, d_{yz})$ of Fe. This band agrees with ARPES results. 
    (e)-(h): same as (a)-(d) but for the kagome surface. The definition of the two surface termination is given in \cref{fig: structure-bands}. }
    \label{fig: CDW-surface-bands-fold}
\end{figure}

\begin{figure}[htbp]
    \centering
    \includegraphics[width=0.5\textwidth]{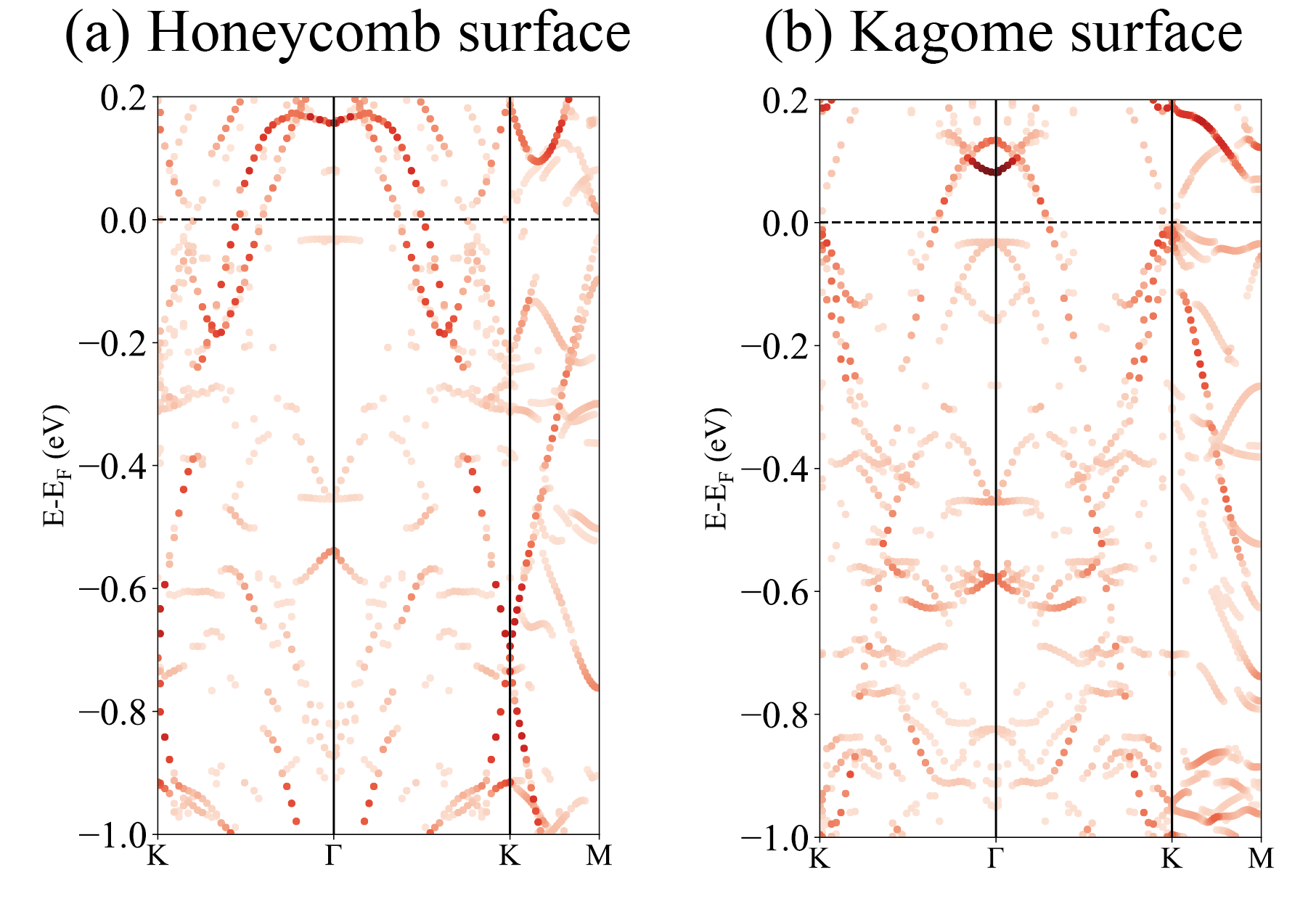}
    \caption{The unfolded surface bands in the CDW phase, for both honeycomb surface (a) and kagome surface (b).}
    \label{fig: CDW-surface-bands-unfold}
\end{figure}

In this section, we discuss the surface bands in the CDW phase. In \cref{fig: CDW-surface-bands-fold}, the folded surface bands for both honeycomb and kagome terminations are given, together with orbital weights. In \cref{fig: CDW-surface-bands-unfold}, we also give the unfolded surface bands. Here the folded bands denote the bands in the CDW BZ, while unfolded bands denote those in the non-CDW BZ. 

We observe the major difference between folded and unfolded surface bands is a U-shaped band centered at $\Gamma$ at about -0.3 eV at the honeycomb surface. This U-shaped band only appears in the folded bands, which mainly comes from $(d_{xz}, d_{yz})$ of Fe. Since this band is not seen in the unfolded bands near $\Gamma$, it is folded from the $M$ point due to the $2\times 2$ CDW order. In the CDW bulk bands, there exist bands with a similar shape from $(d_{xz}, d_{yz})$ near $L=(\frac{1}{2},0,\frac{1}{2})$, as shown in \cref{fig: CDW-AFM-orbital-weight}(g). Thus this surface U-shaped band can be seen as $k_z$-projected bulk bands with surface reconstructions.

This U-shaped band matches well with APRES results. We conjecture that there are strong disorder effects near the surface that break the translational symmetry. Thus the observed bands in ARPES can be explained by the folded surface bands.

\clearpage

\section{Diffuse scattering, DS}

Single crystal's diffuse scattering contains information about local (short-range, pairs of sites) ordering in the sample and the deviations from the average (different types of disorder), and gives information at scales beyond the average unit cell. In this section, we present a comprehensive study of the different DS cuts of as-grown FeGe, annealed and Ge-deficient FeGe and FeSn. 

\begin{figure*}[h!]
    \centering
    \includegraphics[width=0.85\linewidth]{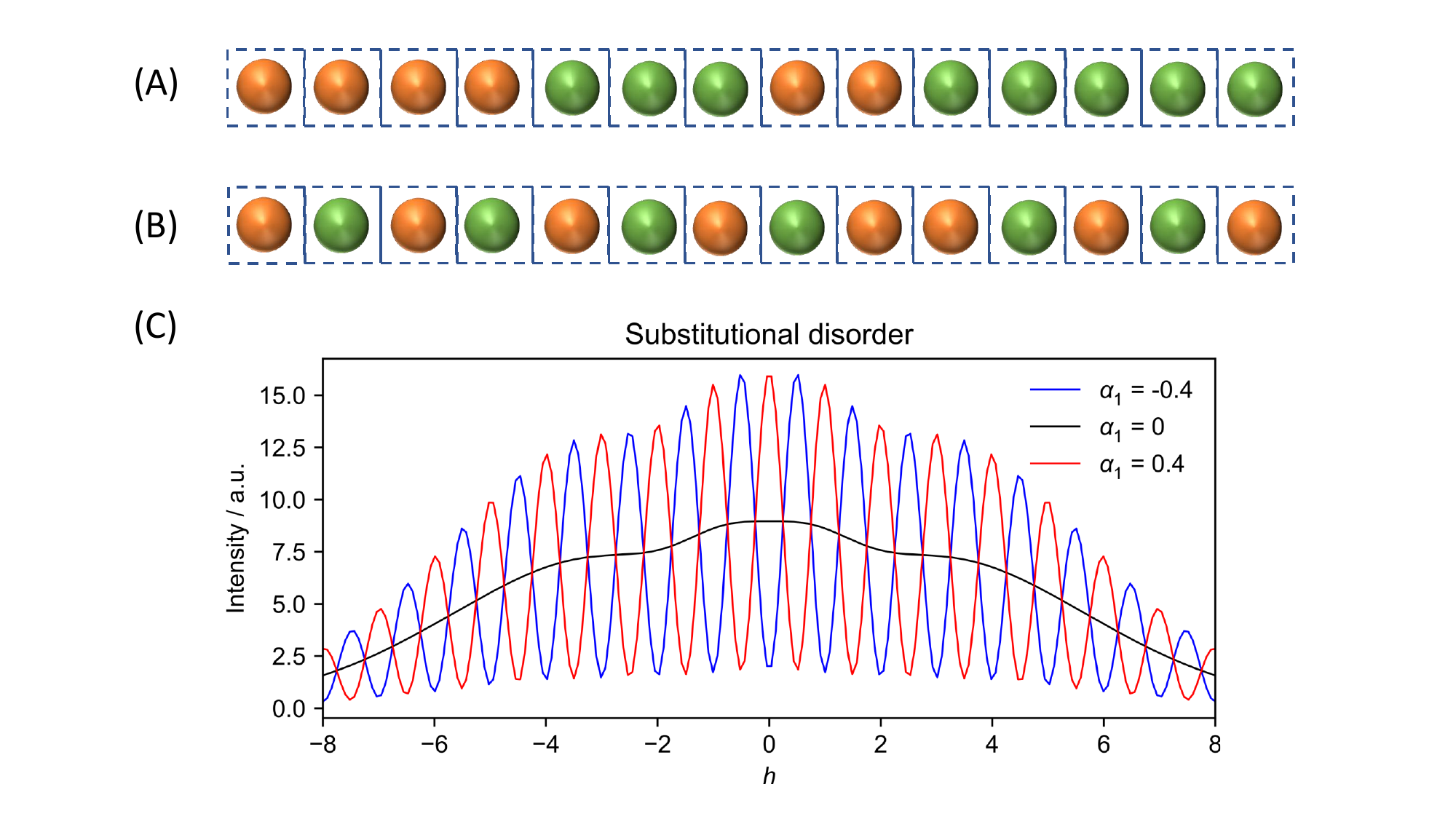}
    \caption{(A) 1D-chain of Fe and Ge atoms with substitutional disorder, where (A) the same type of atoms (Fe or Ge) cluster together and (B) Fe and Ge type of atoms alternate in the chain. (C) Diffuse scattering pattern for substitutional disordered pattern. Note how the DS decreases with \textit{h}.}
    \label{Fig:disorder}
\end{figure*}

First of all, we will give a brief introduction about how to identify substitutional from displacement (phonon driven) disorder \cite{Welberry2022diffuse}.
A system with substitutional disorder can be characterized by a unit cell where a site is occupied by one of several atom types or a missing atom, thus the chemical environment in neighbouring unit cells is not identical. Figure \ref{Fig:disorder} (A) and (B) shows an example of a one dimensional substitutional disordered chain with a random occupation per unit cell and two different types of atom types; i.e. Fe and Ge. 
If Fe and Ge are present in a fraction $m_{\mathrm{Fe}}$ and $m_{\mathrm{Ge}}$, respectively, then:

\begin{equation}
    m_\mathrm{Fe}+m_\mathrm{Ge}=1,
\label{eq._disor}
\end{equation}

In a system with positive correlated disorder, Fe and Ge coherently cluster together; e.g. neighboring atoms are of the same type, figure \ref{Fig:disorder} (A). If disorder is negatively correlated if Fe and Ge alternate along the chain, e.g. Fe tends to have Ge as its neighbor, figure \ref{Fig:disorder} (B).
Starting from a toy model, a binary disordered system is commonly characterized by the Warren-Cowley short-range order parameters $\alpha_{\vec{v}}$;

\begin{equation}
    \alpha_{\vec{v}}=1-\frac{P^{\mathrm{FeGe}}_{\vec{v}}}{m_\mathrm{Fe}m_\mathrm{Ge}},
\label{eq._WC}
\end{equation}

-where P$^{\mathrm{FeGe}}_{\vec{v}}$ describes the probability to find a Ge atom from a Fe atom at a vector $\vec{v}$. The positive and negative correlations in a certain direction are, thus, parametrized by Warren-Cowley short-range order values. 

\begin{equation}
\alpha_{\vec{v}} =
    \begin{cases}
        > 0 & \text{positive correlation,}\\
        = 0 & \text{no correlation,}\\
        < 0 & \text{negative correlation.}\\
    \end{cases}
    \label{eq:cases}
\end{equation}

The Warren-Cowley parameters are equivalent to the interaction energies obtained by DFT in the main text.

The diffraction patterns for a positive, negative and uncorrelated one dimensional lattice are displayed in figure \ref{Fig:disorder} (C). The main observation is a decrease of intensity as a function of the momentum transfer. Moreover, positive correlation gives large diffraction intensity at integer values of \textit{h}, while a negative $\alpha_{\vec{v}}$ localizes the intensity at half-integer \textit{h} (in between Bragg peaks).

In a system with pure displacement disorder the atoms are displaced from their average position within the unit cell. This is the case for thermal disorder or the condensation of a particular phonon mode associated with a CDW phase transition. Therefore, it is strongly dependent on the phonon eigenvectors and polarization and the electron-phonon interaction. The displacement vector of an atom in a unit cell \textit{t} on site \textit{i} from its average position is given by $\vec{\delta}_{t,i}$ and follows a Gaussian probability with a covariance matrix $\underline{u}$:

\begin{equation}
p(\vec{\delta}_{t,i})=\frac{1}{\sqrt{(2\pi){^3}}\mathrm{det}(\underline{u})}\exp\left(-\frac{1}{2}(\vec{\delta}_{t,i})^{T}\underline{u}^{-1}(\vec{\delta}_{t,i})\right),
\label{Gaussian}
\end{equation}

\begin{figure*}[h!]
    \centering
    \includegraphics[width=0.85\linewidth]{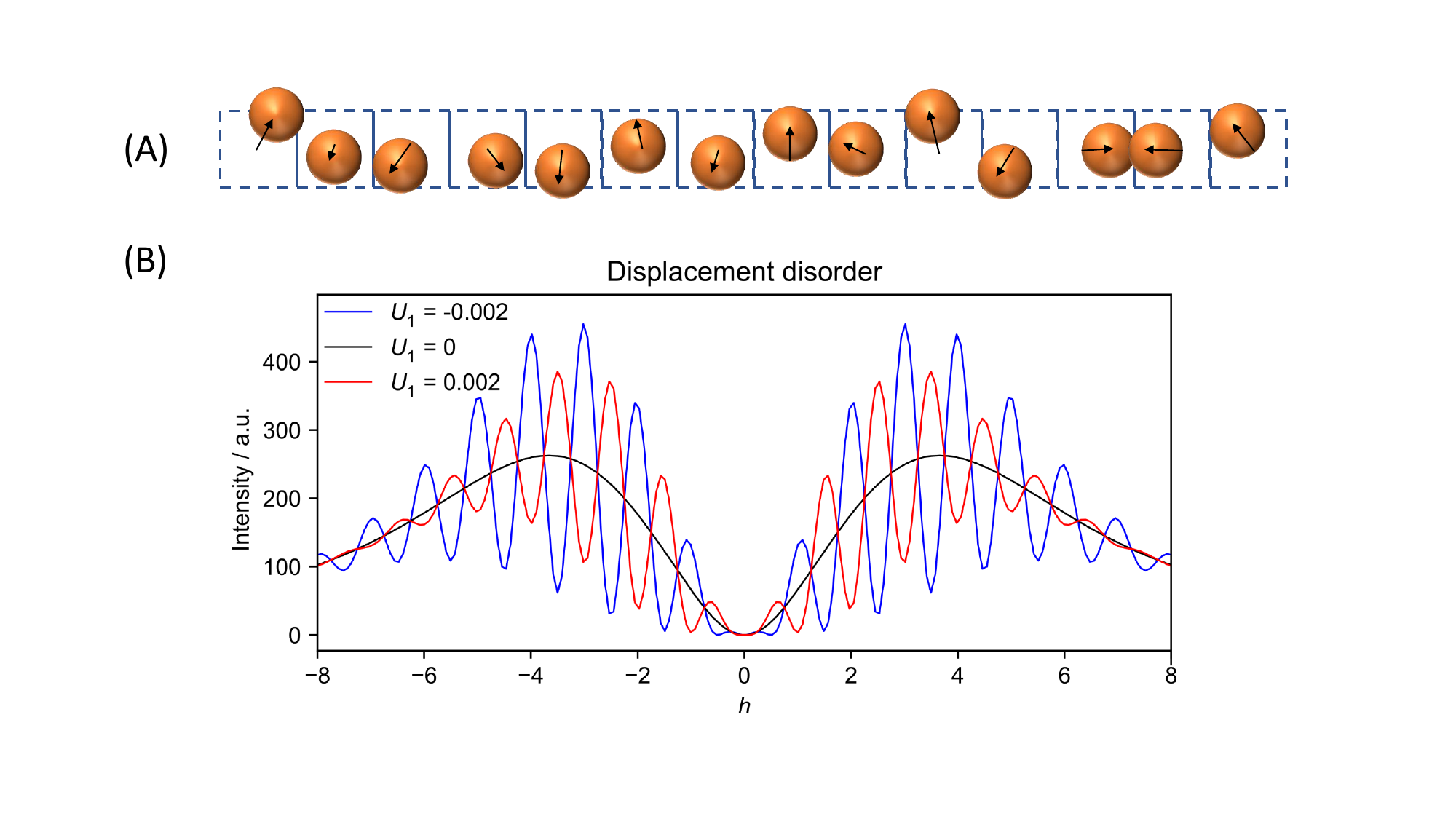}
    \caption{(A) Diffuse scattering form a one-dimensional Ge crystal with displacement disorder. In (A), the arrows stand for the atomic displacements. (B) Calculated DS for correlated (\textit{U} positive, atoms move in-phase) and uncorrelated (\textit{U} negative, atoms move out-of-phase). The DS is negligible at \textit{h}$\rightarrow$=0, hence differentiating from the substitutional disorder case. Note that positive correlation give maxima of DS at the Bragg positions, while negative correlation gives DS at half-integer \textit{h}.}
    \label{Fig:disorder_disp}
\end{figure*}

The displacement disorder is random if the displacement of an atom is independent of the displacement of its neighbouring sites, positively correlated if the displacement of neighbouring atoms is preferably along the same direction or along opposing directions (negative correlation). The diffuse diffraction pattern of a positive, a negative and an uncorrelated system driven by displacement disorder is shown in figure \ref{Fig:disorder_disp}.

The signature of diffuse scattering caused by displacement disorder is a decrease of the maximum intensity at \textit{h}$\rightarrow$0, where the substitutional disorder develops its maximum intensity. Therefore, the momentum dependence of the diffuse scattering allows to directly distinguish between displacement and substitutional disorder. This is the case observed in FeGe, where the DS follows the typical trend characteristic of substitutional disorder instead of phonon-driven.

\clearpage

\subsection{FeGe}

\begin{figure*}[h!]
    \centering
    \includegraphics[width=1.0\linewidth]{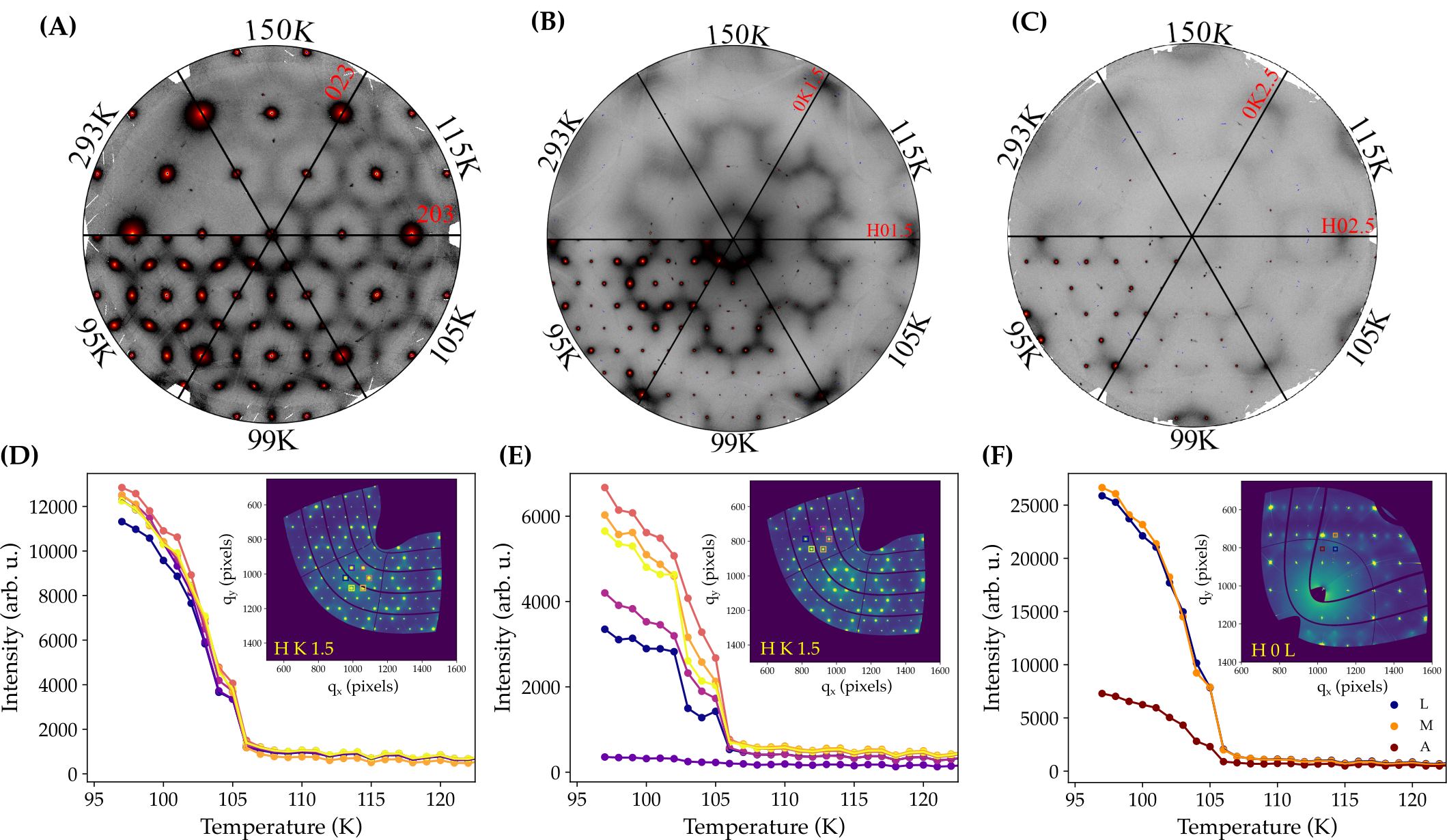}
    \caption{DS maps of FeGe and integrated intensities of CDW peaks. (A) DS map of (h k 3). (B) DS map of (h k 1.5). (C) DS map of(h k 2.5). (D) Temperature evolution of the integrated intensities of the CDW peaks around the position (0 0 1.5). (E) Temperature evolution of the integrated intensities of the CDW peaks around the position (2 0 1.5). (F) Temperature evolution of the integrated intensities of the CDW peaks at A, M and L in the h 0 l plane. In the inset of the panels (D-F), the squares are the integrated areas for each plot.}
    \label{Fig:SI_DS_FeGe}
\end{figure*}

In figure \ref{Fig:SI_DS_FeGe} (A-C), we show the (h k) DS cuts for L=3, 1.5 and 2.5 planes. Signatures of DS driven by substitutional disorder, namely hexagonal diffuse rings, are also visible at the (h k 3) plane. Similarly to the (h k 1.5) plane, the (h k 2.5) shows a complex diffuse pattern, presumably as a result of the small in-plane atomic displacements not considered in the MC simulations.

Figures \ref{Fig:SI_DS_FeGe} (D-F) display the integrated intensity of the CDW peaks and the corresponding diffuse precursors defined by the region of interest (ROI) in the insets. The intensities are strongly modulated in reciprocal space surrounding the Bragg points.  

\subsection{Annealed FeGe}
\begin{figure*}[h!]
    \centering
    \includegraphics[width=0.85\linewidth]{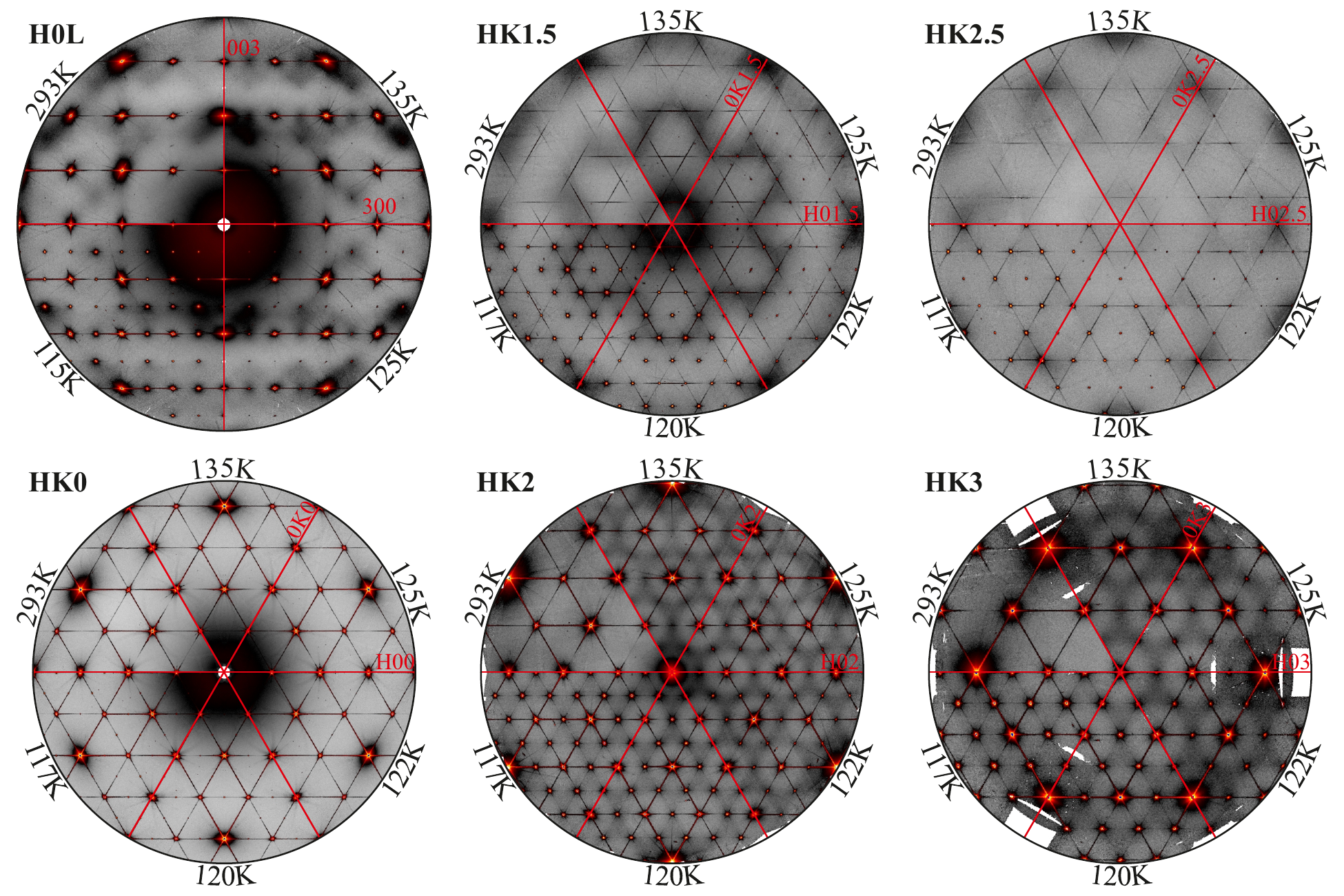}
    \caption{ DS maps of the annealed FeGe. The hexagonal diffuse rings are still present at T$>$T$_\mathrm{CDW}$ but, nevertheless, absent for T$<$T$_\mathrm{CDW}$. In addition, no anisotropic DS is observed at the M point.
    }
    \label{Fig:DS_Ann}
\end{figure*}

As-grown single crystals of FeGe (hereafter FeGe(a)) were annealed for 3 days at 300ºC to increase the CDW correlation length \cite{wu2024annealing}. The spatial correlation of the CDW extends to $30.0\pm0.7\,\mathrm{nm}$ along M,  $33.9\pm0.6\,\mathrm{nm}$ along L and  $43.3\pm0.6\,\mathrm{nm}$ along A.
As we can see in figure \ref{Fig:DS_Ann} (A), the diffuse scattering, although with less signal to background ratio, is again localized along the M-L directions and also develops a hexagonal pattern at T$>$T$_\mathrm{CDW}$.   

Furthermore, at integer L and half integer L-values, streaks of diffuse intensity, characteristic of orthorhombic domains, are visible between Bragg peaks. On the other hand, the hexagonal diffuse pattern is no longer visible at low temperature, suggesting that the annealing reduces the frustration between dimerized and undimerized phases at T$<$T$_{CDW}$. 

\begin{figure*}[h!]
    \centering
    \includegraphics[width=0.85\linewidth]{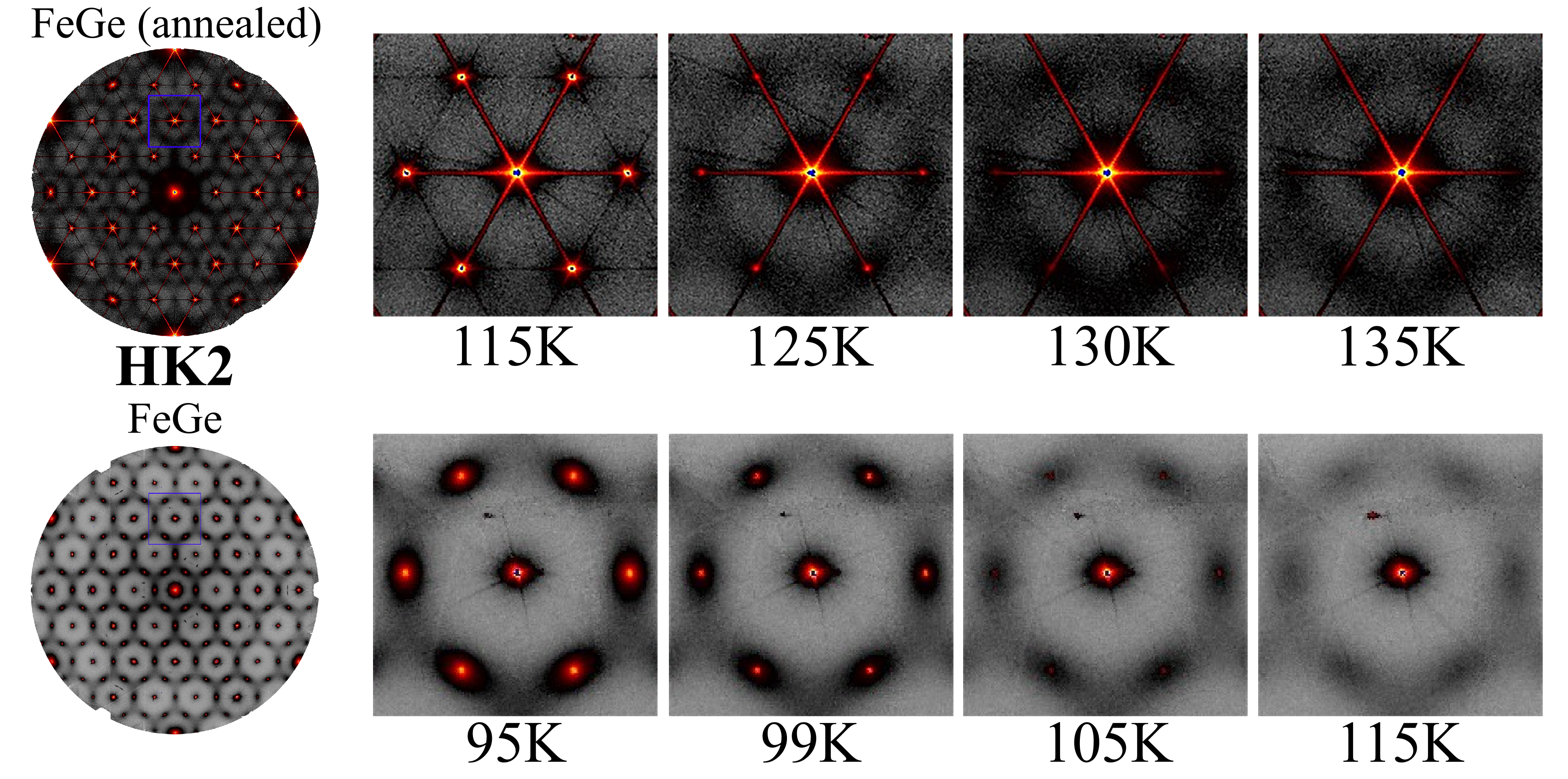}
    \caption{Comparison of the (h k 2) annealed (A) and as-grown (B) FeGe crystals. (A) The DS of the annealed sample is at the background level and the diffuse signal is isotropic around the CDW reflection. 
    }
    \label{Fig:SI_DS_comparison}
\end{figure*}

Figure \ref{Fig:SI_DS_comparison} compares the (h k 2) DS of the annealed and as-grown FeGe. Three features are clearly visible: (1) the DS of the as-grown samples present larger signal-to-noise ratio, (2) streaks of diffuse intensity appear in between Bragg peaks and cross the CDW DS at M and (3) the as-grown CDW peak width is more anisotropic than the annealed FeGe. We note that due to the lower intensity of the DS of FeGe(a) and the presence of orthorhombic domains precludes us to reach a reliable comparison the anisotropic width ratio between FeGe(a) and as-grown crystals.

\clearpage

\subsection{FeGe$_{0.9}$}
\begin{figure*}[h!]
    \centering
    \includegraphics[width=0.8\linewidth]{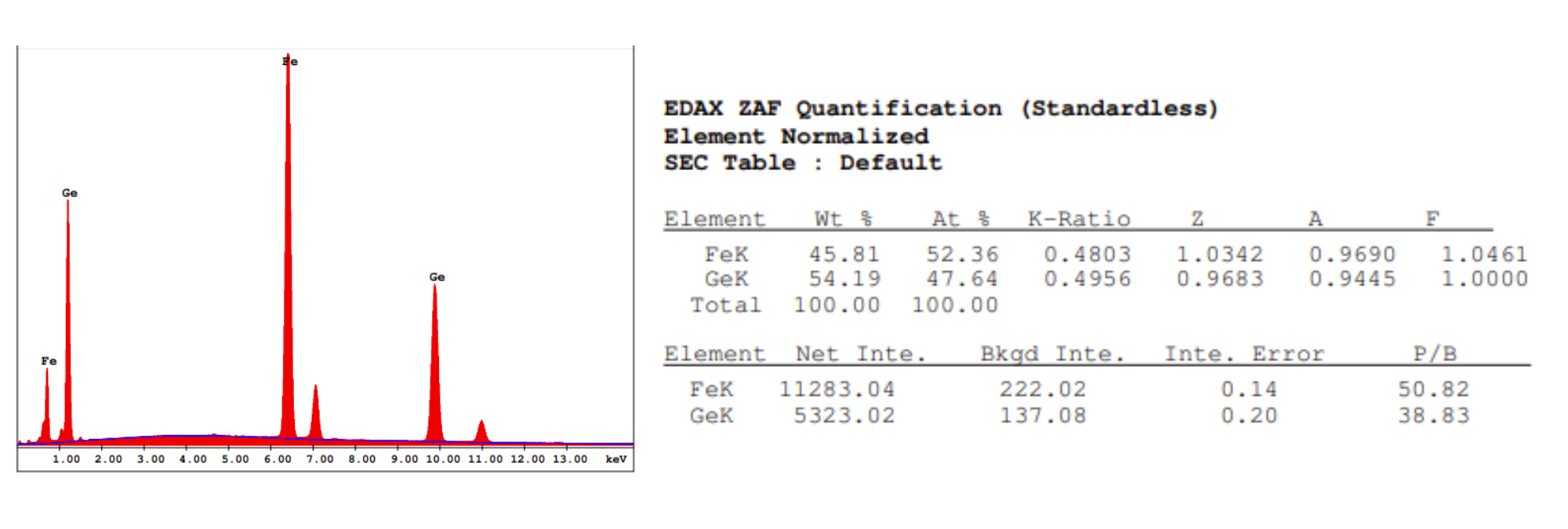}
    \caption{Energy dispersive analysis of Ge-deficient FeGe (FeGe$_{0.9}$). 
    }
    \label{EDX_FeGe0p9}
\end{figure*}

\begin{figure*}[h!]
    \centering
    \includegraphics[width=0.5\linewidth]{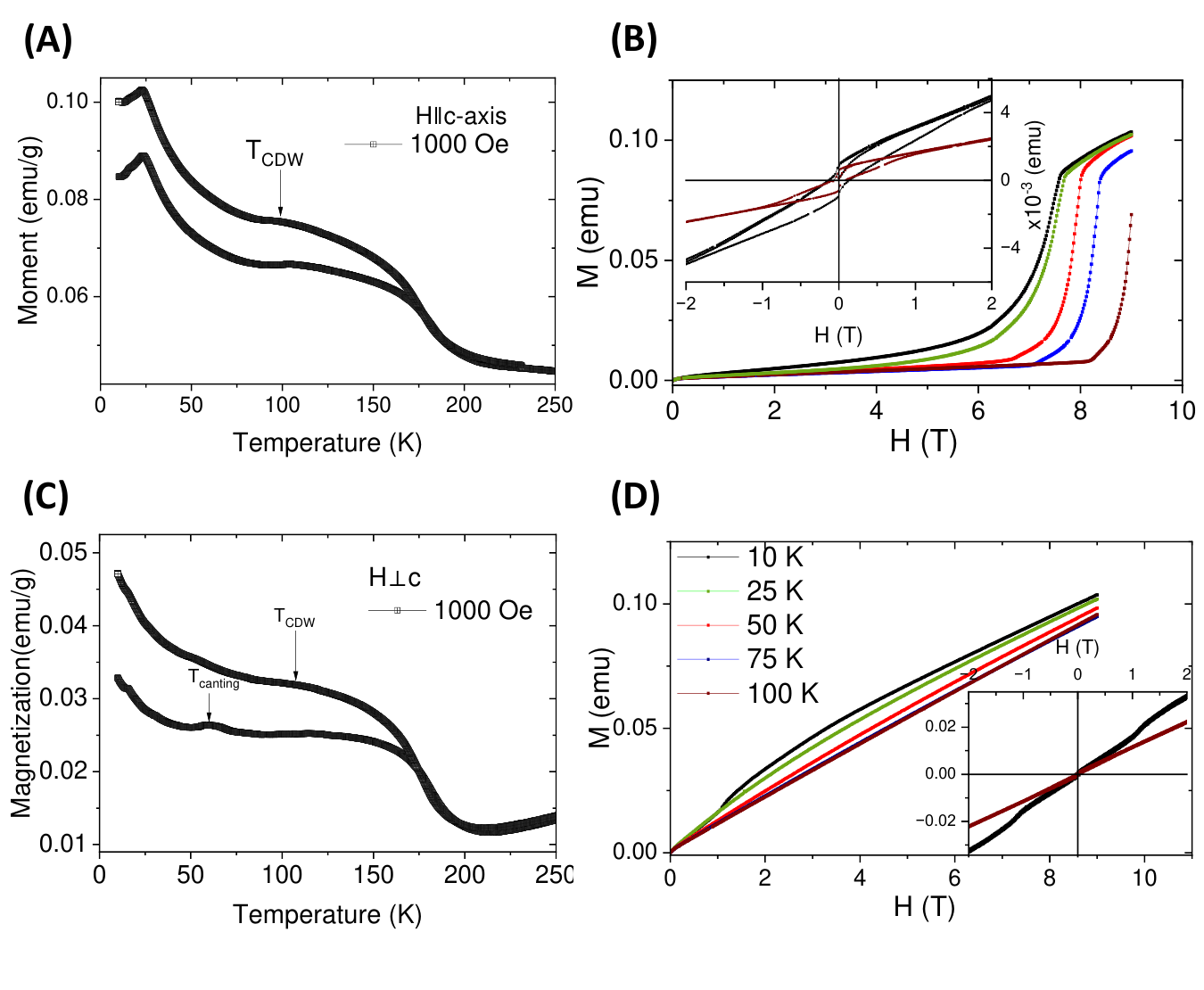}
    \caption{Magnetic characterization of Ge-deficient FeGe (A) Temperature dependence of the magnetization for applied magnetic field (H) parallel to the \textit{c}-axis (B) Magnetic field dependence of the magnetization for H parallel to \textit{c}. Temperature (C) and magnetic field (D) dependence of the magnetization parallel to the \textit{ab}-plane.}
    \label{Mag_FeGe0p9}
\end{figure*}

In this section, we present the results of Ge-deficient FeGe (FeGe$_{0.9}$). The Ge concentration was estimated from the energy dispersive analysis (EDX), figure \ref{EDX_FeGe0p9}. The 10\% Ge deficiency slightly modifies the magnetic behavior of the sample, but the CDW and the AFM canting transition are still visible, figure \ref{Mag_FeGe0p9}. 

\begin{figure*}[h!]
    \centering
    \includegraphics[width=0.85\linewidth]{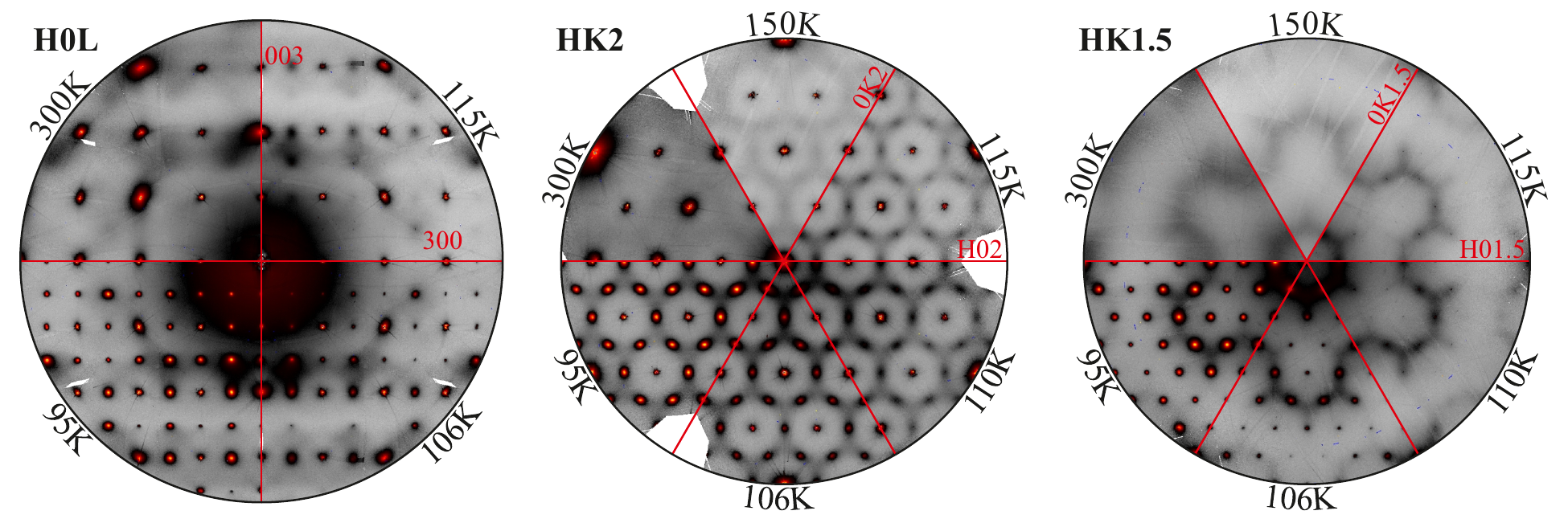}
    \caption{Reciprocal space reconstructions of the Ge-deficient FeGe$_{0.9}$ diffuse maps. 
    }
    \label{Fe0p9_DS}
\end{figure*}

\begin{figure*}[h!]
    \centering
    \includegraphics[width=0.85\linewidth]{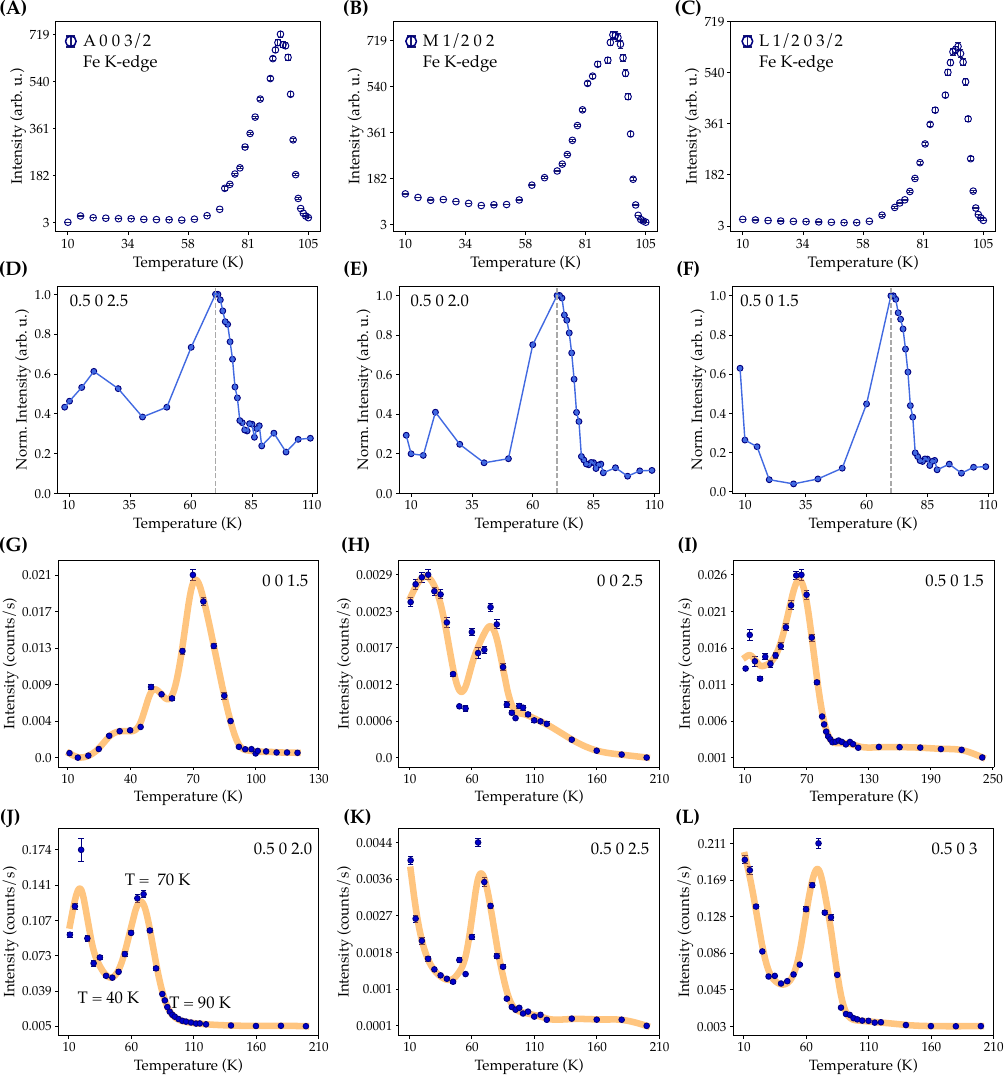}
    \caption{ (A-C) Temperature dependence of the CDW peaks $\mathrm{A},\,\mathrm{M}$ and $\mathrm{L}$, respectively, measured at the Fe K-edge, 7.11 keV. (D-F) Temperature dependence of the DS integrated intensity of three different CDW peaks. (G-L) Temperature dependence of the integrated intensity of the elastic central peak (CP) of IXS for the Ge-deficient FeGe$_{0.9}$. 
    }
    \label{Fe0p9_CP}
\end{figure*}

The DS maps of FeGe$_{0.9}$ show similar hexagonal diffuse patterns at integer and half integer L-values as the stoichometric FeGe. However, the intensity of the CDW peaks sharply drops below the T$_\mathrm{CDW}$, see figure \ref{Fe0p9_DS}, with an upturn for some propagation vectors and even the complete disappearance of the CDW reflection at q$_\mathbf{A}$=(0\ 0\ $\frac{3}{2}$). This behavior was not observed in FeGe$_{1.0}$. We have carried out energy resolved inelastic x-ray scattering (XS) experiments to disentangle the elastic and inelastic contributions of the diffuse signal. As shown in figure \ref{Fe0p9_CP}, the drop of intensity is mostly driven by the temperature dependence of the elastic central peak (CP) of the IXS spectrum.

The temperature dependence of the CP, and in particular the absence of the elastic signal at (0 0 $\frac{3}{2}$) at low temperature, is reminiscent to the reports charge ordered nickelates \cite{Ricci_2021}. Although not explored in detail, we speculate with the role of quenched disorder or fluctuating charge order below T$_\mathrm{CDW}$ that mat hint at a competition between CDW and magnetism. 

\clearpage

\subsection{FeSn}
We have searched for diffuse scattering in the antiferromagnetic FeSn with the in-plane spin polarization \cite{Kang_2019}. The absence of any type of diffuse pattern as a function of temperature demonstrates that the out-of-plane spin polarization of FeGe is responsible for the out-of-plane displacement of the trigonal Ge.   

\begin{figure*}[h!]
    \centering
    \includegraphics[width=0.85\linewidth]{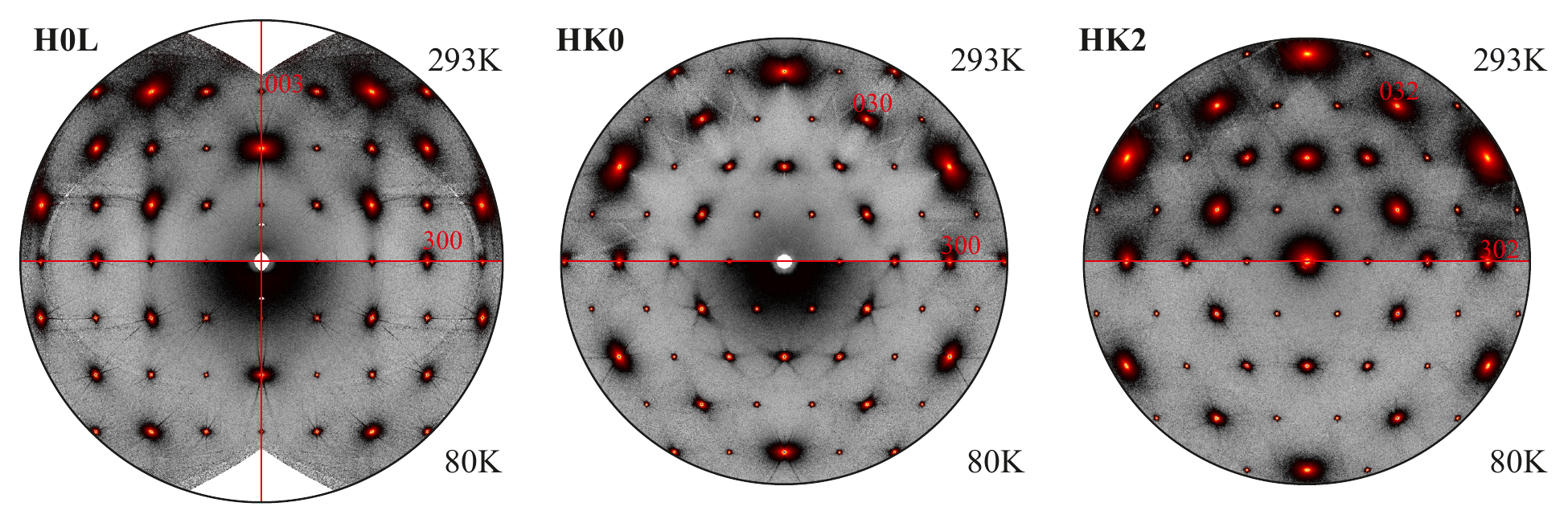}
    \caption{Temperature dependence DS of FeSn. No diffuse signal is observed at either M, L or A points. 
    }
    \label{FeSn}
\end{figure*}

The DS around the Bragg peaks at high temperature is a a result of the thermal excitations of phonons; i.e. thermal diffuse scattering (TDS).

\clearpage

\section{Monte Carlo simulation of DS}
In this section, we use Monte Carlo to simulate the DS based on an effective Ising model that describes the dimerization of triangular Ge.  

\subsection{Ising model for Ge dimerization}

We first build an Ising model to describe the dimerization of triangular Ge and to achieve a certain microscopic realisation via the Monte Carlo simulations. We use an Ising variable $\sigma_i=\pm 1$ to denote the dimerized ($\sigma_i=-1$) and undimerized ($\sigma_i=+1$) triangular Ge pair in the (non-CDW AFM phase) unit cell $\mathbf{R}_i$. By considering in-plane nearest neighbor (NN) coupling $c_1$, next NN (NNN) coupling $c_2$, 3rd-NN (3NN) coupling $c_3$, $z$-direction NN coupling $c_4$, and an effective magnetic field $h$, we build a model with the form
\begin{equation}
\begin{aligned}
H=\sum_{\langle ij \rangle:NN} c_1 \sigma_i\sigma_j + 
\sum_{\langle ij \rangle:NNN} c_2 \sigma_i\sigma_j + 
\sum_{\langle ij \rangle:3NN} c_3 \sigma_i\sigma_j + 
\sum_{\langle ij \rangle:z-NN} c_4 \sigma_i\sigma_j + 
\sum_{i}h\sigma_i + E_0.
\end{aligned}
\label{Eq: Ising-model}
\end{equation}
where $E_0$ is a constant and $\langle ij \rangle$ means each $ij$ pair counts only once. The magnetic field term $h$ is added because the fully dimerized and undimerized configurations have different energies.

\begin{figure}[htbp]
    \centering
    \includegraphics[width=0.4\linewidth]{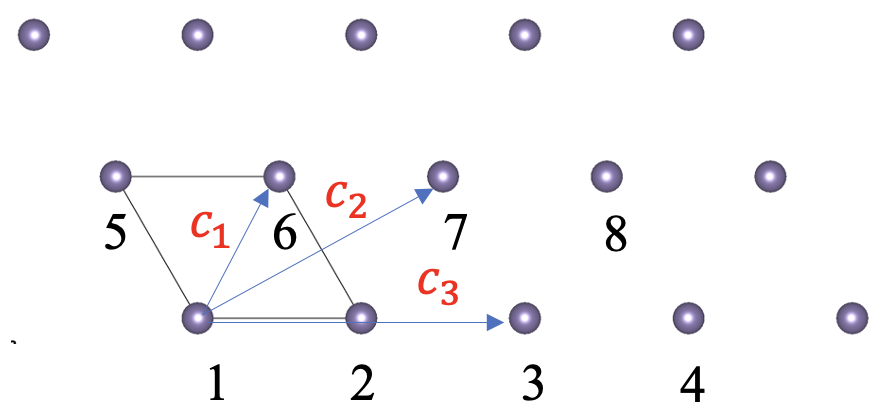}
    \caption{The illustration of the coupling parameters used in the effective Ising model in \cref{Eq: Ising-model}. In the figure, $c_{1,2,3}$ are in-plane NN, NNN, and 3NN coupling parameters. The numbers 1-8 label the eight Ge pairs in a $4\times 2$ in-plane supercell of the non-CDW unit cell. The black lines mark the non-CDW unit cell. 
    }
    \label{Fig:def-Ising-variable}
\end{figure}

We then fit the parameters from DFT. We choose a $4\times 2\times 2$ supercell of the non-CDW AFM unit cell, which contains $4\times 2\times 2$ Ge pairs. As shown in \cref{table: DFT-dimer-config-energies}, we consider 7 inequivalent dimer configurations and compute their averaged magnetic moments and total energies in DFT. The parameters in the Ising model \cref{Eq: Ising-model} are fitted using the DFT data, with their values summarized in \cref{table: value-Ising-parameters}.

\begin{table}[htbp]
\centering
\begin{tabular}{c|c|c|c}
\hline\hline
Dimer configuration       & $\bar{\mu}_{Fe}/\mu_B$ & DFT total energy (meV) & Fitted energy (meV) \\ \hline
$-+-+, ++++,\quad -+-+, ++++$ & 1.44                    & 0.0                & 0.0                 \\ \hline
$++++, ++++,\quad ++++, ++++$ & 1.39                    & 835.4249         & 772.5727          \\ \hline
$-+++, -+++,\quad -+++, -+++$ & 1.44                    & 997.3268         & 997.3268          \\ \hline
$-+++,++-+,\quad -+++, ++-+$  & 1.43                    & 667.4307         & 667.4307          \\ \hline
$-+++, ++++,\quad -+++, ++++$ & 1.41                    & 517.8678         & 517.8678          \\ \hline
$-+-+, ++++,\quad ++++, ++++$ & 1.41                    & 614.8985         & 740.6027          \\ \hline
$-+-+, ++++,\quad ++++, +-+-$ & 1.44                    & 771.4848         & 708.6327          \\ \hline\hline
\end{tabular}
\caption{\label{table: DFT-dimer-config-energies}
The computed total energies from DFT for different dimer configurations. For each dimer configuration, $+$ ($-$) denotes the dimerized (undimerized) triangular Ge pair. The first 8 $\pm$ denotes the 8 Ge pairs marked in \cref{Fig:def-Ising-variable} on the first layer in the $4\times 2\times 2$ supercell, and the second 8 $\pm$ denotes the second layer. The second column of $\bar{\mu}_{Fe}$ is the averaged magnitude of magnetic moment on Fe atoms. The third column is the computed DFT total energy, while the last column is the fitted energy using \cref{Eq: Ising-model}. Remark that the configurations with more dimerized Ge have larger magnetic moments and much higher total energies, and are not used in the fitting. 
}
\end{table}

\begin{table}[htbp]
\centering
\begin{tabular}{c|c|c|c|c|c}
\hline\hline
$c_1$ & $c_2$ & $c_3$ & $c_4$ & $h$ & $E_0$ \\\hline
45.885 & 25.267 & -8.224 & -44.290 & -330.340 & 3746.100 \\
\hline\hline
\end{tabular}
\caption{\label{table: value-Ising-parameters}
The fitted value of parameters in the Ising model \cref{Eq: Ising-model} based on the \textit{ab initio} data in \cref{table: DFT-dimer-config-energies}. All numbers are given in meV. 
}
\end{table}

\subsection{Monte Carlo simulation}

With the effective Ising model derived based on DFT, we perform Monte Carlo simulations to obtain simulated diffuse scattering patterns. In the Monte Carlo simulation, moves are accepted if a random number (0-1) is less than $exp(-\Delta E/k_{B}T)$ with $T = 80K$.
The low-temperature unit cell ($10\times 10\times 8$ \r{A}) was expanded to a $16\times 16\times 16$ supercell. $\sim$5$\%$ dimers were added randomly and the MC simulations were run until convergence.

\begin{figure*}
    \centering
    \includegraphics[width=0.75\linewidth]{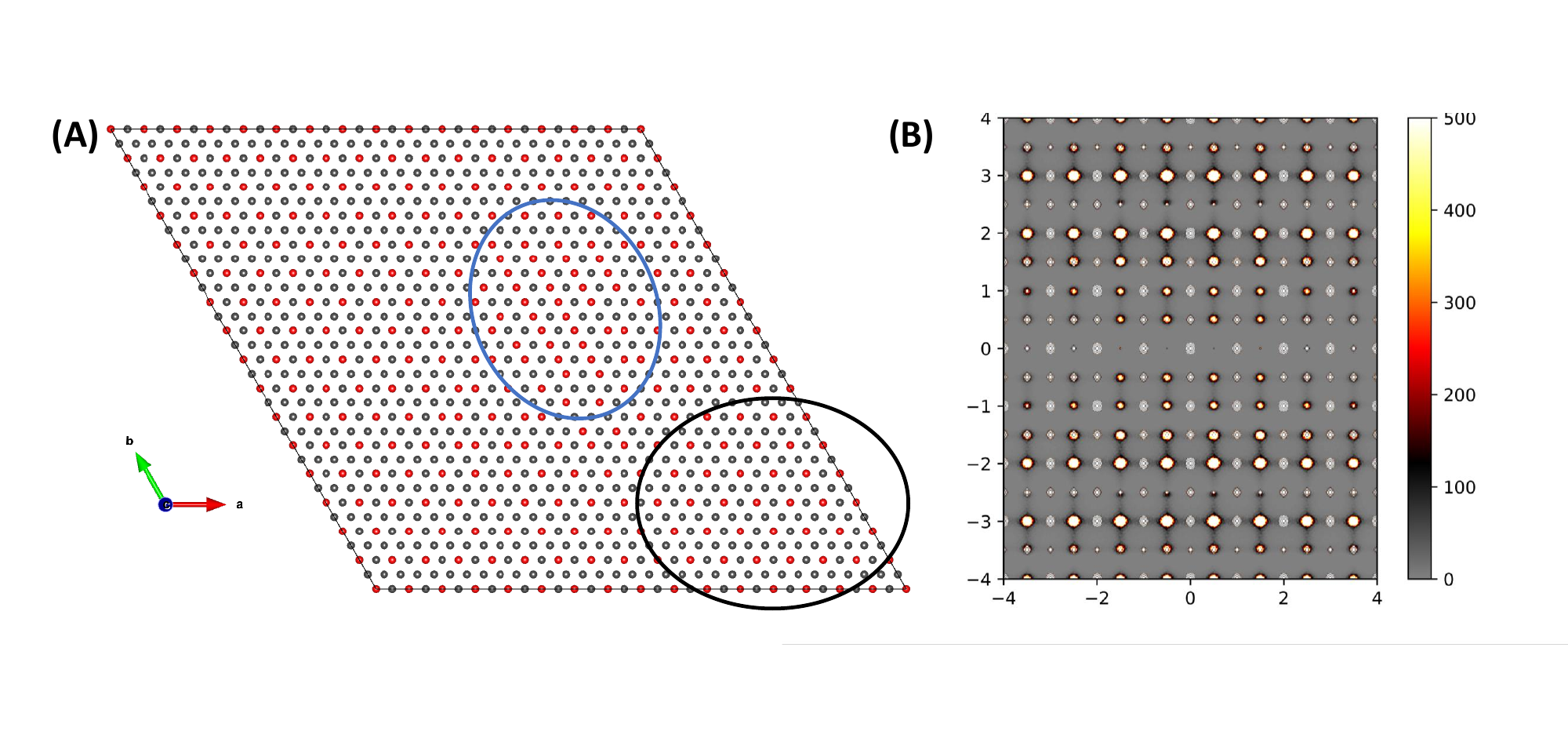}
    \caption{(A) Converged MC simulation of a supercell of FeGe using the Ising model defined in \ref{Eq: Ising-model}. The circles highlight the 2 typical atomic arrangement. (B) (h 0 l) calculated DS map of FeGe showing the diffuse signal at the A, M and L points. Bragg peaks are omitted.  
    }
    \label{Fig:SI_MC}
\end{figure*}

The real space configuration of the minimum energy supercell is depicted in \cref{Fig:SI_MC},(A) where the red and grey balls stand for dimerized and non-dimerized phases. 
Most of the atoms are highly ordered (circled in black) but there are some defect areas where the ordering is not complete (circled in blue). These configurations converged to 37.5$\%$ non-dimers and 62.5$\%$ dimers. 

\begin{figure*}
    \centering
    \includegraphics[width=0.8\linewidth]{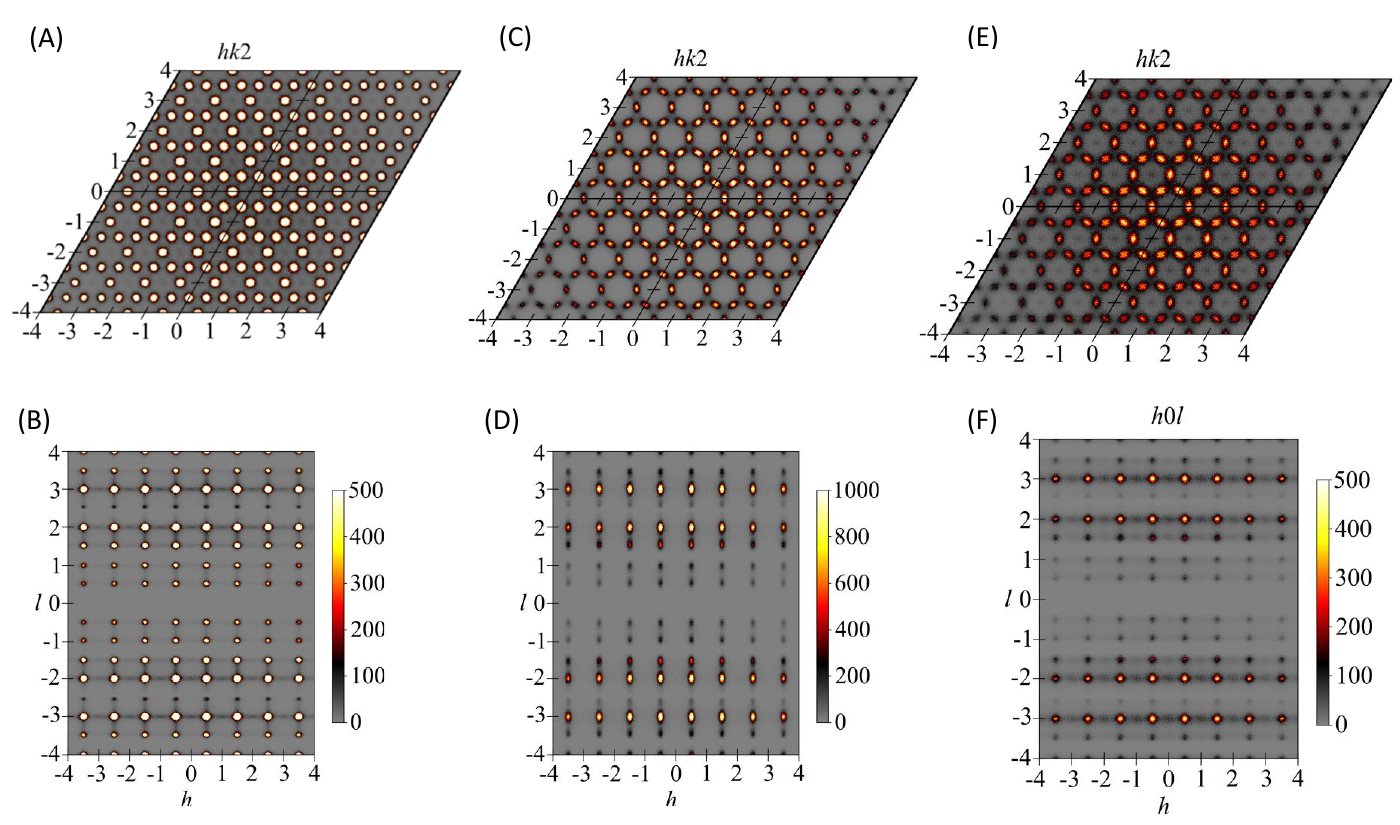}
    \caption{Simulated h k 2 and h 0 l DS maps. (A-B) h k 2 and h 0 l DS obtained by minimising the Ising hamiltonian using the c$_i$ values obtained from DFT, c$_1$= 45.885 meV, c$_2$= 25.267 meV, c$_3$= -8.224, meV c$_4$= -44.290 meV, $h$ = -330.340 meV (T=80 K). (C-D)  h k 2 and h 0 l DS for c$_1$= 165.886 meV, c$_2$= 25.267 meV, c$_3$= -8.224, meV c$_4$= -44.290 meV (T=80 K). (E-F)  h k 2 and h 0 l DS c$_1$= 165.886 meV, c$_2$= 0 meV, c$_3$= -8.224, meV c$_4$= -44.290 meV (T=100 K). The simulations do not include the in-plane atomic displacements, which are responsible for the absence of DS at l=integer+$\frac{1}{2}$. See main text. 
    }
    \label{Fig:SI_MC_sim}
\end{figure*}

The diffuse scattering calculated from these configurations using DFT parameters (Bragg peaks have been removed) is plotted in figures \ref{Fig:SI_MC} (B) and \ref{Fig:SI_MC_sim} (A). The diffuse scattering is calculated using the program Scatty by Fourier transforming of all the atomic coordinates of the atomic positions. In figure \ref{Fig:SI_MC_sim} (A-F), the high intense Bragg nodes at the A point are removed.

The DFT values reproduce the DS at the M and L points, figure \ref{Fig:SI_MC_sim} (A-B). Moreover, upon manually tuning the \textit{c}$_1$ and \textit{c}$_2$ values, we can also simulate the shape of the anisotropic DS at the M point at 80 K and 100 K. This shows that the \textit{c}$_i$'s parameters that describe the nearest neighbour interaction between dimerized and undimerized trigonal Ge are temperature dependent.   

\clearpage

\section{Bond-order correlation function analysis}

In this section, we describe the analysis of the bond-order correlation function analysis. For a discrete set of particles in the real-space, one can define a local ordering between nearest neighbors as

\begin{equation}
    {\Psi}_{6}(\textbf{r}_k)=\frac{1}{N_{k}} \sum_{j=1}^{N_{k}}\mathrm{e}^{i6\theta_{kj}},
    \label{eq:order_parameter}
\end{equation}

where \textit{N}$_{k}$ is the number of nearest neighbors of the $k-$particle at position \textbf{r}$_k$ and $\theta_{kj}$ defines the angle the \textit{k}-\textit{j} bond (fig. 4(I) in the main text). For a particle at $\bm{r}_k$, the six-bond order correlation function at a distance $|\bm{r}-\bm{r_k}|$ is given by

\begin{equation}
    G_6(|\bm{r}-\bm{r_k}|)=\frac{1}{N_{|\bm{r}-\bm{r_k}|}} \sum_{j}^{N_{|\bm{r}-\bm{r_k}|}}{\Psi}_{6}(\bm{r}_k){\Psi}^{*}_{6}(\bm{r}_j),
    \label{eq:local_g6}
\end{equation}

where the sum goes over all the particle at a distance $|\bm{r}-\bm{r_k}|$ with respect to $\bm{r_k}$. Summing over all the particles in the system, we can define a total $G_6(\bm{r})$, 

\begin{equation}
    G_6(\bm{r})=\frac{1}{N_{r}} \sum_{<j,k>}^{N_{r}}{\Psi}_{6}(\bm{r}_k){\Psi}^{*}_{6}(\bm{r}_j),
    \label{eq:g6}
\end{equation}

- where $N_{r}$ goes over any pair of particles which are at a distance $\bm{r}$. To compute the $G_6(\bm{r})$ correlation function, first we need to get the real-space charge distribution from the diffuse scattering maps. The intensity $I$ of a diffuse scattering map is proportional to the square of the structure factor,

\begin{equation}
    I \propto |S\left(q\right)|^2,
    \label{eq:I_diffuse}
\end{equation}

- where $\bm{q}$ is a vector in the the reciprocal space. The real-space charge distribution is defined as the real part of the Fourier transform of the structure factor,

\begin{equation}
    \rho\left(\bm{r}\right)= \frac{1}{V_{cell}}\sum_{\bm{q}}^{N}|S\left(\bm{q}\right)|\cos{\left(2\pi\left(\bm{q}\cdot\bm{r}\right)+\Phi\left(\bm{q}\right)\right)},
    \label{eq:Rho}
\end{equation}

- where $\bm{r}$ is the position vector in the real space and $\Phi\left(\bm{q}\right)$ is a random phase $S\left(\bm{q}\right)$ for a given $\bm{q}$~\cite{Rcomin_Science,Rcomin_Comment,Rcomin_Answer}.

As Eq.\ref{eq:g6} is defined over a discrete set of particles, we make a discretization of the continuous real-space charge distributions by defining the local maxima as particles. Then, we introduce a Voronoi tessellation in the discrete set of particles to define the concept of neighbor. Given a set of points (particle's positions) $\left\{p_1,\,p_2,...,\,p_N\right\}$, each point $p_i$ has a Voronoi cell associated. This cell consists in any point in the Euclidean space for which $p_i$ is the nearest site of the set of points. All the Voronoi cells together form the Voronoi tessellation and any pair of cells which share a boundary will correspond to a pair of points which are neighbors. The geometrical construction of this diagram is equivalent to a the one used to get a Wigner-Seitz cell. Once we have introduced the concept of neighbor in the discrete set of particles, we can compute Eqs.~\ref{eq:order_parameter},\ref{eq:local_g6},\ref{eq:g6}. 

\begin{figure*}
    \centering
    \includegraphics[width=1\linewidth]{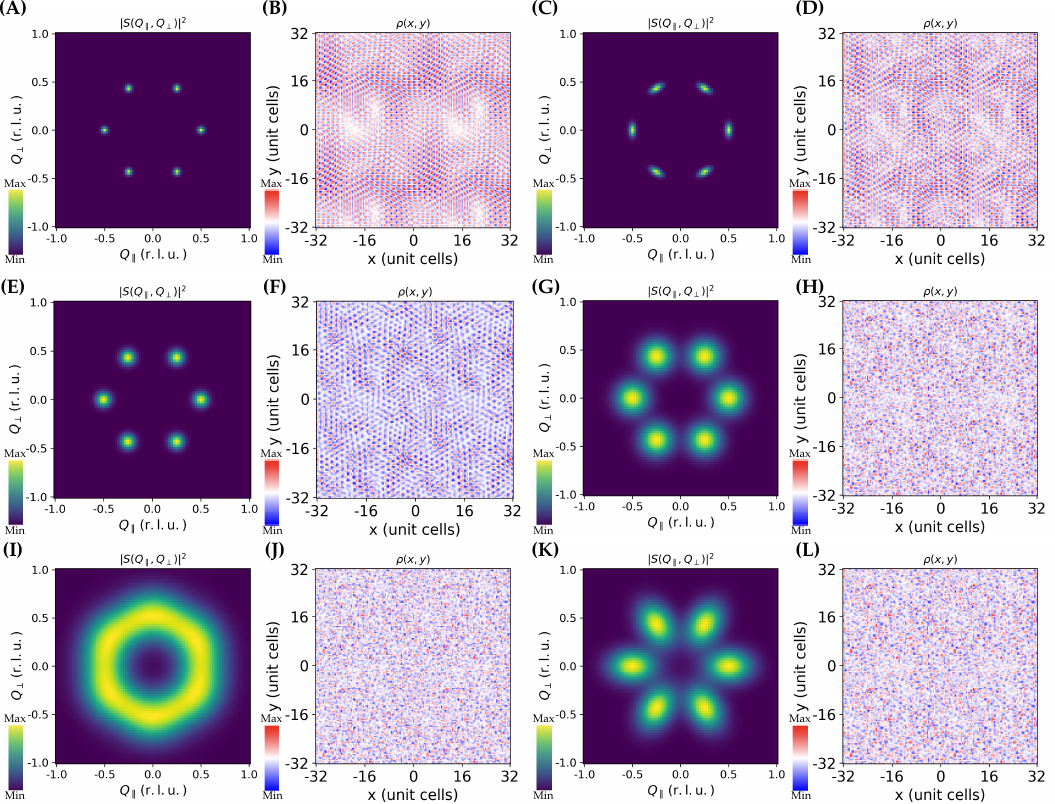}
    \caption{(A) Simulated DS map by using a Gaussian Lineshape centered at each CDW peak and linewidths $[\sigma_x,\,\sigma_y]$ equal to $[0.002,\,0.002]\,\mathrm{r.l.u.}$. (B) Fourier transform of (A). Each pair from (C-L) are the same varying the linewidths of the Gaussians. (C-D) DS and its Fourier transform where the Gaussians have a linewidth of $[0.01,\,0.002]\,\mathrm{r.l.u.}$.  (E-F) DS and its Fourier transform where the Gaussians have a linewidth of $[0.01,\,0.01]\,\mathrm{r.l.u.}$.  (G-H) DS and its Fourier transform where the Gaussians have a linewidth of $[0.05,\,0.05]\,\mathrm{r.l.u.}$.  (I-J) DS and its Fourier transform where the Gaussians have a linewidth of $[0.3,\,0.3]\,\mathrm{r.l.u.}$.  (K-L) DS and its Fourier transform where the Gaussians have a linewidth of $[0.05,\,0.1]\,\mathrm{r.l.u.}$. For each Fourier transform it was used the same random phase matrix.
    }
    \label{Fig:DS_FT_sim}
\end{figure*}

\begin{figure*}
    \centering
    \includegraphics[width=0.85\linewidth]{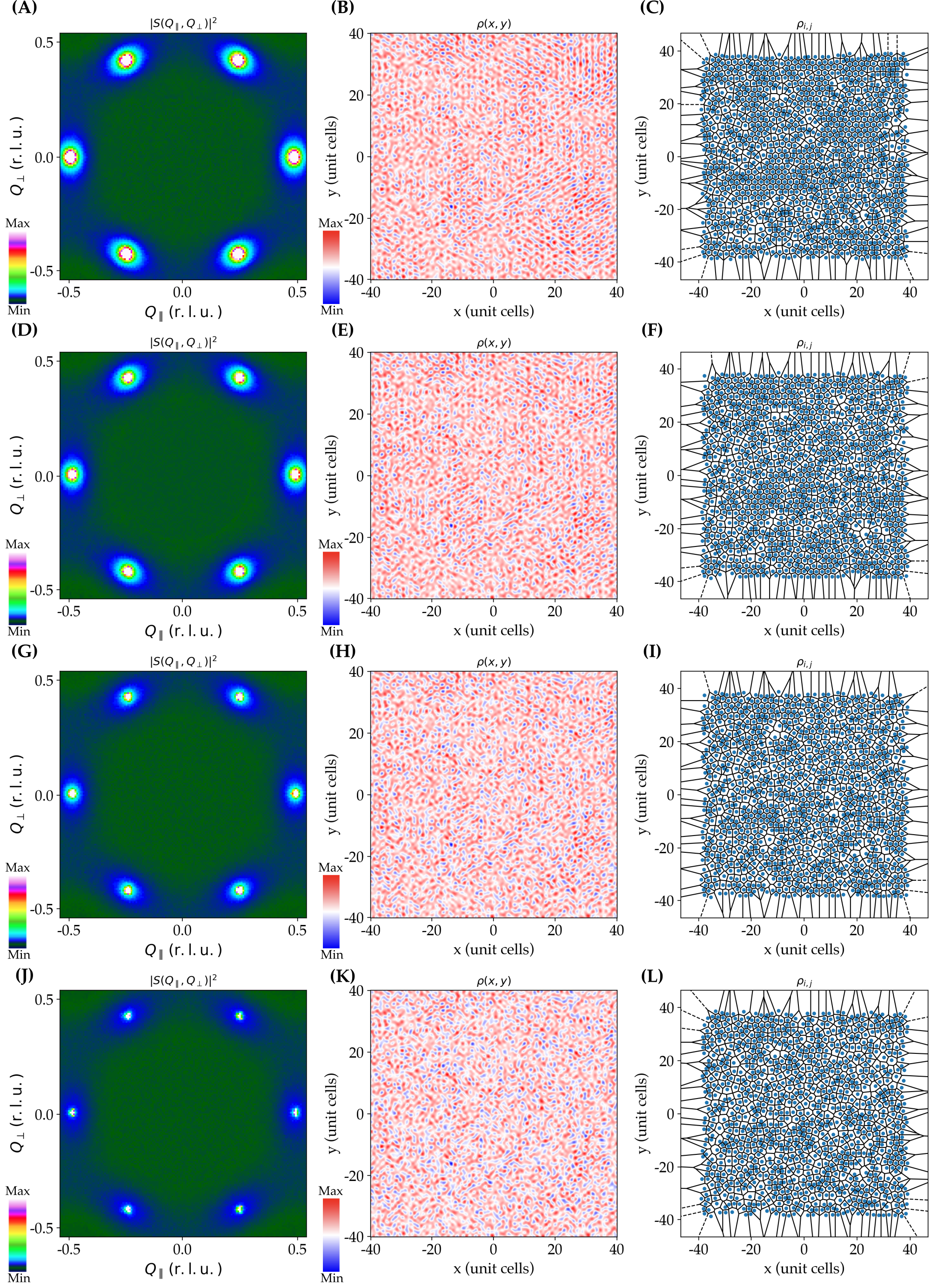}
    \caption{(A,D,G,J) Diffuse Scattering maps of the plane (H K 2.0) around the point (0 0 2.0) for $T = 95,\,97,\,99,\,101\,\mathrm{K}$, respectively. The Bragg point just in the point (0 0 2.0) has been screened by introducing random Gaussian noise with and average and standard deviation equal to the thermal background. (B,E,H,K) Real-space charge distribution for the diffuse scattering maps of (A,D,G,J), respectively. (C,F,I,L) Voronoi tessellation of the local maxima of the charge distribution in (B,E,H,K), respectively.
    }
    \label{Fig:DS_FT_exp}
\end{figure*}

Figure \ref{Fig:DS_FT_sim} summarizes the the real space charge density considering different types of CDW peak shape/anisotropy for a 6-fold symmetry. In figure \ref{Fig:DS_FT_sim} (A,C,E), where the CDW peaks are rather sharp, we can discretize the charge density to perform a Voronoi analysis, however, at high temperature as is the case of figure \ref{Fig:DS_FT_sim} (I), the transformation from a continuous field to a discrete set of particles, where there is no well-defined six-fold symmetry, the charge density is well capture by means of a discretization analysis. Figure \ref{Fig:DS_FT_exp} (A,D,G,J) correspond to the experimental maps of diffuse scattering we have used to perform the $G_6$ analysis in the Figure 4 of main text. For each one of these DS we have performed the Fourier transform to real space (see Figure (B,E,H,K)) via Eq.~\ref{eq:Rho}. After the discretization, we got the Voronoi tessellation and made the bond-order correlation function via  Eq.~\ref{eq:g6}.

\end{document}